\documentclass{advent}

\begin{document}
\setcounter{page}{1}

\chapter*{Preface}

\markboth{Adventures in Theoretical Physics}{Preface}

When I was asked by K. K. Phua to do a book for World Scientific based 
on my work, he suggested a volume of essays or a reprint volume. I have 
decided to combine these two suggestions into one, by preparing 
a reprint volume with commentaries.  Some of the commentaries are  
drawn from historical articles that I have written for publication, others 
are drawn from unpublished historical accounts written
for institutional archives, and yet others have been written 
expressly for this volume.  
In the commentaries,  I try to relate the reprinted articles to 
the time-line of my career, and at the same time to analyze their relations  
with the work of other physicists whose work influenced mine and vice versa.

In keeping with these dual aims, I have arranged the articles 
and the commentaries in approximately chronological order, but 
occasionally deviate from strict chronology in order to group topically 
related articles together.  In choosing which articles to include, I have 
been guided by two generally coinciding measures,  my own estimate of 
significance, and the citation count. However, in occasional cases I have 
included infrequently cited articles where I felt that there was an 
interesting related story to tell.  Often, when finishing a line of 
work, I have written a long summarizing article or review; some of these 
are too long to be included in their entirety, and so I have included in 
the reprints only the sections most relevant to the 
narrative in the commentaries.  Similarly, I have not included among the 
reprints the summer 
school lectures I have given on current algebras, anomalies, and neutrino 
physics, but references to them appear in the commentaries.  
In the last decade, I have published two books related to my work on  
generalized forms of quantum mechanics, and included many research   
results directly in these books in lieu of first writing papers.  It is 
feasible to give only brief descriptions of these projects in the  
commentaries; I have included just a few papers from this period,  
all in the nature of follow-ons to the first book.

In both the texts of  
the commentaries and the reference lists that follow them, reprinted articles 
are identified by a sans serif R, so that for example, {\sf R1} designates 
the first reprinted article.  Numbers in square brackets following each 
reference in the reference lists give the pages in the 
commentaries where that reference is cited.  There is also an index of 
names following the commentaries, and a list of detailed chapter 
subheadings in the Table of Contents.   

I wish to thank Tian-Yu Cao for a critical reading of the commentaries 
and much helpful advice, Alfred Mueller for a helpful conversation on 
renormalon ambiguities, Richard Haymaker for a clarifying email on 
dual superconductivity parameters, and William Marciano, Robert Oakes, and Alberto Sirlin 
for calling my attention to relevant references. I also wish to thank 
the following people for sending me helpful comments on the initial draft 
of the commentaries after it was posted on the archive as hep-ph/0505177: 
Nikolay Achasov, Dimi Chakalov, Christopher Hill, Roman Jackiw, Andrei 
Kataev, Peter Minkowski, Herbert Neuberger, and  Lalit Sehgal. I am grateful to Antonino Zichichi for permission to use the quote from Gilberto 
Bernardini in Chapter 2, to Mary Bell for permission to use the quote 
from John Bell in Chapter 3, to James Bjorken for permission use his 
quote in Chapter 3, and to Clifford Taubes for helpful email correspondence and permission to use his 
quotes in Chapter 7.    

 My editor at World 
Scientific, Kim Tan, has given valuable assistance throughout 
this project.  Miriam Peterson and Margaret Best have patiently assisted  
in the conversion of my TeX drafts to camera-ready copy and with indexing, 
the latter a task that was shared with  Lisa Fleischer and Michelle 
Sage. I am also indebted to Momota Ganguli and Judy 
 Wilson-Smith for bibliographic searches, to 
Christopher McCafferty and James 
Stephens for help with computer problems, and to Marcia Tucker and Herman 
Joachim for assistance, respectively, in scanning and duplicating certain 
of the 
papers to be reprinted.    Finally, I wish 
to express my appreciation to the  
Institute for Advanced Study (abbreviated 
throughout the commentaries as IAS) for its support of my work,  
first from 1966 to 1969, when I was a Long Term Member, and then 
from 1969 onwards, when I have been a member of the  Faculty, in  
the School of Natural Sciences.  My work has also been supported 
by the Department of Energy under Grant No. DE-FG02-90ER40542.  

In addition to the publishers acknowledged on each individual reprint, I
also wish to thank World Scientific for the use of material originally prepared for their volumes commemorating the 50th anniversary of Yang--Mills theory.  Chapter 3 on anomalies is largely based on an essay I contributed to  {\it 50 Years of Yang--Mills Theory}, edited by G. 't Hooft, and the parts of Chapters 7 and 9 dealing respectively with monopoles and projective group representations are based on an essay I wrote for a projected 
companion volume on the influence of Yang--Mills theory on mathematics.  
Also,  some material in Chapters 2 and 3 overlaps with the 
contents of a letter on antecedents of asymptotic freedom  
that I wrote to {\it Physics Today}, which appears 
in the September, 2006 issue.  
 
\newpage
\markboth{Adventures in Theoretical Physics}{Contents}
\vfill\eject
\centerline{\sf TABLE OF CONTENTS FOR THE COMMENTARIES}
\bigskip
\begin{description}
\item[1.]  Early Years, and Condensed Matter Physics
\item[] ~~~~~References for Chapter 1
\medskip
\item[2.]  High Energy Neutrino Reactions, PCAC Relations, and Sum Rules 
\item[] ~~~~~Introduction
\item[] ~~~~~Forward Lepton Theorem
\item[] ~~~~~Soft Pion Theorems
\item[] ~~~~~Sum Rules
\item[] ~~~~~More Low Energy Theorems; Weak Pion Production Redux
\item[] ~~~~~References for Chapter 2
\medskip
\item[3.] Anomalies: Chiral Anomalies and Their Nonrenormalization, 
Perturbative Corrections to Scaling, and Trace Anomalies to all Orders
\item[] ~~~~~Chiral Anomalies and $\pi^0 \to \gamma \gamma$ Decay
\item[] ~~~~~Anomaly Nonrenormalization
\item[] ~~~~~Point Splitting Calculations of the Anomaly 
\item[] ~~~~~The Non-Abelian Anomaly, Its Nonrenormalization and Geometric 
Interpretation
\item[] ~~~~~Perturbative Corrections to Scaling
\item[] ~~~~~Trace Anomalies to All Orders
\item[] ~~~~~References for Chapter 3
\medskip
\item[4.] Quantum Electrodynamics
\item[] ~~~~~Introduction
\item[] ~~~~~Strong Magnetic Field Electrodynamics: Photon Splitting and Vacuum 
Dielectric Constant
\item[] ~~~~~The ``Finite QED'' Program via the Callan--Symanzik Equations
\item[] ~~~~~Compactification of Massless QED and Applications
\item[] ~~~~~References for Chapter 4
\medskip
\item[5.] Particle Phenomenology and Neutral Currents
\item[] ~~~~~Introduction
\item[] ~~~~~Visits to Fermilab
\item[] ~~~~~Neutral Currents
\item[] ~~~~~References for Chapter 5
\medskip
\item[6.] Gravitation
\item[] ~~~~~Introduction
\item[] ~~~~~First Papers 
\item[] ~~~~~Einstein Gravity as a Symmetry Breaking Effect
\item[] ~~~~~References for Chapter 6
\medskip
\item[7.] Non-Abelian Monopoles, Confinement Models, and Chiral 
Symmetry Breaking
\item[] ~~~~~Introduction
\item[] ~~~~~Non-Abelian Monopoles 
\item[] ~~~~~Confinement Models
\item[] ~~~~~Chiral Symmetry Breaking
\item[] ~~~~~References for Chapter 7
\medskip
\item[8.] Overrelaxation for Monte Carlo and Other Algorithms
\item[] ~~~~~Introduction
\item[] ~~~~~Overrelaxation to Accelerate Monte Carlo
\item[] ~~~~~Image Normalization
\item[] ~~~~~References for Chapter 8
\medskip
\item[9.] Quaternionic Quantum Mechanics, Trace Dynamics, and Emergent 
Quantum Theory
\item[] ~~~~~Introduction
\item[] ~~~~~Quaternionic Quantum Mechanics
\item[] ~~~~~Quaternionic Projective Group Representations
\item[] ~~~~~Trace Dynamics and Emergent Quantum Theory
\item[] ~~~~~References for Chapter 9
\medskip
\item[10.]  Where Next?
\end{description}
\vfill\eject

\chapter*{1. Early Years, and Condensed Matter Physics}

\markboth{Adventures in Theoretical Physics}{Early Years, and Condensed Matter Physics}


A brief synopsis of my career appears in an article that I wrote recently 
for the Abdus Salam International Centre for Theoretical Physics (Adler, 2004, {\sf R1}),  
which includes a description of events when I was young that led to my 
becoming a theoretical physicist.  The focus of this article is on the   
career path that led to my work in high energy physics. However, before 
I published anything in high energy theory, I spent several summers working 
in industrial research laboratory jobs in condensed matter physics, and 
it was this work that led to my first scientific publications.  

By the end of my junior year at Harvard, I had taken courses in quantum 
mechanics and also in condensed matter physics (then called 
solid state physics).  
With this background, during the summer of 1960, 
I got a job working for Joseph Birman, who at that time (before going on to  
Professorships at New York University and then  
City College of the City University of New York) headed a section 
studying electroluminescence at the General Telephone 
and Electronics (GT\&E)  
Research Laboratory. This industrial research laboratory, formerly the  
Sylvania Research Laboratory, was conveniently located a few miles from 
where my family lived in Bayside, Queens.  I had a desk in an office 
looking out over 
the entrance to the Long Island Sound, from which I could see sections of 
roadway being hoisted 
into place on the Throgs Neck Bridge, then under construction.  

During my first weeks at GT\&E, Joe got me started learning   
some basic group theory as applied to crystal structures, and then suggested 
the problem of using these group theory methods to check a formula that Hopfield
(1960) had given relating band theory structures in 
hexagonal and 
cubic variants of zinc sulfide (ZnS) and related compounds, substances that   
Joe had been studying (Birman, 1959) with an eye to  
electroluminescence applications.  
This turned out to be basically a technical  
exercise and confirmed Hopfield's results.  
In the course of this work, which I finally wrote up a  
year later (Adler, 1962a, {\sf R2}),  I also attempted an {\it a 
priori} estimate of a parameter determined by experimental fits to 
the Hopfield formula. This 
got me interested in the Ewald sum method for doing crystal lattice sums, 
on which I wrote a paper (Adler, 1961) giving generalized results for sums  
over lattices of functions $f(r) Y_{\ell m}(\theta,\phi)$, with 
$Y_{\ell m}$ a spherical harmonic and $f(r)$ a radial function representable 
as a transform by $f(r)=r^{\ell} \int_0^{\infty}\exp(-r^2t)g(t) dt$.   
These two 
pieces of work stemming from my summer at GT\&E 
were my first  scientific publications.  With Joe's 
encouragement, I also gave a 10 minute contributed paper (Adler and Birman, 1961) on the 
ZnS work at the New York meeting of the American Physical Society the 
following winter, while I was back home on inter-term break from college.  
Since this was my first conference talk, I typed out a text and went  
over it so many times that I knew it by heart. After my talk, Joe said words 
to the effect, ``That was fine, but next time you give a talk don't 
sound like it was memorized'', wisdom that I have taken to heart on many 
subsequent occasions!  

When I returned to Harvard for my senior year I was told by some of the 
faculty that Henry Ehrenreich from the General Electric (GE) 
Research Laboratory 
was on leave at Harvard that year, and was giving the graduate course 
on solid state physics, covering substantially different material from what 
I had heard the year before.  I attended Henry's lectures, which included  
a calculation of the energy and wave-number dependent dielectric constant 
in isotropic solids, using the self-consistent field or energy-band 
approximation, along 
the lines of the treatment given in Ehrenreich and Cohen (1959). 
I got to 
know Henry outside the classroom as well, and he invited me to work at the 
GE Research Laboratory in Schenectady, NY the following summer, 
after my graduation from college in June 1961.  
This was appealing in a number of ways, 
since my family had moved to Bennington, VT the year before, about an 
hour's drive away from Schenectady, and so I was able to drive home for a visit 
on weekends.  At GE, Henry suggested that I generalize the treatment of the  
dielectric constant 
that he and Cohen had given so as  
to include various effects of interest in real solids.  In 
the paper that resulted (Adler, 1962b, {\sf R3}), I calculated 
the full frequency 
and wave-number dependent dielectric tensor in the energy-band approximation, 
including tensor components that couple 
longitudinal and transverse electromagnetic disturbances, which are absent 
in the isotropic approximation but are present even in solids with cubic 
symmetry.  The longitudinal to longitudinal component of the general 
dielectric tensor reduces to the result obtained by Ehrenreich and Cohen when  
various identities (reflecting charge conservation and gauge invariance, as well as symmetries) 
are used.  I also gave a method, based on an analysis of 
``Umklapp'' processes that couple wave numbers differing by a reciprocal 
lattice vector, together with  use of a  multipole expansion, 
for calculating local field corrections to the dielectric constant, giving 
a modified Lorenz--Lorentz formula.  \big(Local field corrections were also 
studied by Cohen's student Nathan Wiser (1963) by a 
different method.\big)  My paper on the dielectric constant in real solids 
has been widely 
cited in the subsequent condensed matter literature, reflecting its relevance 
for spectroscopic studies of solids, as well as its generalizations to 
nonlinear dielectric behavior.

Although I had 
decided to focus on elementary particle theory for my graduate study in 
Princeton, I retained an interest in solid state physics, and  
returned to GE for half of the summer of 1962 to work again with 
Henry Ehrenreich, this  time publishing a paper (Adler, 1963) in which I 
applied the dielectric 
constant results of the previous summer to the theory of hot electron energy
loss in solids. Not long after this visit, Henry left GE to take a 
Professorship at Harvard, where our paths crossed again during my   
postdoctoral years.  
After finishing my PhD at Princeton in 1964, I spent  
the summer working at Bell Telephone Laboratories in Murray Hill, under 
the supervision of Phil Anderson and Dick Werthamer.  However, aside from informal notes on the application of raising and lowering operators 
to the vortex structure in type II superconductors, my principal 
publication resulting from this 
final industrial summer job was a writeup of my work on PCAC consistency 
conditions, which I will discuss in the next chapter. 

\section*{References for Chapter 1}

\begin{itemize}

\item Adler, S. (1961).  A Generalized Ewald Method for Lattice Sums.  {\it Physica} {\bf 27}, 1193-1201. 

\item Adler, S. L. (1962a) {\sf R2}.  Theory of the Valence Band Splittings at $k=0$ in Zinc-Blende and Wurtzite Structures. {\it Phys. Rev.} {\bf 126}, 118-122.

\item  Adler, S. L. (1962b) {\sf R3}.  Quantum Theory of the Dielectric Constant in Real Solids. {\it Phys. Rev.} {\bf 126}, 413-420. 
 
\item  Adler, S. L. (1963).  Theory of the Range of Hot Electrons in Real Metals. {\it Phys. Rev.} {\bf 130}, 1654-1666. 

\item  Adler, S. L. (2004) {\sf R1}.  From Elements of Radio to Elementary Particle Physics, in {\it One Hundred Reasons to be a Scientist} (The Abdus Salam International Centre for Theoretical Physics, Trieste), pp. 25-26. 

\item  Adler, S. and J. L. Birman (1961).  An LCAO Theory of the $\vec k=0,0,0$ Valence Band Splittings in Zinc Blende and Wurtzite Structures. {\it Bull. Am. Phys. Society} Series II, Vol. 6, No.1, Part 1, p. 22. 

\item  Birman, J. L. (1959).  Simplified LCAO Method for Zincblende, Wurtzite, and Mixed Crystal Structures. {\it Phys. Rev.} {\bf 115}, 1493-1505. 

\item  Ehrenreich, H. and M. H. Cohen (1959).  Self-Consistent Field Approach to the Many-Electron Problem. {\it Phys. Rev.}  {\bf 115}, 
786-790. 

\item  Hopfield, J. J. (1960).  Fine Structure in the Optical Absorption Edge of Anisotropic Crystals. {\it J. Phys. Chem. Solids} {\bf 15}, 97-107. 

\item  Wiser, N. (1963).  Dielectric Constant with Local Field Effects Included. 
{\it Phys. Rev.} {\bf 129}, 62-69. 
 
\end{itemize}

\chapter*{\centerline{2. High Energy Neutrino Reactions, PCAC Relations,}
\centerline{and Sum Rules}}

\markboth{Adventures in Theoretical Physics}{High Energy Neutrino Reactions, PCAC Relations, and Sum Rules}

\section*{Introduction} 

By the end of my undergraduate years at Harvard (1957-1961), I had gone 
through most of the graduate
course curriculum, as well as a senior year reading course organized by
Paul Martin for my classmate Fred Goldhaber and me. This course gave me an 
introduction to
quantum field theory, or more precisely, to quantum electrodynamics, through
some of the seminal papers appearing in the reprint volume edited by 
Schwinger (1958).  
Although as a result of my summer research jobs I could have
gone on relatively easily to a PhD in solid state physics, I wanted 
to enter particle physics, and moreover wanted 
exposure
to styles of theoretical physics different from those I had seen already at
Harvard.  Hence I decided on Princeton for my graduate work (with strong 
encouragement from Harvard faculty member Frank Pipkin, who was an 
enthusiastic Princeton graduate alumnus), and enrolled there 
in the fall of 1961.

My first year there was spent preparing for general exams, mostly by reading.
I also participated in a seminar organized by the graduate students, which
surveyed many aspects of dispersion relations and covered 
some topics in Feynman
diagram calculations as well. The only formal course I took was one given
by Sam Treiman, which gave an introductory survey to elementary particle physics.
I was impressed by the clarity of his approach, and both because of this
and because Murph Goldberger was planning a sabbatical leave the following year,
I asked Treiman to take me on as a thesis student.  

This turned out to be a fortunate choice.  Treiman proposed that I do a
thesis in the general area of high energy neutrino reactions, which was
just then emerging as an area of phenomenological interest.  After doing
a survey of the literature in the field, I first did a ``preliminary problem''
of calculating the final lepton and nucleon
polarization effects in the quasielastic neutrino reaction $ \nu_{\ell} + N
\rightarrow \ell + N$, with all induced form factors retained in the vector
and axial-vector vertices (Adler, 1964a).  I did this calculation 
in two ways, first by 
using the covariant form of the matrix element and Dirac $\gamma$ matrix 
algebra, then by using the 
center of mass form and Pauli matrix algebra, and directly checked the 
equivalence of the two forms of the answer.  
This convinced Treiman that I could calculate, and
incidentally introduced me to the axial-vector current and coupling $g_A$ 
which were to be central to my work for many years.

After this calculation was completed, I decided to make the main focus
of my thesis a calculation of the simplest inelastic high energy neutrino
reaction, that of pion production in the (3,3) or $\Delta(1232)$ 
resonance region.  
This problem had the appeal of having as a paradigm the beautiful dispersion
relations calculation of pion photoproduction of Chew, Goldberger, Low, and
Nambu (1957), which was one of the classics of the dispersion relations
program.  An extension to electroproduction had already been carried
out by Fubini, Nambu, and Wataghin (1958), but they had done no 
numerical work,
and on closer examination their matrix element turned out to be divergent
at zero hadronic momentum transfer  $\nu_B$ when the lepton four-momentum
transfer squared denoted by $q^2$ (or $k^2$) is nonzero.  There were 
similar problems (surveyed in 
my thesis) with the other
papers then available dealing with pion electroproduction or weak production,
so doing a complete and careful calculation, including numerical evaluation
of the cross sections, seemed a good choice of thesis topic.  It was also a
demanding one; although I wrote my thesis and got my degree in 1964, my goal
of a complete calculation, including the necessary computer work, was not
achieved until 1968. 

Much of the delay though, was a result of the fact that weak
pion production turned out to be a marvelous theoretical laboratory for
studying the implications of conservation hypotheses for the weak vector and
axial-vector currents, and this became a parallel part of my research program,
as reflected in the title of my thesis ``High Energy Neutrino Reactions and
Conservation Hypotheses'' (Adler, 1964b).  From Treiman and from my reading,
I had learned about the Feynman--Gell-Mann (1958) proposal of a hadronic 
conserved vector current (CVC), and I had also learned about the 
Goldberger--Treiman (1958) relation for the charged pion decay constant, which 
they had discovered through
a pioneering dispersion theoretic calculation of the weak vertex.  A 
simplified
derivation of this relation
had already been achieved through the suggestion
of Nambu (1960), Bernstein, Fubini, Gell-Mann, and Thirring (1960), 
Gell-Mann and L\'evy (1960), and Bernstein, Gell-Mann, and Michel (1960),
that the axial-vector current is
partially conserved, in the sense that the divergence of the
axial-vector current behaves at small squared momentum transfer 
as a good approximation
to the pion field, or equivalently, is pion pole dominated.  
\big(Much later on, after contacts with China 
resumed, I learned that Chou (1960) 
had given a similar simplified derivation of the Goldberger--Treiman relation,
as well as
further applications to decay processes.\big)
The partial conservation hypothesis was an appealing one,
but as Treiman kept emphasizing, it was supported by ``only one number'' and
therefore had to be regarded with caution.  So a second goal of my thesis
work ended up being to keep an eye out for other possible tests of the
conservation hypotheses for the weak vector and axial-vector currents.

Before going on to discuss how these emerged from my weak pion production 
calculation, let
me first recall what I knew when I started the thesis work. The first chapter
of the thesis (written in the spring of 1964) 
was a theoretical survey; in the section  headed ``Partially Conserved
Axial Vector Current (PCAC)'' I referred only 
to the papers of Goldberger and Treiman, of Nambu, of Bernstein et al., and
of Gell-Mann and L\'evy
cited in the preceding
paragraph.  In the final section of the first chapter, entitled ``Survey
of Computations Relating to Specific Reactions''  there is the following
reference to the paper of Nambu and Shrauner (1962), 
which was my reference 37:  ``An entirely different
approach to weak pion production in the low pion-energy region has been 
pursued by Nambu and Shrauner.$^{37}$  These authors assume that the weak 
interactions are approximately $\gamma_5$ invariant 
(``chirality conservation'').  They then obtain formulas for production of low
energy pions, in the approximation in which the pion mass is neglected, in
analogy with the treatment of low energy bremsstrahling (sic) in electron
scattering.''  At the time I started my calculations, neither Treiman nor I
understood 
the relation between the Nambu--Shrauner work and the issue of partial
conservation of the axial-vector current.  
This was partly because we were suspicious
of the assumption of zero pion mass, and partly 
because the Nambu--Shrauner paper
makes no reference to the axial-vector coupling $g_A$,
so it was not clear whether their ``chirality'' was 
related to the weak currents I was studying
in my thesis. This second point is particularly significant, 
and I will return to it in 
considerable detail below.  I was not able to determine from 
my files (by finding either a reference in my notes or a Xerox copy) 
when I first read the Nambu--Luri\'e (1962) paper  
on which the Nambu--Shrauner paper
was based, but it was probably a year later,
in early 1965.
\bigskip
\section*{Forward Lepton Theorem}

Roughly the first year and a half of my thesis work on weak pion production 
was spent mastering the
formal apparatus of Lorentz invariant amplitudes (used for writing dispersion
relations) and center of mass multipole expansions (used for implementing
unitarity) and the transformations between them, the Born approximation
structure, cross section calculations, 
etc.  Then in the winter of 1963-1964 or the spring of 1964 (I can
only establish dates approximately by the sequence of folders, 
since I did not date them),
I began noticing things that
transformed a hard and often dull calculation into a very interesting one
(just in the nick of time, since I was due to finish in June of 1964 and had
already accepted a postdoctoral position at the Harvard Society of Fellows
starting in the fall semester.)

The first thing I noticed was that at zero squared leptonic four
momentum transfer, my expression for the weak pion production 
matrix element reduced to
just the hadronic matrix element of the divergence of the axial-vector current,
which by the partial conservation hypothesis is proportional to the amplitude
for pion-nucleon scattering.  I then tried to abstract something more general
from this specific observation, and soon had a neat theorem showing that
in a general inelastic high energy neutrino 
reaction, when the lepton emerges forward
and the lepton mass is neglected, the leptonic matrix element is proportional
to the four momentum transfer; hence when the leptonic matrix element is 
contracted with the hadronic
part, the vector current contribution vanishes by CVC, and the 
axial-vector current contribution reduces by partial conservation (for which
I coined the parallel acronym PCAC, which has become 
standard terminology) to the corresponding matrix
element for an incident pion.  Thus inelastic neutrino reactions 
with forward leptons
can be used as potential tests of CVC and PCAC; this became a chapter of
my thesis and was written up as a paper (Adler, 1964c, {\sf R4}) as soon 
as my thesis was completed.  
The paper on CVC and PCAC tests was the first of three papers in which I 
found connections between high energy neutrino scattering reactions and 
properties of the weak currents; the other two were my long paper on 
the $g_A$ sum rule, and a paper on neutrino reaction tests of the local  
current algebra, both of which are reprinted in this volume and  
will be discussed shortly.  

To determine whether the CVC/PCAC test  
could be implemented experimentally, I wrote a letter to the neutrino 
experimentalists at CERN.    After a few months 
I received a charming
reply from Gilberto Bernardini, who commented ``The delay of this answer, for which 
I apologize very much, is due to two facts.  The first is the known time 
diagram of the `modern physicist'.  In case you do not know it yet, I plot 
it here: (Diagram with a vertical time axis and an upwards pointing arrow; 
`work' at the bottom, 
`travel \& meetings' in the middle, and `dinners \& ceremonies' at the top.)
Unfortunately, according to my age, I am already very much in the central 
region and even higher.''  Bernardini then went on to say that Antonino 
Zichichi had 
brought my paper to his attention a couple of weeks before, and then continued 
with an analysis of technical problems in executing my proposal.  There 
followed a further exchange of letters with Bernardini, with theorist John 
Bell, and with experimentalists Guy von Dardel and Carlo Franzinetti.  Of  
particular note, von Dardel wrote me a long letter after he read my 
paper, remarking that the care with which he read it was partly due to a 
skiing accident that had kept him in bed with a broken leg and nothing 
better to do, and giving a formula that he had worked out, during 
his enforced time away from  
experimental activities,  for corrections to 
my theorem when the lepton emerges at a small angle to the forward direction.  
This formula turned out to be not quite right (there was an incorrect energy 
factor), but started me thinking 
about the issue, which I discussed with John Bell when I attended an 
Informal Conference on Experimental Neutrino Physics at CERN, January 
20-22, 1965.  Bell had redone the calculation of the pion exchange 
contribution to the small angle correction by splitting the amplitude into 
spin-flip and non-spin-flip parts, getting a result that turned out also 
to be not 
quite right (there was a factor of 2 off in one term).  When I got back 
to Harvard I repeated the calculation, according to my notes, by the  
``Bell method'', and also by a covariant method, and got a formula  that 
I never published, but conveyed in letter of Feb. 10, 1965 to Bell 
(with copies to Bernardini, Block, von Dardel, Faissner, Franzinetti, 
and Veltman, most of whom I had talked with when I was at CERN).  
The corrected 
small angle formula states that the first factor on the second line of 
Eq.~(16) of {\sf R4} should be replaced by 
\begin{equation}
\left[1-\frac{m_{\ell}^2 k_0}{2k_{20}(k^2 + M_{\pi}^2)}\right]^2
+\left[\frac{m_{\ell}k_0\theta}{2 (k^2+M_{\pi}^2)}\right]^2~~~,\nonumber
\end{equation}
with $k^2= m_{\ell}^2 k_0/k_{20} + k_{10}k_{20}\theta^2$ the leptonic
four-momentum transfer squared and with 
$\theta$ the lepton-neutrino polar angle, assuming that the lepton-neutrino 
azimuthal angle has been averaged over.  

Even before my visit to CERN,  Bell (1964) had noted 
that when one considers my 
forward lepton formula in the context of nuclei, ``the following difficulty 
presents itself:  Because of absorption, pion cross sections depend on the 
size of large nuclei roughly as $A^{2/3}$.  But neutrinos penetrate to all 
parts of nuclei; for them cross sections should contain at least a part 
proportional to $A$. This indicates for large nuclei a critical dependence 
of $\sigma(W, -q^2)$ on $q^2$.''  Bell proceeded to use optical model methods
to discuss this ``shadowing effect'', which has continued to be of interest 
over the years.  It took many years for my forward lepton formula, and Bell's
shadowing observation, to be experimentally verified; for   
a survey of the status of both, and further references, see the recent 
conference talk by Kopeliovich (2004).  An earlier review of Mangano et al.    
(2001) also discusses the experimental status of shadowing, and a good    
exposition of the theory is given in the review of Llewellyn Smith (1972).
For specific applications of the forward lepton formula to exclusive 
channels, see Ravndal (1973) and Rein and Sehgal (1981) for 
$\Delta(1232)$ production, and Faissner et al. (1983) for coherent $\pi^0$ 
production (which was used to determine the coupling strength of the isovector 
neutral axial-vector current).  Also, Sehgal (1988) and Weber and 
Sehgal (1991) discuss an interesting analog of the forward lepton theorem for 
purely leptonic neutrino-induced reactions.  
 
\bigskip 
\section*{Soft Pion Theorems} 

Returning now to my thesis work in the spring of 1964, 
the second thing that I noticed, again working from my explicit expression
for the weak pion production amplitude, was that when I imposed the PCAC
condition at zero values of the hadronic energy variable $\nu$ and the
hadronic momentum transfer variable $\nu_B$, only the Born approximation
pole term coming from the nucleon intermediate state 
contributed; all of the model dependent 
parts of the weak amplitudes dropped out.  Thus I got what I called a
``consistency condition'' on the pion-nucleon 
scattering amplitude $A^{\pi N (+)}$, implied
by PCAC, taking the form  
$$g_r^2/M=A^{\pi N (+)}(\nu=0,\nu_B=0,k^2=0)/K^{NN\pi}(k^2=0),$$
with $g_r$ the pion-nucleon coupling constant, $M$ the nucleon mass, 
$-k^2$ the squared mass of the initial pion  (the final pion is still
on mass shell), and with $K^{NN\pi}(0)$ the pionic form factor 
of the nucleon, normalized so that $K^{NN\pi}(-M_{\pi}^2)=1$.  
This seemed absolutely remarkable, and I immediately proceeded to do a 
dispersion relation evaluation of the pion-nucleon amplitude on the right, 
using the Roper (1964) phase shift analysis as input, and assuming that the  
effects of off-shell continuation in $A^{\pi N (+)}$ (as well as in 
$K^{NN\pi}$) were small.  In setting up this calculation, I used several  
theoretically equivalent ways of writing the subtracted dispersion relation 
to get an estimate of the errors in the analysis.  
The Christenson--Cronin--Fitch--Turlay (1964)  
experiment on CP violation had a substantial block of 
computer time reserved  
for analysis, and courtesy of them I was able to use a small amount of their 
time to run my programs, a few days before I was scheduled to give a talk 
at Columbia. I recall staying up all night to get the job done, and at one 
point, in the wee hours of the morning, dropping my deck of cards and then 
having to spend precious time getting them back in the proper order.  But I 
did get my calculation done by morning  (and never again attempted an 
``all-nighter''.)       
The relation worked very well, and
as Treiman later said, ``now there is a second number''; PCAC was starting 
to look interesting.  This work became the final chapter of my thesis.

Immediately after finishing my thesis I took a summer job at Bell Laboratories
at Murray Hill, nominally working for Phil Anderson.  I wanted to learn about
superconductivity, and Phil assigned me to work for Dick Werthamer.  I did
learn about the BCS and Ginzburg--Landau theories, and Abrikosov vortices in  
type-II superconductors, 
but I did not succeed in
my project with Dick, which was to try to understand the resistance to  
vortex line motion using thermal Green's functions.  A few weeks before
the end of the summer, I asked for and got Phil's permission to spend some
time writing a paper on the pion-nucleon consistency condition 
(Adler, 1965a, {\sf R5}),
which I also then extended to pion-pion and pion-lambda scattering. 
In the pion-pion case, since there are no pole terms, the consistency condition takes
the form that the pion-pion scattering amplitude with one zero mass pion, 
evaluated  at the symmetric point
$s=t=u=M_{\pi}^2$, is {\it zero}. This was the first example of a 
soft pion zero or, as termed in the literature, ``Adler zero'', 
in non-baryonic amplitudes, that I will return to shortly.     
Knowing that
I was planning to go on in particle theory, Phil told me one day that
he had an interesting  paper
to show me, which had just been submitted to the journal {\it Physics} 
which he was
editing.  It was Gell-Mann's (1964) paper on current algebra; Phil let me
read it, but not Xerox it.  This was to prove decisive for my work on sum 
rules nine months later.   My interactions with Phil however were brief, and 
never touched on the subject of symmetry breaking in superconductivity  
and particle physics, on which Phil had written a paper  (Anderson, 1963) 
that I learned of only many years later, that was a forerunner 
of work on the ``Higgs mechanism'' for giving masses to vector bosons.  

In the fall of 1964 I moved to Harvard as Junior Fellow in the Society of 
Fellows, sharing a postdoc office next to the
office occupied by Henry Ehrenreich in the Applied Physics division.  
(Henry had recently left General Electric to accept a 
Professorship at Harvard.)
In principle I was going to do solid state physics as well as particle
theory, but that never happened.  I spent the fall term working on numerical
aspects of my weak pion production calculation, 
and also reading papers on attempts to
calculate the axial-vector renormalization constant $g_A$, including the
papers of Gell-Mann and L\'evy (1960) and Bernstein, Gell-Mann and 
Michel (1960).
I had a hunch that
the fact that $g_A$ is near one was somehow connected with PCAC, but I 
did not see 
a concrete way of exploiting PCAC in a calculation.  I also was 
starting to think
about how to make the PCAC consistency condition calculations independent
of the cumbersome Lorentz invariant amplitude apparatus that I 
had used to get them.  I soon found
that the relevant terms could be isolated directly from the Feynman diagrams 
without invoking all the formal kinematic
apparatus of my thesis, and this approach extended to a 
general matrix element
as well; the strategy was the same one that I had used in the paper on 
CVC and PCAC tests, of going from a particular observation 
in the context of my 
weak pion production calculation to something more general.  
The result was a formula for soft pion production, in terms of external 
line insertions on
the hadronic amplitude for the same process in the absence of the pion (Adler, 
1965b, {\sf R6}).
For baryons of nonzero isospin, the insertion factors are nonzero, while 
for isospin zero baryons, and mesons such as the pion or kaon, the insertion 
factor vanishes.  This latter result generalizes the soft-pion zero or 
``Adler zero'' to the emission of a soft pion in any reaction involving 
only incoming and outgoing mesons, but no external baryons.  These zeros 
continue to play a role in the analysis of experimental results on 
mesonic resonances; for recent discussions, see Bugg (2003, 2004) and   
Rupp, Kleefeld, and van Beveren (2004).  

The soft pion zeros are an indication that 
according to PCAC, the pion coupling to other hadrons is effectively 
pseudovector, and not pseudoscalar.   When I visited CERN in late January 
of 1965, while in the midst of work on the Feynman diagram approach to the 
PCAC consistency conditions, I found that Veltman had been thinking in a  
similar direction, but had not reached the point of writing down external 
line insertion rules.  Veltman gave me a one page memo to file 
that he had written, which pointed out   
that my PCAC consistency conditions are equivalent to pseudovector coupling, 
which implies the vanishing of invariant amplitudes for soft pion emission 
after singular terms are split off.      
Veltman also noted that Feynman had briefly remarked on the 
relation between the Goldberger--Treiman relation  and pseudovector 
coupling in his conference summary talk at Aix-en-Provence in 1962, and gave 
me a copy of the relevant page.   
Feynman did not, however, report agreement with experiments on pion-nucleon 
scattering, apparently because he did not recognize the necessity of 
splitting off the singular Born terms before concluding that pion emission 
amplitudes vanish in the soft pion limit.   

In the course of my work on the insertion rules 
I remembered the paper of Nambu and Shrauner (1962) which I had
briefly mentioned in the Introduction to my thesis; I now looked this up, 
as well as the paper of Nambu and Luri\'e (1962) on which 
it was based, and saw
that my final formula, when specialized to the 
case of an ingoing and outgoing nucleon line,  
was substantially the same as the pion bremsstrahlung
formula of Nambu and Luri\'e.  I noted this in my paper,  
and consistently referred to the Nambu papers from this point on.
In recognition of Nambu's work, I used his notation $\chi$ and term
``chirality'' to refer to the integrated
axial-vector charge in my next two papers, which dealt 
with the $g_A$ sum rule;
however, in modern terms this is a misnomer, since chirality is now used
to mean the left- or right-handed sums of vector and axial-vector charges.  
Gell-Mann's notation for the axial-vector charges has become the standard
one, and after these two papers I followed the Gell-Mann notation.

The comparison with Nambu's approach also raised the issue of the role of 
the pion mass: do the PCAC results limit smoothly to the zero pion mass ones,
for which the soft pion theorem derivations appear quite different?  
This point
was dealt with in footnote 6 of my paper {\sf R6}, 
where I showed that the limits,  (1) pion mass approaches zero, and (2) pion
four momentum transfer squared approaches zero, can be taken in either order;
the same soft pion theorem results, although the contribution which comes
from the massless pion pole when the limit (1) is taken first, comes instead
from the axial-vector divergence when the limit (2) is taken first.  
This point is now taken for granted, but in the early years it caused 
me (and others) considerable confusion.
After this paper I almost immediately got involved with sum rules, and 
so I did
not publish the detailed connection between my second PCAC paper and the
Nambu--Luri\'e approach until a few years later, when I included it as
``Appendix A'' of Chapter 2 of the book on {\it Current Algebras} which I
put together with Roger Dashen (Adler and Dashen, 1968). This appendix  
is reprinted here as {\sf R7}.  At the end of
Appendix A, I again discussed the relationship between the zero pion
mass and nonzero pion mass calculations.  The analysis of Appendix A also 
shows how the PCAC approach to soft pion theorems that I had developed 
fixes the undetermined renormalization constant appearing in the 
chirality approach of Nambu--Luri\'e.  In the formulas of
Appendix A, there are factors of  $g_A$ that 
are missing in the formulas of the papers of Nambu, Luri\'e, and Shrauner.
Correspondingly, in the paper of Nambu and Luri\'e, in the discussion 
associated with their Eq. (2.7), they noted that a renormalization constant $Z$ appears, but 
didn't observe that this can be precisely identified as $g_A$. 
Instead, they redefined 
their chirality as $Z^{-1} \chi$, that is as $(g_A)^{-1} \chi$.  They then 
made a compensating adjustment in the pion decay constant in their 
Eq. (4.5), where
they dropped the $g_A$ factor which appears in the 
Goldberger--Treiman relation.
Nambu and Luri\'e say there, ``$1/\lambda$ is more or less the 
conventional pion
coupling constant $1/\lambda=f=g/2m.~~~(4.5)$ It is not proven, however, that
this agrees with the coupling constant defined in the dispersion theory.  For 
the time being, we assume it to be the case.''  In the subsequent paper of 
Nambu and Shrauner (1962), an 
identification of $f$ with the standard pion-nucleon coupling was established,  
but the issue of where to include factors of $g_A$ was not addressed.    
My impression from this was that there was some 
uncertainty in the minds of Nambu and his students about how 
the chirality is to be normalized,   
and this impression was reinforced by a conversation I later 
had with Nambu about their work and my Appendix A derivation of their result.

In the low energy theorem for {\it one} soft pion which Nambu and Luri\'e had
derived, and which I had obtained 
from PCAC and the Feynman rules in my second PCAC paper, the
$g_A$ factor drops out, and so the normalization of the axial-vector 
charge or
``chirality'' is irrelevant.  The applications discussed in the papers of 
Nambu, Luri\'e, and Shrauner all involved only one 
soft pion; Nambu and Luri\'e looked,
for example, at $\pi + N \rightarrow \pi +N +\pi$, with the final pion soft
but with the other pions ``hard''; in fact, what they actually did was to
calculate single soft pion emission in the reaction $\pi + N \rightarrow
\Delta(1232)$.  Similarly, Nambu and Shrauner (1962) analyzed 
single soft pion
electroproduction and weak production, relating them to the form factors
of the vector and axial-vector currents. In this paper they included 
current commutator terms by analogy with the classic  Low (1958) paper on 
bremsstrahlung; their answer for electroproduction is
correct because the $g_A$ factor drops out there anyway, but their answer for
weak axial-vector production lacks $g_A$ factors in places, for reasons 
explained in the next paragraph.
A follow-up paper of
Shrauner (1963) dealt with single soft pion production in pion-nucleon 
scattering, with the scattering pions ``hard''.
My ``PCAC consistency condition'' was likewise a single soft pion theorem 
which gives a relation between the amplitude for 
$\pi + N \rightarrow N + \pi$, with the
final pion soft, and the amplitude $\pi + N \rightarrow N$, which is just the
pion-nucleon coupling constant, and involves no factors of $g_A$.    

The factors of $g_A$ and the explicit identification of the 
``chirality'' with the charge associated with the axial-vector 
current become important, however, 
if one wants to discuss {\it multiple} soft pion production, and also weak
axial-vector pion production, since one then
encounters commutators of an axial-vector charge with an axial-vector 
charge or current, 
which are evaluated by the
Gell-Mann current algebra. If one defines the relevant chirality as
$(g_A)^{-1}$ times the axial-vector charge, as is implicit 
in the Nambu--Luri\'e paper when one
identifies their $Z$ with $g_A$, then the
relevant commutator is $(g_A)^{-2}$ times a vector charge, which at zero
momentum transfer just gives $(g_A)^{-2}$.  This is in fact the origin of
the $(g_A)^{-2}$ term in the $g_A$ sum rule, where the difference between
$(g_A)^{-2}$ and 1 is highly significant.  The point, then, is that while
Nambu and Luri\'e gave a correct formula for single soft pion production, it
in fact cannot be generalized to multiple soft pion production (or soft pion
production by the weak axial-vector current) without first 
dealing carefully with the question of normalization, as I did in my second
PCAC paper {\sf R6} and in Appendix A of the 
book on current algebras {\sf R7}.  

Another difference between the work of Nambu and his students, and what 
I did in my first PCAC consistency condition paper {\sf R5}, related to 
the method of comparison with experiment, and the level of accuracy claimed 
for soft pion predictions. 
The Goldberger--Treiman relation is good to about 7\% accuracy, and 
my comparison of the PCAC consistency condition with experiment also 
indicated that the relation was satisfied  to about 10\%, thus reinforcing 
the idea that PCAC could be used as a quantitative tool for studying the 
strong interactions, with the residual errors arising from the   
extrapolation of the pion four-momentum squared $k^2$ from $M_{\pi}^2$ to 0.  
Given that the pion mass is much smaller than all other hadron masses, an 
extrapolation error $\sim M_{\pi}^2/M_{\rm hadron}^2 \leq 0.1$ is reasonable.    
The success of the $g_A$ sum rule shortly afterwards gave further support to 
the idea that PCAC gives quantitatively accurate predictions.  
Nambu, Luri\'e, and Shrauner, however, argued only for 
qualitative agreement between their soft pion results and experiment based   
on comparisons of rescaled angular distributions, but did not find anything close to 
$\sim 10\%$ agreement for absolute cross sections.
For example, for the relation between the cross sections 
for pion-nucleon scattering with production of an additional pion, and 
pion-nucleon scattering, Nambu and Luri\'e (1962) showed agreement with 
their predictions to 
within roughly a factor of three (giving a predicted cross section of 0.2 mb
versus experimental values in the range 0.6 to 0.7 mb).  Similarly, 
for the same reaction Shrauner (1963) found that 
``the magnitudes of the cross 
sections seem to be significantly underestimated by a factor of about 7''. 
The source of these discrepancies is not clear. 
 They may be  due, in part,  to the fact that, instead of 
testing the soft pion predictions at the kinematic point of zero pion four 
momentum (such as the point $\nu=\nu_B=0$ used in my PCAC consistency 
condition work), Nambu, Luri\'e, and Shrauner did the comparisons in energy intervals above 
scattering threshold. (However, Shrauner argues, on the basis of branching 
ratios,that the discrepancy is probably not attributable to an overlap 
of the  $\Delta(1232)$ resonance with the comparison region.)
I think that a combination of lack of clarity 
about how their chiral current was related to the physical axial-vector 
current, as reflected in the normalization problems noted above, 
together with the  lack of striking quantitative 
comparisons with experiment, were responsible for the work of Nambu and 
his students being largely unnoticed by the community.  It was only    
after the quantitative 
successes of PCAC in my consistency condition paper and in the $g_A$ sum 
rule that followed shortly afterwards, and my demonstration 
of the equivalence 
between the PCAC insertion rules and the chirality conservation approach, 
that the significance of the work of the Nambu group became clear.  

Finally, as an historical footnote to this discussion of soft pion theorems, 
Touschek (1957) appears to have been the first to introduce continuous 
$\gamma_5$ symmetry transformations, as applied to the neutrino field, 
and to observe that invariance under these transformations requires 
that the neutrino mass be zero.  
Nishijima (1959) (in work submitted for publication in late 1958)  
considered continuous $\gamma_5$ symmetry transformations in theories of   
massive fermions; to preserve $\gamma_5$ invariance he gauged the 
transformations with a massless pseudoscalar boson, transforming as $B \to 
B + \lambda$ under a $\gamma_5$ transformation with parameter $\lambda$. 
The action written in Nishijima's paper is just the effective action 
one would now write for a singlet Nambu--Goldstone boson (such as 
an axion) coupled to a massive fermion.  Nambu (1959), in remarks at  
the Kiev Conference, noted the analogy between $\gamma_5$ symmetry in 
particle physics and gauge invariance in superconductivity, and related this 
to his suggestion that a nucleon-antinucleon pair in a 
pseudoscalar state could be the pion. This idea was further developed  
in the well-known paper Nambu and Jona-Lasinio (1961), that  
laid the basis for the modern theory of Nambu--Goldstone bosons associated 
with spontaneous symmetry breaking, and for the fact that most of the mass   
of the nucleon comes from chiral symmetry breaking.  In the meantime, 
G\"ursey (1960) had introduced isovector $\gamma_5$ transformations, as 
an extension 
of the similar isoscalar transformations used by Nishijima, and constructed   
a precursor to nonlinear pion effective Lagrangians.  These papers all 
contained important seeds of our present-day understanding of chiral 
symmetries.    
\bigskip
\section*{Sum Rules}

I have now gotten ahead of the chronological story; a lot of things
happened very fast in 1965. In the fall of 1964 I started thinking about the  
question of the renormalization of the nucleon axial-vector coupling $g_A$,  
and accumulated a file of papers on the subject.  However, my attempts at a
calculation  were based on the commutator of the nucleon field with the 
weak axial-vector charge, giving results identical to those already obtained 
by Bernstein, Gell-Mann, and Michel (1960), which expressed $g_A$ in terms 
of unmeasurable off-shell form factors, but achieving no further progress.  
In early 1965 I saw a preprint of Fubini and
Furlan (published as Fubini and Furlan, 1965) 
which applied the commutator of vector current charges, together
with the ingenious idea of going to an infinite momentum frame, to calculate
the radiatively induced renormalization of the vector current.  
(Harvard did not have a preprint library in those days, but Schwinger's 
secretary Shirley 
would let me into his office from time to time to look through the unread 
preprints that were stacked on his desk.  This presented no difficulty since   
Schwinger was a night-owl who mainly worked at home, and used his office only 
a few hours a week, when he came in to lecture and to see students.   
That is how I became aware of the Fubini--Furlan paper. As a result of this 
experience, one of the first things 
I did when I arrived at the  Institute for Advanced Study eighteen months 
later was to start a preprint library for the particle physicists.)  
I immediately
thought about applying this to the axial-vector current, using the Gell-Mann
current algebra that I'd seen the previous summer at Bell Labs.  However,
because of other things I was working on I didn't get around to it until
a few months later, when in a chance encounter Arthur Jaffe told me that
he had heard a talk by Roger Dashen about work he and Gell-Mann had been
doing on sum rules. I decided I had better stop delaying (although it turned
out that Dashen and Gell-Mann were working on fixed momentum transfer sum
rules), dropped my weak pion production computer work, and spent spring
break working out the consequences of combining the Gell-Mann current algebra,
PCAC, and the Fubini--Furlan method.  It turned out to be surprisingly easy, 
with
the infinite momentum frame solving a problem I had encountered in earlier
attempts to calculate $g_A$, which is that the axial-vector 
charge matrix element
is proportional to the nucleon velocity, and vanishes for nucleons at rest.
I soon had a formula relating the difference between 1 and $(g_A)^{-2}$ to
a convergent integral over a difference of pion-nucleon cross sections, 
\begin{equation}
1-\frac{1}{g_A^2}=\frac{4 M_N^2}{g_r^2 K^{NN\pi}(0)^2}\frac{1}{\pi} 
\int_{M_N+M_{\pi}}^{\infty}\frac{W dW}{W^2-M_N^2}[\sigma_0^+(W)-\sigma_0
^-(W)]~~~,\nonumber
\end{equation}
with $M_{\pi}$ and $M_N$ the pion and nucleon masses, $\sigma_0^{\pm}(W)$ 
the total cross section for scattering of a zero-mass $\pi^{\pm}$ on a proton 
at center-of-mass energy $W$, and again 
with $K^{NN\pi}(0)$ the pionic form factor 
of the nucleon, normalized so that $K^{NN\pi}(-M_{\pi}^2)=1$.  
I first tried to saturate the integral in the narrow $\Delta(1232)$ approximation,
and the result was a disappointing $g_A=3$.  I then pulled out the computer
deck I had used for the consistency condition 
numerical work the previous year,
did the integral carefully, and got $g_A=1.24$.  I also observed that
the relation for $g_A$ could be equivalently recast as a two-soft pion low
energy theorem, 
\begin{eqnarray}
1-\frac{1}{g_A^2}&=& \frac{-2 M_N^2}{g_r^2 K^{NN\pi}(0)^2} G(0,0,0,0)~~~,\nonumber\\
G(\nu,\nu_B,M_{\pi}^i,M_{\pi}^f)&=& \nu^{-1} 
A^{\pi N (-)}(\nu,\nu_B,M_{\pi}^i,M_{\pi}^f) + 
B^{\pi N (-)}(\nu,\nu_B,M_{\pi}^i,M_{\pi}^f)~~~.\nonumber
\end{eqnarray}
Here $A^{\pi N (-)}$ and $B^{\pi N (-)}$ are the isospin-odd pion-nucleon 
scattering amplitudes, $\nu$ and $\nu_B$ are again the energy and momentum 
transfer variables, and $M_{\pi}^{i,f}$ are the initial and final pion 
masses, which are now both off shell.  
A few days after I submitted a letter to {\it Physical Review
Letters}, Sidney Coleman returned from a trip to SLAC and when I described
my results to him, he told me that he had just heard about a similar 
calculation being done there by Bill Weisberger, whose points of departure  
were the same as mine:  the Gell-Mann current algebra, the Fubini--Furlan 
paper, and my paper on PCAC consistency conditions.  I talked to Weisberger 
by phone, and
then called PRL and asked them to delay publication of my letter until they
received the manuscript Weisberger was preparing.  My paper  
(Adler, 1965c, {\sf R8}) and Weisberger's (Weisberger, 1965) appeared as 
back-to-back letters in the June 21 issue.  They 
give substantially identical derivations; Weisberger's numerical result of
1.16 differed from mine of 1.24 because I had included a correction for the
off-pion-mass-shell extrapolation of the threshold phase space factor
associated with the $\Delta(1232)$
resonance, which I knew from my work on weak pion production could be
reliably estimated.  At the time, this correction made agreement with 
experiment
worse (the experimental value for $g_A$ was then 1.18), but the best value 
now has settled down to $g_A=1.257 \pm .003$, in gratifyingly good agreement 
with the value I got when I included the kinematic extrapolation correction.
Weisberger and I both submitted longer papers to {\it Physical Review} describing
our work (Adler, 1965d, {\sf R9});  Weisberger (1966). These emphasized the 
low energy theorem approach to the relation for $g_A$, giving historically 
the first two-soft pion low energy theorem. 
In my paper I also gave an analog for pion-pion scattering, and
then in the final section (Adler, 1965, {\sf R9}, Section V), I
returned to the observation that I had made a year
earlier about forward lepton scattering, and showed that the $g_A$ sum rule
could be converted to an {\it exact} relation, involving no off-shell PCAC 
extrapolation, for forward inelastic high energy neutrino reactions.  This
relation, which provided a test of the Gell-Mann current algebra of axial-vector charge commutators, 
was another indication of a deep connection between the structure
of currents on the one hand, and inelastic lepton scattering on 
the other.

The $g_A$ sum rule provided  yet a third result supporting the use of 
PCAC as a method for calculating soft pion processes. Simultaneously, it was  
a stunning 
success for Gell-Mann's brilliant idea of abstracting the current 
algebra from the naive quark model, with the hope that it would prove to be a 
feature that would also be valid in the 
then unknown theory of the strong interactions.  At this point the 
whole community took notice, and a string of current algebra/PCAC 
applications appeared in rapid succession.  To mention just a few, 
Weinberg (1966a) and Tomozawa (1966) reexpressed the soft pion theorems for 
pion-nucleon scattering, coming from my consistency condition papers and 
the $g_A$ sum rule papers, in the form of formulas 
for the pion-nucleon scattering lengths, and Weinberg in the same 
paper also used 
my result of a PCAC zero in pion-pion scattering, plus a symmetry argument,  
as inputs for a derivation of pion-pion scattering lengths.   
Weinberg (1966b) also 
generalized the two-soft pion low energy form of the $g_A$ sum rule to 
a general formula for multiple soft pion production.   Finally, 
in another striking 
application of soft pion theorems, Callan and Treiman (1966) gave a series 
of important results for $K$ meson decays, in which the role of rapidly 
varying pole terms was clarified in  Weinberg (1966c).  

In connection with the $g_A$ sum rule, I have an interesting Feynman 
anecdote to relate.  I spent the spring term of 1966 as a member of Murray 
Gell-Mann's postdoctoral group at Cal Tech.  
A few weeks after I arrived,  Feynman asked 
me to stop by his office to look at some pages in his notebook, in which   
he had almost derived the $g_A$ sum rule, before Weisberger and I did it. 
The whole expression was there (including the kinematic correction that 
I had included for the off-mass-shell extrapolation),  except that, 
where the Gell-Mann algebra had dictated a 1 coming from the commutator 
of two axial-vector charges giving an unrenormalized vector charge, Feynman 
had put 0!  So numerically the relation did not work, and Feynman had given 
up on it and gone on to other things.  He evidently had not paid attention 
to Gell-Mann's current algebra, or at least not realized, from his heuristic 
way of doing things, that it was essential for this calculation.  

Returning again to events in 1965, as
soon as the long paper on $g_A$ was completed, I departed to be a summer
visitor at CERN.  There I met Murray Gell-Mann for the first time, and had
 long conversations with him.  Murray was particularly interested in
the Section V relation between the current algebra of vector and axial-vector charges and forward high energy 
neutrino reactions, and urged me to try to extend it to a test of the 
{\it local} current algebra which he had given in his {\it Physics} paper
(Gell-Mann, 1964).  I spent the summer working on this, and found that 
I could do it; as I recall, the crucial bits came together when I spent a 
day working at a kitchen table during a week off for holiday at Lake Garda.  
The results were written up in the late summer of 1965 
at CERN and/or Harvard, and appeared in Adler (1966), {\sf R10}.  
This article gave the first detailed working out
of the structure of deep inelastic high energy neutrino scattering \big(the 
electroproduction case was given independently in the review of  
de Forest and 
Walecka (1966)\big), with both the electroproduction and neutrino cases 
specific examples of general local lepton coupling theorems given by 
Lee and Yang and by Pais, as referenced in my article {\sf R10}.   However, 
the $\alpha, \beta, \gamma$ notation that I used for the structure functions
did not become the standard one; the now standard $W_{1,2,3}$ 
structure functions, which follow the notation of de Forest 
and Walecka and were further popularized by Bjorken, are linearly 
related to the ones I used.  
[Specifically, I separated the cross section into strangeness-conserving and 
strangeness-changing pieces, whereas the current convention is to define 
the structure functions as the sum of both.  At zero Cabibbo angle, the 
relation between my $\alpha,\beta,\gamma$ and the conventional $W_{1,2,3}$ is 
$\alpha=W_1,~ \beta=W_2,~2M_N\gamma=W_3$, with $M_N$ the nucleon mass.   For  
general Cabibbo angle $\theta_C$, one has $\cos^2\theta_C \beta^{(+,-)}_
{\Delta S=0} + \sin^2\theta_C \beta^{(+,-)}_{|\Delta S|=1}=W_2^{\nu,\overline
\nu}$, with similar relations for the other two structure functions.]  
The article
actually gave three sum rules; two for the $\alpha$ and $\gamma$ structure
functions which subsequent analysis by Dashen showed to be divergent and
hence useless, and one for the $\beta$ deep inelastic amplitude which is
a convergent and useful relation.  The beta sum rule divides into axial-vector
and vector parts, which are separately given as Eqs. (53a) and (53b) 
respectively of Adler (1966), {\sf R10}, and which when added to give the 
total $\Delta S=0$ cross section yield  
\begin{equation}
2=g_A(q^2)^2+F_1^V(q^2)^2+ q^2 F_2^V(q^2)^2 
+\int_{M_N+M_{\pi}}^{\infty}\frac{W}{M_N} dW [\beta^{(-)}(q^2,W) - 
\beta^{(+)}(q^2,W)]~~~.\nonumber
\end{equation}
This sum rule (and the ones for the separate vector and axial-vector 
contributions) has the notable feature that the left-hand side is independent 
of $q^2$, even though the Born term contributions and the continuum  
integrand on the right are $q^2$-dependent.   
At zero squared momentum transfer $q^2$, the 
axial-vector part of the $\beta$ sum rule
reduces to the relation I gave in my long paper on $g_A$, which had
prompted Gell-Mann's question about a generalization; the first derivative 
of the vector part with respect to $q^2$ at $q^2=0$ gives the sum rule
also derived by Cabibbo and Radicati (1966) using moments of currents.
Because the neutrino and antineutrino differential cross 
sections $d^2 \sigma 
/ d(q^2) dW $are 
dominated by the $\beta$ structure function in the limit of large neutrino 
energy, by integrating over $W$ one gets the limiting cross section relation   
(at zero Cabibbo angle) 
\begin{equation}
\lim_{E_{\nu} \to \infty}\left[ \frac{d\sigma(\overline \nu + p)}{d(q^2)}-
 \frac{d\sigma( \nu + p)}{d(q^2)} \right]=\frac{G^2}{\pi}~~~,\nonumber\end{equation}
with $G$ the Fermi constant.   Similar relations at non-zero Cabibbo 
angle are 
given in Eq.~(27) of {\sf R10}, and it is easy to obtain  
analogous relations for the vector   
and axial-vector contributions to the cross sections taken separately.  

In late October of 1965 I spoke on ``High Energy Semileptonic 
Reactions'' at the International Conference on Weak Interactions held at 
Argonne National Laboratory (Adler, 1965e), in which I gave 
the first public presentation of the local current algebra sum rules for the 
$\beta$ deep inelastic neutrino structure functions, and the limiting  
relations for the differential cross sections that they imply.  
In the published 
discussion following this talk, in answer to a question by Fubini, 
I noted that  the $\beta$ sum rule had been rederived by Callan (unpublished) 
using the infinite momentum 
frame limiting method, but that the $\alpha$ and $\gamma$ sum rules could 
not be derived this way, reinforcing suspicions that ``the integral for 
$\beta$ is convergent, while the other two relations (for $\alpha$ and  
$\gamma$) really need subtractions.''  
Bjorken was in the audience and 
was intrigued by the $\beta$ sum rule results, and soon afterwards 
converted them into a differential cross section  
inequality (Bjorken, 1966, 1967)  for deep inelastic 
electron-nucleon  scattering, for which there was the 
prospect of experimental tests  
relatively soon.  To see why the neutrino cross section relation 
given above 
implies an inequality for electron scattering, one notes 
that since the $\nu +p$  
differential cross section is positive, the right-hand side $G^2/\pi$ 
gives a lower bound for the $\overline \nu +p$ differential cross section,  
with a similar lower bound holding for the vector current contribution alone.
But noting that according to CVC, the vector weak 
current is in the same isospin multiplet as the isovector part of the 
electromagnetic current, and using the Wigner--Eckart 
theorem, one gets a corresponding lower bound for the inelastic differential 
cross section induced by an isovector virtual photon scattering on a nucleon.
One then notes that in the  scattering of a virtual photon on 
a target containing equal numbers of neutrons and protons, the isovector 
and isoscalar currents add incoherently, and so the isovector current 
contribution alone gives a lower bound.  Combining the two bounds,   
and including 
an extra $1/(k^2)^2$ for the virtual photon propagator,  replacing $G$ by  
the fine structure constant $\alpha$,  and keeping track of numerical factors, 
one gets Bjorken's electron scattering result   
\begin{equation}
\lim_{E_e \to \infty}\frac{d[\sigma(e+p) +\sigma(e+n)]}{d(k^2)} 
>\frac{2\pi \alpha^2}{(k^2)^2}~~~,\nonumber
\end{equation} 
which was testable in the experiments soon to begin at SLAC.  
Verification of my neutrino sum rule, on the other hand, took two decades and 
more; see Allasia et al. (1985) for the first reported test, and Conrad, 
Shaevitz, and Bolton (1998) for more recent high precision results. For a 
recent study of my neutrino sum rule, in comparison with the  
Gottfried (1967) sum rule for electron-proton scattering, within 
the framework of the large $N_c$ expansion of QCD with $N_c$ 
colors, see Broadhurst, Kataev, and Maxwell (2004) and Kataev (2004).  

Although not directly tested until many years after it was derived in 1965, 
my neutrino sum rule had important conceptual  
implications that figured prominently in developments over the next few years.  
To begin with, it gave the first indications that 
deep inelastic lepton scattering would give information about the local 
properties of currents, a fact that at first seemed astonishing, but which 
turned out to have important extensions.  Secondly, as noted by Chew in 
remarks at the 1967 Solvay Conference (Solvay, 1968), the 
closure property tested in 
the sum rules, if verified experimentally, would suggest the presence of 
elementary constituents inside hadrons.  In a Letter (Chew, 1967) published 
shortly  after this conference, Chew argued that my sum rule, if verified, 
would 
rule out the then popular ``bootstrap'' models of hadrons, in which all 
strongly interacting particles were asserted to be equivalent (``nuclear 
democracy'').  In his words, ``such sum rules may allow confrontation 
between an underlying local spacetime structure for strong interactions 
and a true bootstrap.  The pure bootstrap idea, we suggest, 
may be incompatible with closure.''  In a similar vein, Bjorken, 
in his 1967 Varenna lectures (Bjorken, 1968),  
argued that the neutrino sum rule was strongly suggestive of the presence  
of hadronic constituents, and this was also noted in the review of   
Llewellyn Smith (1972).  

These conceptual developments still left undetermined the 
mechanism by which the neutrino sum rule, and Bjorken's electron scattering  
inequality, could be saturated at large $q^2$.  During my visit to Cal Tech 
in 1966, I renewed my graduate school acquaintance with  
Fred Gilman and worked with him on two projects.  One was
an analysis of the saturation of the neutrino sum 
rule for small $q^2$ (Adler and Gilman, 1967, {\sf R11}), in which we 
concluded that SLAC (soon to start operating) 
would have enough energy to confront the saturation of 
the nonzero $q^2$ sum rules in a meaningful way.  
In this paper, we noted that the $\beta$ sum rule 
posed what at the time was a puzzle:  the left-hand 
side of the sum rule is a constant, while the Born terms on the right
are squares of nucleon form factors, which vanish rapidly as the momentum
transfer $q^2$ becomes large.  The low lying nucleon resonance contributions
on the right were expected to behave like the $\Delta(1232)$ contribution, 
which is form factor dominated and also falls off 
rapidly with $q^2$.  Hence it was clear that something
new and interesting must happen in the deep inelastic region if the sum rule
were to be satisfied for large $q^2$:  ``to maintain a constant sum at 
large $q^2$, the high
$W$ states, which require a large $E$ to be excited, must make a much more
important contribution to the sum rules than they do at $q^2=0$''. 
We were cautious, however (too cautious, as it turned out!), and did not 
attempt to model the structure of the deep inelastic component 
needed to saturate the sum rule at large $q^2$. 
Bjorken became interested in the 
issue of how the sum rule could be saturated, and formulated several 
preliminary models that (in retrospect) already had hints of the dominance 
of a regime where the energy transfer $\nu$ grows proportionately 
to the value of $q^2$.  
I summarized these pre-scaling 
proposals of Bjorken in the discussion period of the 1967 Solvay 
Conference  (Solvay, 1968), which Bjorken did not attend, 
in response to  questions from Chew and others 
as to how the neutrino sum rule could be saturated.  The precise saturation 
mechanism was clarified (to a very good first approximation) 
some months after the Solvay conference 
with the proposal by Bjorken (Bjorken, 1969)  of scaling, 
and soon afterwards, with 
the experimental work at SLAC on deep inelastic electron scattering, that  
confirmed Bjorken's intuition.   For a very clear exposition of the relation 
between scaling and the neutrino sum rule, see Sec. 3.6B of Llewellyn Smith 
(1972), who notes that when the sum rule is rewritten in terms of Bjorken's 
scaling variable $\omega$, ``The simplest way to ensure the $Q^2$ [my $q^2$] 
independence of the left-hand side as $Q^2 \to \infty$ is to assume 
that the limit in eq.(3.71) 
[in my notation, $\lim_{Q^2 \to \infty, \omega\, {\rm fixed}}\beta^{(\pm)}
(\omega,Q^2/M_N^2) $] exists''.  
\bigskip
\section*{More Low Energy Theorems; Weak Pion Production Redux}

In the fall of 1965 I received an invitation from Oppenheimer, which 
I accepted,
to come to the Institute for Advanced Study as a long term member with a five
year appointment, starting
in the fall of 1966.  Roger Dashen, whom I had met briefly when he visited 
Harvard earlier in 1965, received a similar invitation.  The intent   
behind our appointments was that we would reinvigorate 
high energy theory at the Institute, which had fallen into a decline 
with the departures of Lee, Yang, and Pais to professorships 
elsewhere, and with a turn of Dyson's research interests 
towards astrophysics.  

Before going to Princeton, as mentioned above, I spent the spring term of
1966 as a postdoc in Murray Gell-Mann's group at Cal Tech.  By this time
the successes of PCAC and current algebra had attracted a lot of attention 
and stimulated an outpouring of papers, the more important ones of which 
appear
in the volume which Dashen and I put together a year later.  My own work
in the spring of 1966 was focused on two issues.  The first involved using
PCAC to get small momentum expansions of matrix elements of the axial-vector
current, in analogy with the paper of Low (1958) on soft
photon bremsstrahlung.  With Joe Dothan, I wrote a long paper (Adler and
Dothan, 1966, {\sf R12}) applying these ideas to the weak pion 
production amplitude  and to radiative muon capture.  The weak pion results  
figured in my later comprehensive paper on the subject (see below), while 
the radiative muon capture work was incorporated into later chiral 
perturbation theory treatments of radiative muon capture; for a review  
of the current theoretical and experimental status of muon capture, 
including a discussion of discrepancies between theory 
and experiment in the radiative capture case, see Gorringe and Fearing (2004).  
The other direction of work involved two 
phenomenological studies done with Fred Gilman.  One of these dealt with  
saturation of the neutrino sum rule, as described in the preceding section.
The other dealt with a detailed phenomenological study of the 
PCAC predictions for pion photo- and electro-production (Adler and Gilman,
1966, {\sf R13}), including a saturation analysis for 
the Fubini--Furlan--Rossetti (1965) sum rule;   
for a recent update on this, see Pasquini, Drechsel, and Tiator (2005).  

My first year at the Institute was largely devoted to writing 
the book on
{\it Current Algebras} with Roger Dashen (Adler and Dashen, 1968).  The book 
consisted of selected reprints grouped by categories with commentaries 
that we supplied, plus some general 
introductory material.  I was responsible for writing the introductory 
sections and the commentaries for Chapters 1-3, which included 
Appendix A, reprinted here as {\sf R7}.  Roger was responsible for the 
commentaries for Chapters 4-7, which included an original and very  
detailed analysis of precisely which sum rules could be derived by the  
infinite momentum frame method, or in different language,  when 
a naive assumption of unsubtracted dispersion 
relations would (and would not) give correct results.  This analysis 
confirmed earlier suspicions that my $\beta$ neutrino sum rule was correct, 
but that the $\alpha$ and $\gamma$ sum rules should have subtractions, and 
so were not useful.  The book on {\it Current Algebras} was completed, and sent off 
to the publisher, in the fall of 1967.  

During this period 
I also worked with Bill Weisberger, who was then at Princeton,  
on sorting out the tricky pion pole structure in two pion
photo- and electro-production, which had to be handled carefully to get  
a fully  gauge-invariant expression (Adler and Weisberger, 1968, {\sf R14}).
Our interest in this process, as noted in the title of the paper, was 
motivated by the fact that it gives an alternative, indirect method of 
measuring the nucleon axial-vector form factor $g_A(k^2)$.  An experiment 
to measure $g_A(k^2)$ by this method was carried out by Joos et al. (1976) 
giving a value $m_A=1.18\pm 0.07$ GeV for the mass in the dipole formula 
$g_A(k^2)=g_A(0)(1+k^2/m_A^2)^{-2}$.  This value is in good agreement 
with the value $m_A=1.07\pm 0.06$ GeV given in the quasielastic scattering 
$\nu_{\mu}+n \to \mu^{-}+p$ experiment of Baker et al. (1981), and also in 
reasonable agreement 
with values of $m_A$ obtained from single pion electroproduction at threshold 
using the low energy theorem of Nambu and Shrauner (1962) 
\big(for which experimental references are 
given in both the Joos et al. and Baker et al. articles\big).  
At the 1968 Nobel Symposium on Elementary Particle Theory, I gave a 
brief talk (Adler, 1968a) reviewing 
various methods that had been proposed to measure the nucleon axial-vector 
form factor:  quasielastic neutrino scattering, neutrino   
production of the $\Delta(1232)$,  electroproduction of a single soft
pion (the Nambu--Shrauner proposal), and electroproduction of the 
$\Delta(1232)$ plus an additional soft pion (the proposal of my 
paper {\sf R14} with Weisberger).  Over the years since then,  
all of these methods have been  carried out.  

I also returned, after completion of the book on {\it Current Algebras}, to the 
repeatedly delayed project of completing the numerical work   
associated with my thesis calculation of weak pion production, and 
this kept me busy until the spring of 1968, when I finished a comprehensive 
article on photo-, electro-, and weak 
single-pion production in the $\Delta(1232)$, or as it was then termed, the 
(3,3) resonance 
region (Adler, 1968b, {\sf R15}).  This paper is so long (123 pages) 
that it is not feasible to reprint it all here, so I have included only 
the introduction (Sec. 1) and part of the discussion of implications of 
PCAC (Secs. 5A and 5B).   
The basic approximation used in this paper 
consisted of using the Born approximation for all nonresonant multipoles, 
augmented by terms coming from the PCAC low energy theorems,
together with  a unitarized Born approximation for the dominant 
resonant (3,3) multipoles, giving predictions for weak pion production 
in the (3,3) region in terms of the vector and axial-vector form factors of 
the nucleon.  By 1968 there were experimental results on
pion electroproduction which were in satisfactory agreement
with my theory, except for values of the momentum transfer  
$k^2$ significantly larger than roughly  $0.6({\rm GeV}/c)^2$, where 
in retrospect 
one can see effects from the scaling regime showing up. For neutrino  
pion production, preliminary comparison of my results with CERN data 
showed an axial-vector form factor $g_A(k^2)$ that falls off more slowly  
with $k^2$ than the vector form factors, with a dipole mass of  
$m_A \sim 1.2$GeV. A subsequent comparison of my model with high-statistics    
neutrino data from Brookhaven by Kitagaki et al. (1986) gave good fits 
with a dipole mass of $m_A = 1.28 \pm 0.11$ GeV, somewhat high compared 
to values obtained by other methods described above.  
Reasonable fits of my model to the $\Delta$ cross section and density matrix elements 
measured in the hydrogen 
bubble chamber at Argonne were also reported in papers of Schreiner and von 
Hippel (1973a,b), and a comparison with other models and data was given 
by Rein and Sehgal (1981).    
\big(For a recent alternative approach to 
$\Delta(1232)$ weak 
production, and extensive references to earlier theoretical and 
experimental studies of this reaction, see Paschos et al. (2004).\big)     
After 1968 I did not work again on weak pion 
production until 1974-75, when the subject became important because it 
was an avenue for exploring weak neutral currents, as discussed 
in Chapter 5 below.  

To conclude this section on low energy theorems, let me address the  
question of the extent to which the modern viewpoint, of pions as 
Nambu--Goldstone bosons, entered into my work.      
The earliest reference that
I could find in my research notes to the ``Goldstone theorem'' 
(and specifically to the derivations
given in the paper of Goldstone, Salam, and Weinberg, 1962)
dates from
the spring of 1967, in other words, after nearly all the work on 
soft pion theorems
was completed.  (This reference was in the context of calculations on the
axial-vector vertex in QED  that were the starting point of 
my work on the axial anomaly, to be discussed in the next chapter.)
I fully appreciated the role of pions as 
Nambu--Goldstone bosons only after
hearing seminars that referred to Nambu--Goldstone versus 
Wigner--Weyl
representations of $\gamma_5$ symmetry, which were connected 
(as best I recall) with the  
work of Gell-Mann, Oakes, and Renner (1968) and Dashen (1969)   
on chiral $SU(3) \times SU(3)$ as a strong interaction symmetry. 
This may at first seem surprising, but now that the tapestry
of the standard model is completed, we see clearly the interrelations of 
its many threads;  at the time when these threads were being laid down, those
working from one direction were often unaware or only dimly aware of progress
from another.  

Perhaps this is also a good point to say that the elucidation
of the chiral structure of the strong interactions was only {\it one} of 
the results flowing from the successes of current algebra methods and 
PCAC; something
that was perhaps even more significant at the time was the demonstration that
quantum field theory methods were really valid, after all, in dealing with
hadronic interactions.  When I entered graduate school, the prevailing view
was that the strong interactions would be understood through some kind of
dispersion theoretic ``reciprocal bootstrap'', and nearly every particle
physics talk I heard began with a Mandelstam diagram on the blackboard.  By
1967, this view had changed; it was clear that field theory could 
produce results which could not be obtained from the dispersion relations
program, and this strongly influenced subsequent developments. \vfill\eject

\section*{References for Chapter 2}

\begin{itemize}

\item Adler, S. L. (1964a). Polarization Effects in High-Energy Weak 
Interactions.  {\it Nuovo Cimento} {\bf 30}, 1020-1039. 

\item Adler, S. L. (1964b). High Energy Neutrino Reactions and Conservation 
Hypotheses, Princeton University dissertation, on deposit with University 
Microfilms. 

\item Adler, S. L. (1964c) {\sf R4}. Tests of the Conserved Vector Current and  
Partially Conserved Axial-Vector Current Hypotheses in High-Energy Neutrino
Reactions. {\it Phys. Rev.}  {\bf 135}, B963-B966. 

\item Adler, S. L. (1965a) {\sf R5}.  Consistency Conditions on the Strong 
Interactions Implied by a Partially Conserved Axial-Vector Current. 
{\it Phys. Rev.} {\bf 137}, B1022-B1033. 

\item Adler, S. L. (1965b) {\sf R6}.  Consistency Conditions on the Strong 
Interactions Implied by a Partially Conserved Axial-Vector Current. II.
{\it Phys. Rev.} {\bf 139}, B1638-B1643. 

\item Adler, S. L. (1965c) {\sf R8}.  Calculation of the Axial-Vector Coupling   
Constant Renormalization in $\beta$ Decay.  {\it Phys. Rev. Lett.} 
{\bf 14}, 1051-1055. 

\item Adler, S. L. (1965d) {\sf R9}.  Sum Rules for the Axial-Vector 
Coupling-Constant Renormalization in $\beta$ Decay.   
{\it Phys. Rev.} {\bf 140}, B736-B747. 

\item Adler, S. L. (1965e). High Energy Semileptonic Reactions, in {\it Proceedings of 
the International Conference on Weak Interactions}, held at Argonne National 
Laboratory, October 25-27, 1965,  ANL-7130, pp. 285-303 (talk) 
and pp. 304-309 (discussion). 

\item Adler, S. L. (1966) {\sf R10}.  Sum Rules Giving Tests of Local Current  
Commutation Relations in High-Energy Neutrino Reactions.  
{\it Phys. Rev.} {\bf 143}, 1144-1155. 
  
\item Adler, S. L. (1968a). Measurement of the Nucleon 
Axial-Vector Form Factor, 
in {\it Elementary Particle Theory, Relativistic Groups and Analyticity},  
Proceedings of the Eighth Nobel Symposium, held May 19-25, 1968, 
N. Svartholm, ed.
(Almqvist \& Wiksell, Stockholm, 
and John Wiley \& Sons, New York), pp. 263-268. 

\item Adler, S. L. (1968b) {\sf R15}.  Photo-, Electro-, and Weak Single-Pion   
Production in the (3,3) Resonance Region.  {\it Ann. Phys.} {\bf 50}, 
189-311.  Pages 189-192 and 255-266 are reprinted here. 

\item Adler, S. L. and R. F. Dashen (1968), Appendix A {\sf R7}. 
{\it  Current 
Algebras and Applications to Particle Physics} (W. A. Benjamin, 
New York), pp. 139-146. 

\item Adler, S. L.  and Y. Dothan (1966) {\sf R12}.   Low-Energy Theorem for the  
Weak Axial-Vector Vertex.  {\it Phys. Rev.} {\bf 151}, 1267-1277. 

\item Adler, S. L. and F.J. Gilman (1966) {\sf R13}.  Partially Conserved 
Axial-Vector Current Restrictions on Pion Photoproduction and 
Electroproduction Amplitudes.  {\it Phys. Rev.} {\bf 152}, 
1460-1467. 

\item Adler, S. L. and F. J. Gilman (1967) {\sf R11}.  Neutrino or Electron 
Energy Needed for Testing Current Commutation Relations.  {\it Phys. Rev.} 
{\bf 156}, 1598-1602. 

\item Adler, S. L. and W. I. Weisberger (1968) {\sf R14}.  
Possible Measurement of the 
Nucleon Axial-Vector Form Factor in Two-Pion Electroproduction Experiments. 
{\it Phys. Rev.} {\bf 169}, 1392-1397. 

\item Allasia, D. et al. (1985).  $Q^2$ Dependence of the Proton and Neutron 
Structure Functions from Neutrino and Antineutrino Scattering on Deuterium. 
{\it Z. Phys. C. - Particles and Fields} {\bf 28}, 321-333. 

\item Anderson, P. W. (1963).  Plasmons, Gauge Invariance, and Mass.  
{\it Phys. Rev.} {\bf 130}, 439-442. 

\item Baker, N. J. et al. (1981).  Quasielastic Neutrino Scattering: A Measurement 
of the Weak Nucleon Axial-Vector Form Factor.  {\it Phys. Rev. D} {\bf 23}, 
2499-2505. 

\item Bell, J. S. (1964).  Nuclear Optical Model for Virtual Pions. {\it Phys. 
Rev. Lett.} {\bf 13}, 57-59. 

\item Bernstein, J., S. Fubini, M. Gell-Mann, and W. Thirring (1960). On the  
Decay Rate of the Charged Pion. 
{\it Nuovo Cimento} {\bf 17}, 757-766.  

\item Bernstein, J., M. Gell-Mann, and L. Michel (1960). On the Renormalization  
of the Axial Vector Coupling Constant in $\beta$-Decay.  {\it Nuovo Cimento} 
{\bf 16}, 560-568. 

\item Bjorken, J. D. (1966).  Inequality for Electron and Muon Scattering from Nucleons.  {\it Phys. Rev. Lett.} {\bf 16}, 408. 

\item Bjorken, J. D. (1967).  Inequality for Backward Electron- and Muon-Nucleon     
Scattering at High Momentum Transfer. 
Phys. Rev. {\bf163}, 1767-1769. 

\item Bjorken, J. D. (1968).  Current Algebra at Small Distances, in 
{\it Proceedings of the International School of Physics ``Enrico Fermi'' 
Course XLI}, J. Steinberger, ed., Academic Press, New York, pp. 55-81.  
See p. 56 of this proceedings. 

\item Bjorken, J. D. (1969). Asymptotic Sum Rules at Infinite Momentum.   
{\it Phys. Rev.} {\bf 179}, 1547-1553. 

\item Broadhurst, D. J., A. L. Kataev, and C. J. Maxwell (2004).  Comparison of 
the Gottfried and Adler Sum Rules Within the Large-$N_c$ Expansion. 
{\it Phys. Lett. B} {\bf 590}, 76-85.

\item Bugg, D. V. (2003). Comments on the $\sigma$ and $\kappa$.  {\it Phys. 
Lett. B}  {\bf  572}, 1-7. 

\item Bugg, D. V. (2004).  Four Sorts of Meson.  {\it Physics Reports} {\bf 397}, 
257-358; see Sec. 11. 

\item Cabibbo, N. and  L. A. Radicati (1966).  Sum Rule for the Isovector Magnetic  
Moment of the Nucleon.  {\it Phys. Lett.} {\bf 19}, 697-699. 

\item Callan, C. G. and S. B. Treiman (1966).  Equal Time Commutators and 
$K$-Meson Decays.  {\it Phys. Rev. Lett.} {\bf 16}, 153-157. 

\item Chew, G. F. (1967).  Closure, Locality, and the Bootstrap.  {\it Phys. Rev. 
Lett.} {\bf 19}, 1492-1495.  Chew also refers to a more general local 
current algebra sum rule submitted for publication by Fubini after my 
Argonne Conference presentation, the integrand of which is not expressible 
in terms of measurable stucture functions:  
see S. Fubini, Equal-Time Commutators and 
Dispersion Relations, {\it Nuovo Cimento A}  {\bf 43}, 475-482 (1966).  [Chew 
however, through an apparent misunderstanding, gives as the reference 
an earlier paper, Fubini, Furlan, and Rossetti (1965), 
on which the 1966 Fubini 
paper was based.]  More general local current algebra sum rules, and a   
survey  of earlier work, were also given in R. Dashen 
and M. Gell-Mann,  Representation of Local Current Algebra at Infinite 
Momentum, {\it Phys. Rev. Letters} {\bf 17}, 340-343 (1966). 

\item Chew, G. F., M. L. Goldberger, F. E. Low, and Y. Nambu (1957). Relativistic  
Dispersion Relation Approach to Photomeson Production.  
{\it Phys. Rev.} {\bf 106}, 1345-1355. 

\item Chou, K.-C. (1960). On the Pseudovector Current and Lepton Decays of Baryons  
and Mesons {\it J. Exptl. and Theoret. Phys. (U.S.S.R.) }
{\bf39}, 703-712 \big(English translation: {\it Sov. Phys. JETP} {\bf 12}, 492-497 (1961)\big). 

\item Christenson, J. H., J. W. Cronin, V. L. Fitch, and R. Turlay (1964). 
Evidence for the $2\pi$ Decay of the $K_2^0$ Meson. 
{\it Phys. Rev. Lett.} {\bf 13}, 138-140. 

\item Conrad, J. M., M. H. Shaevitz, and T. Bolton (1998).  Precision Measurements 
with High-Energy Neutrino Beams.  {\it Rev. Mod. Phys.} {\bf 70},
 1341-1392. 

\item Dashen, R. F. (1969). Chiral $SU(3) \otimes  SU(3)$ as a Symmetry of the    
Strong Interactions.  {\it Phys. Rev.} {\bf 183}, 1245-1260. 

\item de Forest, T. and J.D. Walecka (1966).  Electron Scattering and   
Nuclear Structure.  {\it Advances in Physics}   
{\bf15}, 1-109. The general formula for inelastic electron scattering is   
given on p. 8 of this review. 

\item Faissner, H. et al. (1983).  Observation of Neutrino and 
Antineutrino Induced Coherent Neutral Pion Production off ${\rm Al}^{27}$. {\it Phys. Lett. B} {\bf 125}, 230-236. 

\item Feynman, R. P. and M. Gell-Mann (1958). 
Theory of the Fermi Interaction {\it Phys. Rev.} 
{\bf 109}, 193-198. 

\item Fubini, S. and G. Furlan (1965). Renormalization Effects for Partially 
Conserved Currents.  {\it Physics} {\bf 1}, 229-247. 

\item Fubini, S., G. Furlan, and S. Rossetti (1965).  A Dispersion Theory of 
Symmetry Breaking.  {\it Nuovo Cimento} {\bf 40}, 1171-1193. 

\item Fubini, S., Y. Nambu, and V. Wataghin (1958). Dispersion Theory Treatment  
of Pion Production in Electron-Nucleon Collisions. {\it Phys. Rev.} 
{\bf 111}, 329-336. 

\item Gell-Mann, M. (1964).  The Symmetry Group of Vector and Axial Vector Currents. 
{\it Physics} {\bf 1}, 63-75. 

\item Gell-Mann, M. and M. L\'evy (1960). The Axial Vector Current in Beta Decay. 
{\it Nuovo Cimento} 
{\bf 16}, 705-726. 

\item Gell-Mann, M., R. J. Oakes, and B. Renner (1968).  Behavior of Current 
Divergences under $SU_3 \times SU_3$.   {\it Phys. Rev.} {\bf 175}, 
2195-2199. 

\item Goldberger, M. L. and S.B. Treiman (1958). Decay of the Pi Meson.  
{\it Phys. Rev.} 
{\bf 110}, 1178-1184. 

\item Goldstone, J., A. Salam, and S. Weinberg (1962). Broken Symetries.  
{\it Phys. Rev.} {\bf 127}, 965-970. 

\item Gorringe, T. and H. W. Fearing (2004).  Induced Pseudoscalar Coupling of 
the Proton Weak Interaction.  {\it Rev. Mod. Phys.} {\bf 76}, 31-91. 
See in particular Sec. V. 

\item Gottfried, K. (1967). Sum Rule for High-Energy Electron-Proton Scattering. 
{\it Phys. Rev. Lett.} {\bf 18}, 1174-1177. 

\item G\"ursey, F. (1960).  On the Symmetries of Strong and Weak Interactions.  
{\it Nuovo Cimento} {\bf 16}, 230-240. 

\item Joos, P. et al. (1976).  Determination of the Nucleon Axial Vector Form-Factor 
from $\pi \Delta$ Electroproduction Near Threshold.  {\it Phys. Lett. B} 
{\bf 62}, 230-232. 

\item Kataev, A. L. (2004).  The Puzzle of the Non-Planar Structure of the QCD  
Contributions to the Gottfried Sum Rule; arXiv: hep-ph/0412369. 

\item Kitagaki, T. et al. (1986).  Charged-Current Exclusive Pion Production in  
Neutrino-Deuterium Interactions.  {\it Phys. Rev. D} {\bf 34}, 2554-2565. 

\item Kopeliovich, B.Z. (2004).  PCAC and Shadowing of Low Energy Neutrinos; arXiv:hep-ph/0409079. 

\item Llewellyn Smith, C. H. (1972). Neutrino Reactions at Accelerator Energies.   
{\it Physics Reports} {\bf 3}, 261-379.  For the forward lepton PCAC test 
and shadowing, see Sec. 2.2B and Sec. 3.8; for the derivation of the 
neutrino sum rule and its suggestion of ``point-like'' constituents, see 
Sec. 2.2C; for the relation between the neutrino sum rule and scaling, 
see Sec. 3.6B.   Llewellyn Smith's review, along with others, was reprinted  
in {\it Gauge Theories and Neutrino Physics, Physics Reports Reprint 
Book Series, Vol. 2}  (in memory of Benjamin W. Lee), M. Jacob, ed., 
(North-Holland, 1978), for 
which I wrote a short general introduction. 

\item Low, F. E. (1958).  Bremsstrahlung of Very Low-Energy Quanta in Elementary   
Particle Collisions.  {\it Phys. Rev.} {\bf 110}, 974-977. 

\item Mangano, M. L. et al. (2001).  Physics at the Front-End of a Neutrino 
Factory: A Quantitative Appraisal;  arXiv: hep-ph/0105155. 

\item Nambu, Y. (1959).  Discussion remarks, in {\it Proceedings of the International 
Conference on High-Energy Physics IX} (1959) (Academy of Science, Moscow, 
1960), Vol. 2, pp. 121-122. 

\item Nambu, Y. (1960). Axial Vector Current Conservation in Weak Interactions.  
{\it Phys. Rev. Lett.} {\bf 4}, 380-382. 

\item Nambu, Y. and G. Jona-Lasinio (1961). Dynamical Model of Elementary Particles  
Based on an Analogy with Superconductivity. I  {\it Phys. Rev.} 
{\bf 122}, 345-358. 

\item Nambu, Y. and D. Luri\'e (1962). Chirality Conservation and Soft Pion 
Production. {\it  Phys. Rev.} {\bf 125}, 1429-1436. 

\item Nambu, Y. and E. Shrauner (1962). Soft Pion Emission Induced by 
Electromagnetic and Weak Interactions.   
{\it Phys. Rev.} {\bf 128}, 862-868. 

\item Nishijima, K. (1959).  Introduction of a Neutral Pseudoscalar Field and a Possible Connection between Strangeness and Parity. {\it Nuovo Cimento} {\bf 11}, 698-710. 

\item Paschos, E. A., M. Sakuda, I. Schienbein, and J. Y. Yu (2004). Comparison  
of a New $\Delta$ Resonance Production Model with Electron and Neutrino 
Data; arXiv: hep-ph/0408185. 

\item Pasquini, B., D. Drechsel, and L. Tiator (2005).  The Fubini-Furlan-Rossetti 
Sum Rule Revisited. {\it Eur. Phys. J. A} {\bf 23}, 279-289.  

\item Ravndal, F. (1973).  Weak Production of Nuclear Resonances 
in a Relativistic Quark Model.  {\it Nuovo Cimento A} {\bf 18}, 385-415. 

\item Rein, D. and L. M. Sehgal (1981).  Neutrino-Excitation of Baryon 
Resonances and Single Pion Production. {\it Ann. Phys.} {\bf 133}, 79-153.
See particularly pp. 117-123. 

\item Roper, L. D. (1964).  Evidence for a $P_{11}$ Pion-Nucleon Resonance at 
556 MeV.  {\it Phys. Rev. Lett.} {\bf 12}, 340-342, and private 
communication.

\item Rupp, G., F. Kleefeld, and E. van Beveren (2004).  Scalar Mesons and 
Adler Zeros;  arXiv: hep-ph/0412078. 

\item Schreiner, P. A. and F. von Hippel (1973a).  $\nu p \to \mu^{-} \Delta^{++}$: 
Comparison with Theory.  {\it Phys. Rev. Lett.} {\bf 30}, 339-342. 

\item Schreiner, P. A. and F. von Hippel (1973b).  Neutrino Production of the 
$\Delta(1236)$.  {\it Nucl. Phys. B} {\bf 58}, 333-362. 

\item Schwinger, J. (1958). {\it Quantum Electrodynamics} (Dover Publications, 
New York). 

\item  Sehgal, L. M. (1988).  Neutrino Tridents, Conserved 
Vector Current, and Partially Conserved Axial-Vector Current.  
{\it Phys. Rev. D} {\bf 38}, 2750-2752. 

\item Shrauner, E. (1963).  Chirality Conservation and Soft-Pion Production in   
Pion-Nucleon Collisions.  Phys. Rev. {\bf 131}, 1847-1856. 

\item Solvay (1968).  Fundamental Problems in Elementary Particle Physics 
(Proceedings of the Fourteenth Conference on Physics at the University of 
Brussels, October 1967).  Interscience, London.  My untitled 
remarks and the ensuing discussion are on pp. 205-214; Chew's comment 
is on pp. 212-213.   In my remarks I attributed the saturation models to 
Bjorken, but there is no preprint reference; I believe I learned of the 
models directly from Bjorken when we were both lecturers at the Varenna 
summer school in July, 1967; see the discussion on page 63 of his lectures 
published as Bjorken (1968). 

\item Tomozawa, Y. (1966).  Axial-Vector Coupling Constant Renormalization and the 
Meson-Baryon Scattering Lengths.  {\it Nuovo Cimento A}  {\bf 46}, 
707-717. 

\item Touschek, B. F. (1957).  The Mass of the Neutrino and the Non-Conservation of Parity.  {\it Nuovo Cimento} {\bf 5}, 1281-1291. 

\item Weber, A. and L. M. Sehgal (1991).  CVC and PCAC in Neutrino-Lepton Interactions.  {\it Nucl. Phys. B} {\bf 359}, 262-282. 

\item Weinberg, S. (1966a).  Pion Scattering Lengths. {\it Phys. Rev. Lett.} 
{\bf 17}, 616-621. 

\item Weinberg, S. (1966b). Current Commutator Theory of Multiple Pion Production.
{\it Phys. Rev. Lett.} {\bf 16}, 879-883. 

\item Weinberg, S. (1966c).  Current-Commutator Calculation of the $K_{\ell 4}$ 
Form Factors. {\it Phys. Rev. Lett.} {\bf 17}, 336-340. 

\item Weisberger, W. I. (1965). Renormalization of the Weak Axial-Vector Coupling    
Constant.  {\it Phys. Rev. Lett.} {\bf 14}, 1047-1051.  

\item Weisberger, W. I.  (1966). Unsubtracted Dispersion Relations and the  
Renormalization of the Weak Axial-Vector Coupling Constants. 
{\it Phys. Rev.} {\bf 143}, 1302-1309. 

\end{itemize}

\chapter*{\centerline{3. Anomalies: Chiral Anomalies and Their Nonrenormalization,}\centerline{ Perturbative Corrections to Scaling, and Trace Anomalies}\centerline{to all Orders}}

\markboth{Adventures in Theoretical Physics}{Anomalies: Chiral Anomalies and Their Nonrenormalization}

\section*{Chiral Anomalies and $\pi^0 \to \gamma \gamma$ Decay} 

I got into the subject of anomalies in an indirect way, through exploration 
during 1967-1968 of the speculative idea that the muon-electron mass 
difference could be accounted for by giving the muon an additional magnetic 
monopole electromagnetic coupling through an axial-vector current, which 
somehow was nonperturbatively renormalized to zero.  After much fruitless 
study of the integral equations for the axial-vector vertex part, I decided 
in the spring of 1968 to first try to answer a well-defined question, which 
was whether the axial-vector vertex in QED was renormalized by multiplication 
by $Z_2$, as I had been implicitly assuming.  At the time when I turned 
to this question, I had just started a 6-week visit to 
the Cavendish Laboratory in 
Cambridge, England after flying to London with my family 
on April 21, 1968 (as recorded by my ex-wife Judith in my oldest daughter 
Jessica's ``baby book'').  In the Cavendish I shared an office 
with my former adviser Sam Treiman, and was 
enjoying the opportunity to try a new project not requiring extensive 
computer analysis; I had only a month before finished my {\it Annals of 
Physics} paper {\sf R15} on weak pion production (see Chapter 2), 
which had required  
extensive computation, not easy to do in those days when one had to wait 
hours or even a day for the results of a computer run.  

My interest in the multiplicative  
renormalization question had been piqued by work of 
van Nieuwenhuizen, in which he had attempted to demonstrate the 
finiteness to all orders of radiative corrections to $\mu$ decay, using 
an argument based on subtraction of renormalization constants that I knew  
to be incorrect beyond leading order.  I had learned about this work during  
the previous summer, when I was a lecturer at the Varenna summer school held 
by Lake Como from July 17-29, 1967, at which van Nieuwenhuizen had given a  
seminar on this topic that was critiqued 
by Bjorken, another lecturer.  \big(For further historical details 
about this, see my review article Adler (2004a)   
on anomalies and anomaly nonrenormalization, from 
which much of this commentary has been adapted.\big)  
Working in the old Cavendish, 
I rather rapidly found an inductive multiplicative renormalizability proof, 
paralleling the one in Bjorken and Drell (1965) for finiteness of $Z_2$ 
times the vector vertex.  I prepared a detailed outline for a paper 
describing the proof, but before writing things up, I decided as a check 
to test whether the formal argument for the closed loop part of the Ward 
identity  worked in the case of the smallest loop diagram.  
This is a triangle diagram with one axial and two vector vertices 
\big(the $AVV$ triangle; see Fig. 1(a)\big),   
which because of Furry's theorem ($C$ invariance) has no analog in the 
vector vertex case.  I knew from a student seminar that I had 
attended during my graduate study at Princeton that this diagram had been 
explicitly calculated using a gauge-invariant regularization 
by Rosenberg (1963), who was interested in the 
astrophysical process $\gamma_V+ \nu \to \gamma + \nu$, with $\gamma_V$ a 
virtual photon emitted by a nucleus.  I got Rosenberg's paper, tested 
the Ward identity, and to my astonishment (and Treiman's when I told him 
the result) found that it failed!  I soon found that the problem was that 
my formal proof used a shift of integration variables inside a linearly 
divergent integral, which (as I again recalled from student reading) had 
been analyzed in an Appendix to the classic text of Jauch and Rohrlich 
(1955), with a calculable constant remainder.  For all closed 
loop contributions to the axial vertex 
in Abelian electrodynamics with larger numbers of vector vertices 
\big(the $AVVVV$, $AVVVVVV$,... loops; see Fig. 1(b)\big),  
the fermion loop integrals for fixed photon momenta are highly 
convergent and the shift of integration variables needed in the Ward 
identity {\it is} valid, so proceeding in 
this fermion loop-wise fashion there were  
apparently no further additional or ``anomalous'' contributions to the   
axial-vector Ward identity.  With this fact in the back of my mind 
I was convinced from the outset that the anomalous contribution to the 
axial Ward identity would come just from the triangle diagram, with no 
renormalizations of the anomaly coefficient arising 
from higher order $AVV$ diagrams with virtual photon insertions.  

\begin{figure}[th]      
\centerline{\epsfxsize=5in\epsfbox{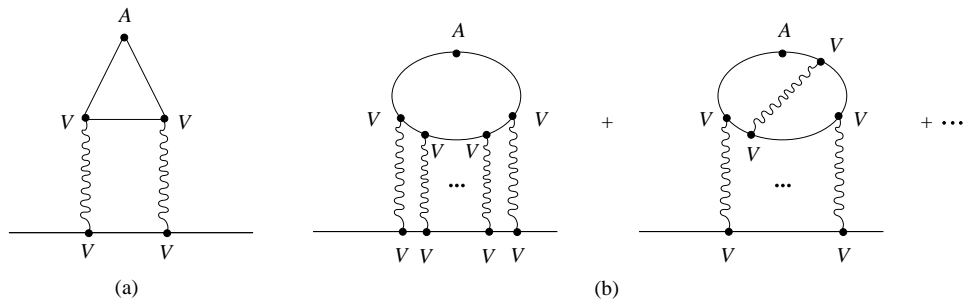}}
\vspace*{8pt}
\caption{Fermion loop diagam contributions to the axial-vector vertex part.  Solid lines are fermions, and dashed lines are photons. (a) The smallest loop, the $AVV$ triangle diagram.  (b) Larger loops with four or more vector vertices, which (when summed over vertex orderings) obey normal Ward identities. \label{fig1a-b}}
\end{figure}
In early June, at the end of my 6 weeks in Cambridge,  
I returned to the US and then went to Aspen, where 
I spent the summer working out a manuscript on the properties of the 
axial anomaly, which became the body (pages 2426-2434) of the final 
published version (Adler, 1969, {\sf R16}).  Several of the things 
done there deserve mention, since they  were important in later applications. 
The first was a calculation of the field theoretic form of the anomaly,
giving the now well-known result 
\begin{equation}
\partial^{\mu} j_{\mu}^5(x) = 2im_0j^5(x) 
+\frac{\alpha_0}{4 \pi} 
F^{\xi \sigma}(x) F^{\tau \rho}(x) \epsilon_{\xi\sigma \tau \rho}~~~,\nonumber 
\end{equation}
with $j_{\mu}^5=\overline \psi \gamma_{\mu}\gamma_5 \psi$ 
the axial-vector current 
(referred to above as $A$), 
$j^5=\overline \psi \gamma_5 \psi$ the pseudoscalar 
current, and with $m_0$ and $\alpha_0$ the 
(unrenormalized) fermion mass and coupling constant.  The second was  
a demonstration that because of the anomaly, $Z_2$ is no longer the 
multiplicative renormalization constant for the axial-vector vertex, as a   
result of the diagram drawn in Fig.~1(a) in which the $AVV$ triangle 
is joined  to an electron line with two virtual photons.   
Instead, the axial-vector vertex is made finite by multiplication by 
the renormalization constant 
\begin{equation}
Z_A=Z_2[1+\frac{3}{4}(\alpha_0/\pi)^2 \log(\Lambda^2/m^2)+...]~~~, \nonumber
\end{equation}
thus giving an answer to the question with which I started my investigation.
Thirdly, as an application of this result, I showed that the anomaly leads,  
in fourth order of perturbation theory, to 
infinite radiative corrections to the current-current theory of 
$\nu_{\mu} \mu$ and $\nu_{e}e$ scattering, but that this 
infinity can be cancelled between   
different fermion species by adding appropriate $\nu_{\mu} e$ and 
$\nu_e \mu$ scattering terms to the Lagrangian.  This result is a forerunner 
of anomaly cancellation mechanisms in modern gauge theories.    
It is related  
to the fact, also discussed in my paper, that the asymptotic behavior of 
the $AVV$ triangle diagram saturates the bound given by the Weinberg power 
counting rules, rather than being one power better as is the case for  
the $AVVVV$ and higher loop diagrams, and has a 
leading asymptotic term that is a function solely of the external momenta.   
Finally, I also showed that a gauge 
invariant chiral generator still exists in the presence of the anomaly.   
Although not figuring in our subsequent discussion here, in its non-Abelian 
generalization this was relevant (as reviewed in Coleman, 1989) to later 
discussions of the $U(1)$ problem in quantum chromodynamics (QCD), leading     
up to the solution given by 't Hooft (1976).  

No sooner was this part of my paper completed than  
Sidney Coleman arrived in Aspen from Europe, and told me that Bell and 
Jackiw (published as Bell and Jackiw, 1969) 
had independently discovered the anomalous behavior of the $AVV$ 
triangle graph, in the context of a sigma model investigation of the
Veltman (1967)--Sutherland (1967) theorem stating that 
$\pi^0 \to \gamma \gamma$ decay is forbidden in a PCAC calculation.  
The Sutherland--Veltman theorem is a kinematic statement 
about the $AVV$ three-point function, which asserts 
that if the momenta associated 
with the currents $A,V,V$ are respectively $q,k_1,k_2$, then the requirement 
of gauge invariance on the vector currents forces the $AVV$ vertex to be 
of order $q k_1 k_2$ in the external momenta.  Hence when one applies a 
divergence to the axial-vector vertex and uses the standard PCAC relation 
(with the quark current ${\cal F}_{3\mu}^5$ the 
analog of $\frac{1}{2}j_{\mu}^5$ )
\begin{equation}
\partial^{\mu} {\cal F}_{3\mu}^5(x)= 
(f_{\pi}M_{\pi}^2/\sqrt{2}) \phi_{\pi}(x)~~~,\nonumber
\end{equation}  
with $M_{\pi}$ the pion mass, 
$\phi_{\pi}$ the pion field, and $f_{\pi}$ the charged pion decay 
constant, one finds that the $\pi^0 \to \gamma \gamma$ 
matrix element is of order $q^2 k_1 k_2$, and hence vanishes in the 
soft pion limit $q^2 \to 0$. Bell and Jackiw analyzed this result by  
a perturbative calculation in the $\sigma$-model, in 
which PCAC is formally built 
in from the outset, and found a non-vanishing result for the 
$\pi^0 \to \gamma \gamma$ amplitude, which they traced back to the fact 
that the regularized $AVV$ triangle diagram cannot be defined to satisfy 
the requirements of both PCAC and gauge invariance. This constituted the  
``PCAC Puzzle'' referred to in the title of their paper.  They then proposed 
to modify the original $\sigma$-model by adding further regulator fields 
with mass-dependent coupling constants 
in such a manner as to simultaneously enforce gauge invariance and PCAC,  
thus enforcing the Sutherland--Veltman prediction of a vanishing 
$\pi^0 \to \gamma \gamma$ decay amplitude.  In the words of Bell and 
Jackiw in their paper,  ``It has to be insisted that the introduction of 
this mass dependence of coupling constants is not an arbitrary step in the 
PCAC context.  If a regularization is introduced to define the theory, it 
must respect any formal properties which are to be appealed to.''  And 
again in concluding their paper, they stated ``To the complaint that we have 
changed the theory, we answer that only the revised version embodies 
simultaneously the ideas of PCAC and gauge invariance.''

It was immediately clear to me, in the course of the conversation with 
Sidney Coleman, that introducing additional regulators to eliminate 
the anomaly would entail renormalizability problems in $\sigma$ meson 
scattering, and was not the correct way to proceed.  However, it was also 
clear that Bell and Jackiw had made an important observation in tying the 
anomaly to the Sutherland--Veltman theorem for $\pi^0 \to \gamma \gamma$ 
decay, and that I could use the sigma-model version of the anomaly 
equation to get 
a nonzero prediction for the $\pi^0 \to \gamma \gamma$ amplitude, with 
the whole decay amplitude 
arising from the anomaly term.  I then wrote an Appendix 
to my paper (pages 2434-2438), clearly delineated from the manuscript that 
I had finished before Sidney's arrival, in which I gave a detailed rebuttal 
of the regulator construction, by showing that the anomaly could not be 
eliminated without spoiling either gauge-invariance or 
renormalizability. \big(In later discussions I added unitarity to this 
list, to exclude the possibility of canceling the anomaly by adding 
a term to the axial current 
with a $\partial_{\mu}/(\partial_{\lambda})^2 $ singularity.\big)  
In this Appendix I also used an anomaly-modified PCAC equation   
\begin{equation}
\partial^{\mu} {\cal F}_{3\mu}^5(x)= (f_{\pi}M_{\pi}^2/\sqrt{2}) 
\phi_{\pi}(x)
+S \frac{\alpha_0}{4 \pi} 
F^{\xi \sigma}(x) F^{\tau \rho}(x) \epsilon_{\xi\sigma \tau \rho}
~~~,\nonumber
\end{equation}   
with $S$ a constant determined by the constituent fermion charges and 
axial-vector couplings, to obtain a PCAC formula for the 
$\pi^0 \to \gamma \gamma$ amplitude $F^{\pi}$    
\begin{equation}
F^{\pi}=-(\alpha/\pi) 2S \sqrt{2} /f_{\pi}~~~.\nonumber
\end{equation}
Although the axial anomaly, in the context of breakdown 
of the ``pseudoscalar-pseudovector equivalence theorem'', had in fact been 
observed much earlier, starting with Fukuda and Miyamoto (1949) and 
Steinberger (1949) and continuing to Schwinger (1951), my paper broke new   
ground by treating 
the anomaly neither as a baffling calculational result, nor as  
a field theoretic artifact to be 
eliminated by a suitable regularization scheme, but instead as a real 
physical effect (breaking of classical symmetries by the quantization 
process)  with observable physical consequences.  

This point of view was not immediately embraced by everyone else.    
After completing my Appendix I sent Bell and Jackiw copies of my longhand 
manuscript, and an interesting correspondence ensued.  In a letter dated 
August 25, 1968, Jackiw was skeptical whether one could extract concrete 
physical predictions from the anomaly, and whether one could 
augment the divergence of the axial-vector current by a definite 
extra electromagnetic contribution, as in the modified PCAC equation above.     
Bell, who was traveling, wrote me on Sept. 2, 
1968, and was more appreciative of the possibility of using a modified 
PCAC to get a formula for the neutral pion decay amplitude, writing  
``The general idea of adding some quadratic electromagnetic 
terms to PCAC has been in our
minds since Sutherland's $\eta$ problem.  We did not see what to do with 
it.''  
He also defended the approach he and Jackiw had taken, writing 
``The reader may be left with the impression that your development is 
contradictory to ours, rather than complementary.  Our first observation 
is that the $\sigma$ model interpreted in a conventional way just does not
have PCAC.  This is already a resolution of the puzzle, and the one which 
you develop in a very nice way.  We, interested in the $\sigma$-model only 
as exemplifying PCAC, choose to modify the conventional procedures, in order 
to exhibit a model in which general PCAC reasoning could be illustrated 
in explicit calculation.''
In recognition of this letter from John Bell, whom I revered, I added a 
footnote 15 to my manuscript saying ``Our results do not 
contradict those of Bell and Jackiw, but rather complement them.  The main 
point of Bell and Jackiw is that the $\sigma$ model interpreted in the 
conventional way, does not satisfy the requirements of PCAC.  Bell and 
Jackiw modify the $\sigma$ model in such a way as to restore PCAC.  We, on 
the other hand, stay within the conventional $\sigma$ model, and try to 
systematize and exploit the PCAC breakdown.''  This footnote, which 
contradicts statements made in the text of my paper, has puzzled 
a number of people; in retrospect, rather than writing it as a paraphrase 
of Bell's words, I should have quoted directly from Bell's letter.  

Following this correspondence, my paper was typed on my return to Princeton 
and was received by the {\it Physical Review} on Sept. 24, 1968.  (Bell and Jackiw's 
paper, a CERN preprint dated July 16, 1968,  
was submitted to {\it Il Nuovo Cimento}, and received by that journal 
on Sept. 11, 1968.)  My paper was accepted along with a signed referee's 
report from Bjorken, stating 
``This paper opens a topic similar to the old controversies on photon mass 
and nature of vacuum polarization.  The lesson there, as I (no doubt 
foolishly) predict will happen 
here, is that infinities in diagrams are really 
troublesome, and that if the cutoff which is used violates a cherished 
symmetry of the theory, the results do not respect the symmetry.  I will 
also predict a long chain of papers devoted to the question the author has 
raised, culminating in a clever renormalizable cutoff which respects chiral 
symmetry and which, therefore, removes Adler's extra term.''
Thus, acceptance of the point of view that I had advocated was 
not immediate, but only followed over time.  In 1999, 
Bjorken was a speaker at my 60th birthday conference at the 
Institute for Advanced 
Study, and amused the audience by reading from his report, and then very 
graciously gave me his file copy, with an appreciative inscription, as a 
souvenir.  

The viewpoint that the anomaly determined the $\pi^0 \to \gamma \gamma$ 
decay amplitude had significant physical consequences.  In the Appendix 
to my paper, I showed that the value  $S=\frac{1}{6}$ implied by the 
fractionally charged quark model gave a decay amplitude that was roughly 
a factor of 3 too small.  More generally, I showed that 
a triplet constituent model  
with charges $(Q,Q-1,Q-1)$ gave $S=Q-\frac{1}{2}$, and so with integrally 
charged constituents ($Q=0$ or $Q=1$) 
one gets  an amplitude that agrees in absolute 
value, to within the expected accuracy of PCAC, with experiment. I noted in 
my paper that $Q=0$, or $S=-\frac{1}{2}$ 
corresponded to the case in which radiative corrections 
to weak interactions had been shown to be finite, but this choice for 
the  sign of the $\pi^0 \to \gamma \gamma$ amplitude was soon to be ruled out.  
Over the next few months Okubo (1969) and 
Gilman (1969) wrote me letters  accompanying preprints which  
demonstrated, by different methods, that the sign corresponding 
to a single positive integrally charged constituent going around the triangle loop 
agrees with experiment.  Okubo also analyzed various alternative models 
for proton constituents,  and pointed out that while some are excluded by 
the experimentally determined value of $S$, the integrally charged 
Maki (1964)--Hara (1964) single triplet model (the model that I had 
considered in my Appendix, but now with $Q=1$),  and the corresponding 
integrally charged three 
triplet model of Han and Nambu (1965) \big(see also  
Tavkhelidze (1965), Miyamoto (1965), and Nambu (1965)\big), are both 
in accord with the empirical value $S \simeq \frac{1}{2}$.  
In a conference talk a year later, in September 1969 
(Adler, 1970a, {\sf R17}) I reviewed the subject of the anomaly 
calculation of neutral pion decay, as developed in the papers that had 
appeared during the preceding year.  

The work just described gave the first 
indications that neutral pion decay provides empirical evidence that can 
discriminate between different models for hadronic constituents. The correct  
interpretation of the fact that $S\simeq \frac{1}{2}$ came only later, 
when what  we now call the ``color'' degree of freedom was introduced in 
the seminal papers of Bardeen, Fritzsch, and Gell-Mann (1972; reprinted 
as hep-ph/0211388) and Fritzsch and Gell-Mann 
(1971/1972; reprinted as hep-ph/0301127). These papers used my calculation of 
$\pi^0 \to \gamma \gamma$ decay as supporting justification for the 
tripling of the number of fractionally charged quark degrees of freedom,  
thus increasing the theoretical value of $S$ for fractionally charged quarks 
from $\frac{1}{6}$ to $\frac{1}{2}$.  
The paper of Bardeen, Fritzsch, and Gell-Mann  
also pointed out that this tripling would show up in a  
measurement of $R$, the ratio of hadronic to muon pair production in 
electron positron collisions, while 
noting that ``Experiments at present are too  
low in energy and not accurate enough to test this prediction, but in the 
next year or two the situation should change.'', as indeed it did.  

Before leaving the subject of the early history of the anomaly and its 
antecedents, perhaps this is the appropriate place to mention the paper 
of Johnson and Low (1966), which showed that the 
Bjorken (1966)--Johnson--Low (1966) (BJL) method of 
identifying formal commutators with an 
infinite energy limit of Feynman diagrams gives, in significant cases, 
results that differ from the naive field-theoretic evaluation of these 
commutators.  This method was later used by Jackiw and Johnson (1969) and by 
Boulware and myself (Adler and Boulware, 1969, {\sf R18})  
to show that the $AVV$ axial anomaly can be reinterpreted 
in terms of anomalous commutators.  This line of investigation, however, 
did not readily lend itself to a determination of anomaly effects beyond 
leading order.  For example, I still have in my files an unpublished 
manuscript (circa 1966) attempting to use the  BJL method to tackle 
a simpler problem, that of proving that the Schwinger 
term in quantum electrodynamics (QED) 
is a $c$-number to all orders of perturbation 
theory. I believe that this result is true (and it may well have been  
proved by now using operator product expansion methods), 
but I was not able at that time to achieve 
sufficient control of the BJL limits of high order diagrams with general  
external legs to give a proof.  (See also remarks on this in Chapter 4.)

\bigskip
\section*{Anomaly Nonrenormalization}    

We are now ready to address the issue of the determination   
of anomalies beyond leading order in perturbation theory. 
Before the neutral pion low energy theorem could be used as evidence for 
the charge structure of quarks, one needed to be sure that there were no 
perturbative corrections to the anomaly and the low energy theorem 
following from it.  As I noted above, the 
fermion loop-wise argument that I used 
in my  original treatment left me convinced that only the lowest order 
$AVV$ diagram would contribute to the anomaly, but this was not a proof. 
This point of view was challenged in the article by Jackiw and Johnson 
(1969), received by the {\it Physical Review} on Nov. 25, 1968, who stated 
``Adler has given an argument to the end that there exist no higher-order 
effects.  He introduced a cutoff, calculated the divergence, and then let 
the cutoff go to infinity. This is seen in the present context to be  
equivalent to the second method above.  
However, we believe that this method may not 
be reliable because of the dependence on the order of limits.''  And in 
their conclusion, they stated  ``In a definite model the nature of the 
modification (to the axial-vector current divergence equation) can be  
determined, but in general only to lowest order in interactions.'' 
This controversy with 
Jackiw and Johnson was the motivation for a more thorough analysis of the 
nonrenormalization issue undertaken by Bill Bardeen and myself in the fall 
and winter of 1968-1969 (Adler and 
Bardeen, 1969, {\sf R19}) and was cited in the ``Acknowledgments'' section of our 
paper, where we thanked ``R. Jackiw and K. Johnson for a stimulating 
controversy which led to the writing of this paper.''

The paper with Bardeen approached the problem of nonrenormalization by two 
different methods.  We first gave a general constructive argument for 
nonrenormalization of the anomaly to all orders, in both quantum 
electrodynamics and in the $\sigma$-model in which PCAC is canonically 
realized, and we then backed this argument up with an explicit calculation 
of the leading order radiative corrections to the anomaly, showing that   
they cancelled among the various contributing Feynman diagrams.  The 
strategy of the general argument was to note that since the anomaly equations  
written above 
involve unrenormalized fields, masses, and coupling constants, these 
equations are well defined only in a cutoff field theory.  Thus, for both 
electrodynamics and the $\sigma$-model, we constructed cutoff versions by 
introducing photon or $\sigma$-meson regulator fields with mass $\Lambda$.  
(This was simple for the case of electrodynamics, but more difficult,  
relying heavily on Bill Bardeen's prior experience 
with meson field theories, in the case of the 
$\sigma$-model.) 
In both cases, the cutoff prescription allows the usual renormalization 
program to be carried out,  expressing the  unrenormalized quantities 
in terms of renormalized ones and the cutoff $\Lambda$.  In the cutoff 
theories, the fermion loop-wise argument 
I used in my original anomaly paper is 
still valid, because regulating boson propagators does not alter the  
chiral symmetry properties of the theory, and thus it is 
straightforward to prove the validity of the anomaly equations involving  
unrenormalized quantities to all orders of perturbation theory.  

Taking 
the vacuum to two $\gamma$ matrix element of the anomaly equations, and 
applying the Sutherland--Veltman theorem, which asserts the vanishing 
of the matrix element 
of $\partial^{\mu}j_{\mu}^5$ at the special kinematic point $q^2=0$, 
Bardeen and I then got exact low energy theorems for the matrix elements 
$\langle 2\gamma|2im_0j^5|0\rangle$ (in electrodynamics) 
and $\langle 2\gamma|(f_{\pi}M_{\pi}^2/\sqrt{2}) 
\phi_{\pi}|0 \rangle$ (in the $\sigma$-model) 
of the ``naive'' axial-vector divergence 
at this kinematic point, which were given by the negative of the 
corresponding matrix element of the anomaly term.    
However, since we could 
prove that these matrix elements are finite in the limit 
as the cutoff $\Lambda$ 
approaches infinity, this in turn gave exact low energy theorems for the 
renormalized, physical matrix elements in both cases.  One subtlety that 
entered into the all orders calculation was the role of photon rescattering 
diagrams connected to the anomaly term,  but using gauge invariance 
arguments analogous to those involved in the Sutherland--Veltman theorem, 
we were able to show that these diagrams made a vanishing contribution to 
the low energy theorem at the special kinematic point $q^2=0$.  Thus, 
my paper with Bardeen provided a rigorous underpinning for the use of the 
$\pi^0 \to \gamma \gamma$ low energy theorem to study the charge structure 
of quarks.  

In our explicit second order calculation, we calculated the leading order 
radiative corrections to this low energy theorem, arising from addition  
of a single virtual photon or virtual $\sigma$-meson to the lowest order 
diagram.  We did this by two methods, one involving a direct calculation 
of the integrals, and the other (devised by Bill Bardeen) using a clever  
integration by parts argument to bypass the direct calculation.  Both 
methods gave the same answer:  the sum of all the radiative corrections 
is zero, as expected from our general nonrenormalization argument.  We also 
traced the contradictory results obtained in the paper of Jackiw and Johnson 
to the fact that these authors had studied an axial-vector current  
(such as $\overline \psi \gamma_{\mu} \gamma_5 \psi$ in the $\sigma$-model) 
that is not made finite by the usual renormalizations in the absence 
of electromagnetism; as a consequence, the naive divergence of this current  
is not multiplicatively renormalizable.  As we noted in our paper, 
``In other words, the axial-vector current considered by Jackiw and Johnson 
and its naive divergence are not well-defined objects in the usual 
renormalized perturbation theory; hence the ambiguous results which these 
authors have obtained are not too surprising.''  Our result of a definite, 
unrenormalized low energy theorem, we noted, came about because ``In each 
model we have studied a {\it particular} axial-vector current:  in spinor 
electrodynamics, the usual axial-vector current ... and in the $\sigma$ 
model the Polkinghorne (1958a,b) axial-vector current ... which, in the absence 
of electromagnetism, obeys the PCAC condition.''  It is these axial-vector 
currents that obey a simple anomaly equation to all orders in perturbation 
theory, and which give an exact, physically relevant low energy theorem 
for the naive axial-vector divergence.  

This paper with Bill Bardeen should have ended the controversy over whether 
the anomaly was renormalized, but it didn't.  Johnson pointed out in an 
unpublished report that since the anomaly is mass-independent, it should 
be possible to calculate it in massless electrodynamics, for which the naive 
divergence $2im_0j^5$ vanishes and the divergence of the axial-vector 
current directly gives the anomaly.  Moreover, in massless electrodynamics 
there is no need for mass renormalization, and so if one chooses Landau gauge 
for the virtual photon propagator, the second order radiative correction 
calculation becomes entirely ultraviolet finite, with no renormalization 
counter terms needed.  Such a second order calculation was reported by 
Sen (1970), a Johnson student, who claimed to find nonvanishing second 
order radiative corrections to the anomaly.   However, the calculational 
scheme proposed by Johnson and used by Sen has the problem that, 
while ultraviolet finite, there are severe 
infrared divergences, which if not handled carefully can lead to spurious 
results.  After a long and arduous calculation (Adler, Brown, Wong, and 
Young, 1971) my collaborators and I were able to show that the zero mass 
calculation, when properly done, also gives a vanishing second order 
radiative correction to the anomaly.  This  confirmed the result I had found    
with Bardeen, which had by then also been confirmed by different methods 
in the $m_0\not =0$ 
theory in papers of Abers, Dicus, and Teplitz (1971) and Young, Wong, 
Gounaris, and Brown (1971).  

Even this was not the end of controversies over the nonrenormalization 
theorem, as discussed in detail in my review Adler (2004a) that focuses 
specifically on anomaly nonrenormalization.  Suffice it to 
say here that no objections raised have withstood careful analysis, and 
there is now a detailed understanding of anomaly nonrenormalization both by 
perturbative methods,  and by non-perturbative methods proceeding from the 
Callan--Symanzik equations.  There is also a detailed understanding of  
anomaly nonrenormalization in the context of supersymmetric theories, where   
initial apparent puzzles are now resolved.  

\bigskip
\section*{Point Splitting Calculations of the Anomaly}

At this point let me backtrack, and discuss the role of point-splitting 
methods in the study of the Abelian electrodynamics anomaly.  In the present 
context, point-splitting was first used in the discussion 
given by Schwinger (1951) of the
pseudoscalar-pseudovector equivalence theorem, to be described in more 
detail shortly.  Almost immediately 
following circulation of the seminal anomaly preprints 
in the fall of 1968, Hagen (1969, received Sept. 24, 1968, and a
letter to me dated Oct. 16, 1968), Zumino 
(1969, and 
a letter to me dated Oct. 7, 1968), and Brandt (1969, received  
Dec. 17, 1968, and a letter to me dated Oct. 16, 1968) 
all rederived the anomaly formula by a point-splitting method.  
Independently,  a point-splitting  
derivation of the anomaly was given by Jackiw and 
Johnson (1969, received 25 November, 1968), who explicitly made the   
connection to Schwinger's earlier work (Johnson was a Schwinger student, and 
was well acquainted with Schwinger's body of work).  
The point of all of 
these calculations is that the anomaly   
can be derived by formal algebraic use of the equations of motion, 
provided one redefines 
the singular product $\overline \psi(x) \gamma_{\mu} \gamma_5 \psi(x)$ 
appearing in the axial-vector current by the point-split expression 
\begin{equation}
\lim_{x \to x^{\prime}} \overline \psi(x^{\prime}) \gamma_{\mu} \gamma_5 
\exp[-ie\int_{x^{\prime}}^x dx^{\lambda} B_{\lambda}] \psi(x)~~~,\nonumber
\end{equation}
and takes the limit $x^{\prime} \to x$ at the end of the calculation. 

Responding to these developments, 
I appended a ``Note added in proof'' to my anomaly paper, mentioning the 
four  field-theoretic, point-splitting derivations 
that had subsequently been 
given, and adding  ``Jackiw and Johnson point out that the 
essential features of the field-theoretic derivation, in the case of 
external electromagnetic fields, are contained in J. Schwinger, {\it Phys. 
Rev.} {\bf 82}, 664 (1951)''.  
What to me was an interesting irony emerged from learning  
of the connection between anomalies and the 
famous Schwinger (1951) paper on vacuum polarization.    
I had in fact read Section II 
and the Appendices of the 1951 paper, when Alfred Goldhaber and I, during  
our senior year at Harvard (1960-61), 
did a reading course on quantum electrodynamics 
with Paul Martin, which focused on papers in Schwinger's reprint volume 
(Schwinger, 1958).  Paul had told us to read the parts of the Schwinger paper 
that were needed to calculate the $VV$ vacuum polarization loop, but 
to skip the rest as being too technical.  Reading Section V of Schwinger's 
paper brought back to mind a brief, forgotten conversation I had had with 
Jack Steinberger, who was Director of the Varenna Summer School in 1967.  
Steinberger had told me that he had done a calculation 
on the pseudovector-pseudoscalar 
equivalence theorem for $\pi^0 \to \gamma \gamma$, but had 
gotten different answers in the two cases; also that Schwinger had claimed 
to reconcile the answers, but that he (Steinberger) couldn't make sense out 
of Schwinger's argument.  Jack had urged me to look at it, which I never did 
until getting the Jackiw--Johnson preprint, but in retrospect everything 
fell into place, and the connection to Schwinger's work was apparent.    

This now brings me to the question, did Schwinger's paper constitute  
the discovery of  
the anomaly?  Both Jackiw, in his paper with Johnson, and I were 
careful to note the 
connection between Schwinger's (1951) paper and the point-splitting 
derivations of the anomaly, once it was called to our attention.  However, 
recently some of Schwinger's former students have gone further, arguing 
that Schwinger was the  discoverer of the anomaly and that my paper and 
that of Bell and Jackiw were merely a ``rediscovery'' of a previously 
known result. I believe that this claim goes beyond the published record 
of what is in Schwinger's paper, as analyzed in  detail in Sec. 2.3 and  
Appendix A of my review Adler (2004a).  Stated briefly, Schwinger's 
calculation was devoted to making the pseudovector calculation give {\it the 
same} non-zero answer as the pseudoscalar one, and what 
Schwinger calls the redefined axial-vector 
divergence is in fact    
{\it not} the divergence of the gauge-invariant 
axial-vector current, but rather the axial-vector current  
divergence {\it minus} the anomaly.   
In other words, Schwinger's calculation effectively transposes the 
anomaly term to the left-hand side of the anomaly equation, so that what he  
evaluates is the effective Lagrangian arising from the left-hand side of 
the equation 
\begin{equation}
\partial^{\mu} j_{\mu}^5(x) 
- \frac{\alpha_0}{4 \pi} 
F^{\xi \sigma}(x) F^{\tau \rho}(x) \epsilon_{\xi\sigma \tau \rho}
= 2im_0j^5(x)   
~~~,\nonumber
\end{equation}
which then necessarily gives the same result as calculation of an 
effective Lagrangian 
from the right-hand side, which is pseudoscalar coupling.  There is no 
gauge-invariant axial-vector current for which the 
combination on the left-hand side is the divergence, but 
as shown in Eqs. (58) and (59) of {\sf R16}, 
there is a gauge-non-invariant axial-vector current which has this 
divergence.  

The use of a point-splitting method was of course 
important and fruitful, and in retrospect, the axial anomaly is hidden 
within Schwinger's calculation.  But Schwinger never 
took the crucial step of observing that the axial-vector current matrix 
elements cannot, in a renormalizable quantum theory, be made to satisfy 
all of the expected classical symmetries.  And more specifically, he   
never took the step of defining a gauge-invariant axial-vector 
current by point splitting, which has a well-defined anomaly term in  
its divergence, with the anomaly term  completely 
accounting for the disagreement 
between the pseudoscalar and pseudovector calculations of neutral pion 
decay.  So I would say that although Schwinger took steps in the right 
direction, particularly in noting the utility of point-splitting in 
defining the axial-vector current, his 1951 paper {\it obscured} the 
true physics and does  not mark the discovery of the anomaly. 
This happened only much later, in 1968, and led to a flurry of activity 
by many people.  My view is supported, I believe, by the 
fact that Schwinger's calculation seemed arcane, even to 
people (like Steinberger)  with whom he had talked about it  
and to colleagues familiar with his work,  
and exerted no influence on the field until after preprints on the seminal   
work of 1968 had appeared.  
\bigskip
\section*{The Non-Abelian Anomaly, Its Nonrenormalization     
 and Geometric Interpretation}

Since in the chiral limit the $AVV$ triangle is identical to an $AAA$ 
triangle (as is easily seen by an argument involving anticommutation of a $\gamma_5$ 
around the loop), I knew 
already in unpublished notes dating from the late summer of 1968 that 
the $AAA$ triangle would also have an anomaly; a similar observation was   
also made by Gerstein and Jackiw (1969).  From fragmentary calculations begun in 
Aspen I suspected that higher loop diagrams might have anomalies as well, 
so after the nonrenormalization work with Bill Bardeen was finished I  
suggested to Bill that he work out the general anomaly for larger 
diagrams. (I was at that point involved in other calculations with  
Wu-Ki Tung, on the perturbative breakdown of scaling formulas such as 
the Callan--Gross relation, to be discussed shortly.)  I showed Bill my notes, 
which turned out to 
be of little use, but which 
contained a very pertinent remark by Roger Dashen that including charge 
structure (which I had not) would allow a larger class of potentially 
anomalous diagrams.   Within a few weeks, Bill carried out a 
brilliant calculation, by point-splitting methods, of the general anomaly 
in both the Abelian {\it and} the non-Abelian cases (Bardeen, 1969).  
Expressed in terms of vector and axial-vector Yang--Mills field strengths 
\begin{eqnarray}
F_V^{\mu\nu}(x)&=&\partial^{\mu}V^{\nu}(x)-\partial^{\nu}V^{\mu}(x) 
-i[V^{\mu}(x),V^{\nu}(x)]-i[A^{\mu}(x),A^{\nu}(x)]~~~,\nonumber\\\cr
F_A^{\mu\nu}(x)&=&\partial^{\mu}A^{\nu}(x)-\partial^{\nu}A^{\mu}(x) 
-i[V^{\mu}(x),A^{\nu}(x)]-i[A^{\mu}(x),V^{\nu}(x)]~~~,\cr
\nonumber
\end{eqnarray}
Bardeen's result takes 
the form 
\begin{eqnarray}
\partial^{\mu}J_{5\mu}^{\alpha}(x)&=&{\rm normal~divergence~term} \nonumber\\
&+&(1/4\pi^2)\epsilon_{\mu\nu\sigma\tau}{\rm tr}_I[\lambda_A^{\alpha}
[(1/4)F_V^{\mu\nu}(x)F_V^{\sigma \tau}(x)
+(1/12)F_A^{\mu\nu}(x)F_A^{\sigma \tau}(x)\nonumber\\
&+&(2/3)iA^{\mu}(x)A^{\nu}(x)F_V^{\sigma \tau}(x) 
+(2/3)i F_V^{\mu \nu}(x) A^{\sigma}(x)A^{\tau}(x)\nonumber\\ 
&+&(2/3)i  A^{\mu}(x)F_V^{\nu \sigma}(x) A^{\tau}(x)
-(8/3)A^{\mu}(x)A^{\nu}(x)A^{\sigma}(x)A^{\tau}(x) ]~~~,
\nonumber
\end{eqnarray} 
with ${\rm tr}_I$ denoting a trace over internal degrees of freedom, and   
$\lambda_A^{\alpha}$ the internal symmetry matrix associated with 
the axial-vector external field.  In the Abelian case, with trivial  
internal symmetry structure, 
the terms involving two or three factors of $A^{\mu,\nu,...}$ vanish by 
antisymmetry of $\epsilon_{\mu\nu\sigma\tau}$, and there are only 
$AVV$ and $AAA$ triangle anomalies.  When there is non-trivial internal 
symmetry or charge 
structure, there are anomalies associated with the box and pentagon 
diagrams as well, confirming Dashen's intuition mentioned earlier.  
Bardeen notes that whereas the triangle and box anomalies result from linear 
divergences associated with these diagrams, the pentagon anomalies 
arise not  from linear divergences, but rather from the definition of the 
box diagrams to have the correct vector current Ward identities.  
Bardeen also notes, in his 
conclusion, another prophetic remark of Dashen, to the effect that the 
pentagon anomalies should add anomalous terms to the PCAC low energy 
theorems for five pion scattering; I shall return to this shortly.

There are two distinct lines of argument leading to the conclusion that 
the non-Abelian chiral anomaly also has a nonrenormalization theorem, and 
is given exactly by Bardeen's leading order calculation.  The first 
route parallels that used in the Abelian case, involving variously a 
loop-wise regulator construction,  
explicit fourth order calculation, and an 
argument using the Callan--Symanzik equations; for detailed  
references, see  Adler (2004a).   
The conclusion in all cases is that the Adler--Bardeen 
theorem extends to the non-Abelian case.  Heuristically, what is happening 
is that except for a few small one-fermion loop diagrams, 
non-Abelian theories, just like Abelian ones, are made 
finite by gauge invariant regularization of the gluon 
propagators.  But this regularization has no effect on the chiral properties 
of the theory, and therefore does not change its anomaly structure, which     
can thus be deduced from the structure of the few small fermion loop 
diagrams for which naive classical manipulations break down.    

The second route leading to the conclusion that the non-Abelian anomaly 
is nonrenormalized might be termed ``algebraic/geometrical'', and consists 
of two steps.  The first step consists of a demonstration that 
the higher order terms in Bardeen's non-Abelian formula are completely 
determined by the leading, Abelian anomaly.  During a summer visit to  
Fermilab in 1971, I collaborated with Ben Lee, Sam Treiman, and Tony Zee 
(Adler, Lee, Treiman, and Zee  1971, {\sf R20}) in a calculation 
of a low energy theorem for the 
reaction $\gamma + \gamma \to \pi + \pi + \pi$ in both the neutral and charged
pion cases.  This was motivated in part by discrepancies in calculations 
that had just appeared in the literature, and in part by its relevance to 
theoretical unitarity calculations of a lower bound on the $K_L^0 \to 
\mu^+ \mu^-$ decay rate.  Using PCAC, we showed that the fact that the 
$\gamma + \gamma \to 3 \pi$ matrix elements vanish in the limit when a 
final $\pi^0$ becomes soft, together with photon gauge invariance, relates  
these amplitudes to the matrix elements $F^{\pi}$ for 
$\gamma +\gamma \to \pi^0$ and $F^{3\pi}$ for $\gamma \to \pi^0 + 
\pi^+ + \pi^-$, and moreover, gives a relation between the latter two 
matrix elements, 
\begin{equation}
eF^{3 \pi} =f^{-2} F^{\pi}~~~,~f=\frac{f_{\pi}} 
{\sqrt 2}~~~.\nonumber
\end{equation} 
Thus all of the matrix elements in question are uniquely determined by 
$F^{\pi}$, which itself is determined by the $AVV$ anomaly calculation.  
An identical result for the same reactions was independently given by 
Terent'ev (Terentiev) (1971).  In the meantime, in a 
beautiful formal analysis, 
Wess and Zumino (1971) showed that the current algebra satisfied by the 
flavor $SU(3)$ octet of vector and axial-vector currents implies 
integrability or ``consistency'' conditions on the non-Abelian 
axial-vector anomaly, which are satisfied 
by the Bardeen formula, and conversely, that these constraints 
uniquely imply the Bardeen structure 
up to an overall factor, which is determined by the Abelian $AVV$ anomaly.   
By introducing an auxiliary pseudoscalar field, Wess and Zumino were able 
to write down a local action obeying the anomalous Ward identities and the  
consistency conditions.  (There is no corresponding local action involving 
just the vector and axial-vector currents, since if there were, the 
anomalies could be eliminated by a local counterterm.)  
Wess and Zumino also gave expressions for the processes 
$\gamma \to 3 \pi$ and $2 \gamma \to 3 \pi$ discussed by Adler et al.   
and Terentiev, as well as giving the anomaly contribution to the five 
pseudoscalar vertex.  The net result of these three simultaneous pieces of 
work was to show that the Bardeen formula has a rigidly constrained 
structure, up to an overall factor given by the $\pi^0 \to \gamma \gamma$ 
decay amplitude.  

The second step in the ``algebraic/geometric'' route to anomaly 
renormalization is a celebrated paper of Witten (1983), which shows  
that the Wess--Zumino action has a  representation as the integral of 
a fifth rank antisymmetric tensor (constructed from the auxiliary 
pseudoscalar field) 
over a five-dimensional disk of which four-dimensional space is the 
boundary.  In addition to giving a new interpretation of the Wess--Zumino 
action $\Gamma$, Witten's argument also gave a constraint on the 
overall factor in $\Gamma$ that was not determined by the Wess--Zumino    
consistency argument.  Witten observed 
that his construction is not unique, because a closed five-sphere 
intersecting a hyperplane gives two ways of bounding the four-sphere along 
the equator with a five dimensional hemispherical disk.  Requiring 
these two constructions 
to give the same value for  
$\exp(i\Gamma)$, which is the way the anomaly enters into a Feynman 
path integral, requires integer quantization of the overall coefficient 
in the Wess--Zumino--Witten action.  This integer can be read off from 
the $AVV$ triangle diagram, and for the case of an underlying color 
$SU(N_c)$ gauge theory turns out to be just $N_c$, the number of colors.     

To summarize, the ``algebraic/geometric'' approach shows that the Bardeen 
anomaly has a unique structure, up to an overall constant, and moreover that 
this overall constant is constrained by an integer 
quantization condition. Hence once the overall constant is fixed by 
comparison with leading order perturbation theory (say in QCD), it is 
clear that this  
result must be exact to all orders, since the presence of renormalizations 
in higher orders of the strong coupling constant would lead to violations 
of the quantization condition.   

The fact that non-Abelian anomalies are given by an overall rigid structure 
has important implications for quantum field theory.  For example,  
the presence of anomalies spoils the renormalizability of non-Abelian 
gauge theories and requires the cancellation
of gauged anomalies between different fermion species \big(see 
Gross and Jackiw (1972), Bouchiat, Iliopoulos, and Meyer (1972), and 
Weinberg (1973)\big), through imposition   
of the condition ${\rm tr} \{T_{\alpha},T_{\beta}\} T_{\gamma}=0$ for all 
$\alpha,\beta,\gamma$, with $T_{\alpha}$ the coupling matrices of gauge 
bosons to left-handed fermions.  
The fact that  
anomalies have a rigid structure then implies that once these anomaly 
cancellation conditions are imposed for the lowest order anomalous triangle 
diagrams, no further conditions arise from anomalous square or pentagon  
diagrams, or from radiative corrections to these leading fermion loop
diagrams.  Other places where the one-loop geometric 
structure of non-Abelian 
anomalies enters are in instanton physics, and in the 't Hooft anomaly 
matching conditions.  These and other chiral anomaly applications are  
discussed in more detail in my review Adler (2004a),   
and also in my {\it Encyclopedia of Mathematical Physics} article  
Adler (2004b). Both of these sources give extensive references to recent 
review articles and books on anomalies, which update the 1970 reviews 
given in my Brandeis lectures (Adler, 1970b) and in 
Jackiw's Brookhaven lectures (Jackiw, 1970).  

\bigskip
\section*{Perturbative Corrections to Scaling}

While finishing the paper with Bardeen on anomaly nonrenormalization, I 
had embarked on a different set of perturbative calculations with Wu-Ki Tung;  
these became a forerunner of a different kind of ``anomaly'', the anomalous 
scaling observed in deep inelastic electron and neutrino scattering.  
Our starting point was the question of whether applications of the 
Bjorken (1966) limit technique, which assumed that the asymptotic behavior 
of time-ordered products is given by the ``naive'' or free field theory        
equal time commutator, would be modified in perturbation theory.  Strong 
hints in this direction had been given in a paper of Johnson and Low (1966), 
which showed that the ``Bjorken--Johnson--Low'' limit can produce anomalous 
commutators,  and related results were also obtained in an earlier paper of 
Vainshtein and Ioffe (1967);  our aim was to do calculations focusing on 
several physically important applications not covered in this previous work.      
These were the calculation by Bjorken (1966) of the  
radiative corrections to $\beta$-decay, the Bjorken (1967)  
backward-neutrino-scattering asymptotic sum rule, 
and the Callan--Gross (1969) relation 
relating the ratio of the longitudinal to transverse deep inelastic electron 
scattering cross sections to the constitution of the electric current, with   
the latter an application both of  the Bjorken--Johnson--Low limit method, and 
of the later proposal by Bjorken (1969) of scaling of the deep inelastic   
structure functions.  

For our test model, we considered an $SU(3)$ 
triplet of spin-1/2 particles bound by exchange of a massive singlet gluon, 
which we took as either a vector, scalar, or pseudoscalar.  The results 
of the vector exchange calculation, to leading order of perturbation theory, 
were reported in Adler and Tung (1969), {\sf R21}, while additional 
leading order results in the scalar and pseudoscalar gluon cases, and 
some fourth order results, were given in the follow-up paper Adler and 
Tung (1970), {\sf R22}.  We concluded that the Callan--Gross relation for   
spin-1/2 quarks, 
which asserts the vanishing of $q^2 \sigma_L(q^2,\omega)$  
for large $q^2$ with fixed scaling variable  $\omega$, breaks down in 
leading order of perturbation theory. A similar conclusion was also 
reached by Jackiw and Preparata (1969a,b), whose first 
paper appears in the same issue of {\it Physical Review Letters} as our   
paper {\sf R21}. Tung and 
I related the breakdown of the Callan--Gross relation 
to a corresponding breakdown of Bjorken's backward  
neutrino sum rule.  We also  showed that the certain current commutators 
receive a systematic pattern of logarithmic asymptotic corrections, and  
calculated the leading perturbative correction to the logarithmically 
divergent part of the radiative corrections to $\beta$ decay.   Tung 
(1969), while still at the Institute, and Jackiw and Preparata (1969c),  
went on to carry out general analyses 
of the range of validity and breakdown of the Bjorken--Johnson--Low limit 
in perturbation theory.  

These papers had a number of implications for subsequent developments.  
The logarithmic deviations from the Callan--Gross relation 
were soon understood in a more systematic  
way through the Wilson (1969) operator product expansion and 
the Callan (1970)--Symanzik (1970) equations, which gave anomalous dimensions 
in accord with the leading order results obtained by Tung and me 
and by  Jackiw and Preparata, and with the fourth order 
results obtained by Tung and me in {\sf R22}; for a discussion of this,   
see B\'eg (1975).  The fact that perturbative field theory gives strong  
violations of scaling led to a skepticism as to whether field theory 
could describe the strong interactions at all.  For example, Fritzsch 
and Gell-Mann (1971/1972), in their long paper on ``Light Cone Current 
Algebra'', remarked that ``The renormalized perturbation theory, taken 
term by term, reveals various pathologies in commutators of currents.  Not 
only are there in each order logarithmic singularities on the light cone, 
which destroy scaling, and violations of the rule that $\sigma_L/\sigma_T 
\to 0$ in the Bjorken limit, but also a careful perturbation theory treatment 
show the existence of higher singularities on the light cone...'' This 
was one of their motivations for introducing the light cone algebra, which 
abstracted from field theory 
algebraic relations that led to scaling and parton model results, with the 
field theory itself being discarded.  

At the same time, there were also 
thoughts that a renormalization group fixed point in field theory might 
provide a remedy.  In the same article, Fritzsch and Gell-Mann noted that 
in the context of a singlet vector gluon theory, ``we must imagine that the 
sum of perturbation theory yields the special case of a `finite vector 
theory'${}^{27}$[reference to Gell-Mann and Low, and Baker and Johnson] 
if we are to bring the 
vector gluon theory and the basic algebra into harmony.''  Quite 
independently, in a conference talk at Princeton that I gave in October  
of 1971 
\big(published considerably later as Adler (1974), {\sf R23}\big),  
in Section 
2.4, on ``Questions raised by the breakdown of the BJL limit'', I 
made the remark ``Can one make a consistent calculational scheme in which 
Bjorken limits, the Callan--Gross relation and scaling are all valid?  This  
is a {\it real challenge} to theorists...Perhaps a successful approach 
would involve summation of perturbation theory graphs plus use of the 
Gell-Mann--Low eigenvalue condition (see sect. 3).'' \big(I made these comments 
at just the time   
when I was working on a possible eigenvalue condition in quantum 
electrodynamics, growing out of the work of Gell-Mann and Low, and 
Johnson, Baker, and Willey, 
as described below in Chapter 4.  The relevance of an   
eigenvalue to power law  behavior was also pointed out 
in the papers  of Callan (1972) and of
Christ, Hasslacher, and Mueller (1972), which I included as references when 
I edited my 1971 conference talk in the fall of 1972.\big)  
However, in 
the field theories then under consideration, there was an obstacle 
to realizing this idea. As I noted in Sec. 3 of my Princeton talk, for  
singlet gluon theories 
the renormalization group methods suggested either no simple 
scaling behavior (if there were no renormalization group fixed point at which 
the $\beta$ function had a zero), or power law deviations 
from scaling of the form $(q^2)^{-\gamma}$ (if there were a fixed point 
at a nonzero coupling value 
$\lambda_0$ where $\beta$ vanished, with $\gamma$ the 
value of the anomalous 
dimension at the fixed point).  Since in a strong coupling theory $\gamma$ 
would be expected to be large at the fixed point, power law deviations from 
scaling looked to be too large to agree with experiment.  

It took another eighteen months for this obstacle to be overcome.  Three
developments were involved:  the introduction of the modern form of ``color''
as a tripling of the fractionally charged quark degrees of freedom by 
Bardeen, Fritzsch, and Gell-Mann (1972), the non-Abelian gauging of this 
form of color by Fritzsch and Gell-Mann (1972), and finally, in line with  
Gell-Mann's dictum ``Nature reads the books of free field theory'',    
a search for field theories that 
would have almost free behavior in the scaling limit.  The conclusion  
of this search, the discovery of the asymptotic freedom of non-Abelian 
gauge theories and its implications by Gross, Politzer, and Wilczek,  
in the end proved a realization of the field-theoretic route that been  
contemplated by various people in 1971.  
In asymptotically free theories, because the renormalization 
group fixed point (the Gell-Mann--Low eigenvalue) 
is at zero coupling, where the anomalous dimension  
$\gamma$ vanishes, the deviations from scaling are not powers of $q^2$, 
but rather only powers of 
$\log q^2$, with exponents that can be calculated in leading order of 
perturbation theory.  Thus the deviations from scaling predicted by 
non-Abelian gauge theories, and specifically by quantum chromodynamics (QCD)  
as the theory of the strong interactions, 
are much weaker than would be expected for 
singlet gluon theories, and are compatible with experiment.  

Returning briefly to the calculations that Tung and I did, our 
results for the 
radiative corrections to $\beta$-decay in the singlet vector gluon model 
turned out later to have applications in the QCD context.  
They can be converted to the realistic case of the octet gluon of  
QCD  by multiplication by a color factor, as discussed in the 
review of Sirlin (1978), and so have become part of the   
technology for calculating radiative corrections to weak processes.  

\bigskip 

\section*{Trace Anomalies to  All Orders}

In an influential paper Wilson (1969) proposed the 
operator product expansion, 
incorporating ideas on the approximate scale invariance of the strong 
interactions suggested by Mack (1968).  As one of the applications of 
his technique, Wilson discussed $\pi^0 \to  2\gamma$ decay and the 
axial-vector anomaly from 
the viewpoint of the short distance singularity of the coordinate space 
$AVV$ three-point function. Using these methods, Crewther (1972) and 
Chanowitz and Ellis (1972) investigated the short distance structure of 
the three-point function  $\theta V_{\mu}V_{\nu}$, 
with $\theta=\theta_{\mu}^{\mu}$ 
the trace of the energy-momentum tensor, and concluded that this 
is also anomalous, thus confirming earlier indications of a perturbative 
trace anomaly obtained in a study of broken scale invariance by 
Coleman and Jackiw (1971).  
Letting $\Delta_{\mu\nu}(p)$ be the momentum space 
$\theta V_{\mu}V_{\nu}$ three point function, and $\Pi_{\mu\nu}$ be 
the corresponding $V_{\mu}V_{\nu}$ two-point function, the naive Ward 
identity $\Delta_{\mu\nu}(p)=(2-p_{\sigma}\partial/\partial p_{\sigma}) 
\Pi_{\mu\nu}(p)$ is modified to 
\begin{equation}
 \Delta_{\mu\nu}(p)=\left(2-p_{\sigma}\frac{\partial}{\partial p_{\sigma}}
\right) \Pi_{\mu\nu}(p)-\frac{R}{6 \pi^2}(p_{\mu}p_{\nu}-\eta_{\mu\nu}p^2) 
~~~~,\nonumber
\end{equation}
with the trace anomaly coefficient $R$ given by 
\begin{equation}
R=\sum_{i,{\rm spin} \frac{1}{2}}Q_i^2 
+\frac{1}{4} \sum_{i,{\rm spin}\,0}Q_i^2~~~.\nonumber
\end{equation} 
Thus, for QED, with a single fermion of charge $e$, 
the anomaly term is $-[2\alpha/(3 \pi)] (p_{\mu}p_{\nu}-\eta_{\mu\nu}p^2)$.
In a subsequent paper, Chanowitz and Ellis (1973) 
showed that the fourth order  
trace anomaly can be read off directly from the coefficient of the leading 
logarithm in the asymptotic behavior of $\Pi_{\mu\nu}(p)$, giving to next 
order an anomaly coefficient $ -2\alpha/(3 \pi) -\alpha^2/(2 \pi^2)$.  
Thus, their fourth order argument indicated a direct connection 
between the trace anomaly and the renormalization group $\beta$ function.

My involvement with trace anomalies began roughly five years later, when 
{\it Physical Review} sent me for refereeing a paper by Iwasaki (1977).  In   
this paper, which noted the relevance to trace anomalies, Iwasaki proved 
a kinematic theorem on the vacuum to two photon matrix element of the  
trace of the energy-momentum tensor, 
that is an analog of the Sutherland--Veltman theorem for the vacuum to two  
photon matrix element of the divergence of the axial-vector current.  
Just as the latter has a kinematic zero at $q^2=0$, Iwasaki showed that 
the kinematic structure of the vacuum to two photon matrix element of the 
energy-momentum tensor implies, when one takes the trace, that there is 
also a kinematic zero at $q^2=0$, irrespective of the presence of   
anomalies (just as the Sutherland--Veltman result holds in the presence of 
anomalies).   Reading this article suggested the idea that just as the 
Sutherland--Veltman theorem can be used as part of an argument to prove 
nonrenormalization of the axial-vector anomaly, Iwasaki's theorem could be 
used to analogously calculate the trace anomaly to all orders.    
(In addition  
to writing a favorable report on Iwasaki's paper, I invited him to 
spend a year at the IAS, which he did during the 
1977-78 academic year.)  During the spring of 1976    
I wrote an initial preprint attempting an all 
orders calculation of the trace anomaly in quantum electrodynamics,  
but this had an error pointed out to me by Baqi B\'eg.  
Over the summer of 1976 
I then collaborated with two local postdocs, John Collins (at Princeton) 
and Anthony Duncan (at the Institute), to 
work out a corrected version (Adler, Collins, and Duncan, 1977, {\sf R24}). 
Collins 
and Duncan simultaneously teamed up with another Institute postdoc,  
Satish Joglekar, to apply similar ideas to quantum chromodynamics, published 
as Collins, Duncan, and Joglekar (1977), and independently the same result 
for QCD was obtained by N. K. Nielsen (1977). Similar results were given 
in a preprint of Minkowski (1976), which grew out of discussions in the 
Gell-Mann group at Cal Tech in which the role of the $\beta$ function in 
the trace anomaly formula, and its implications for generating the scale 
of the strong interactions, were appreciated (C. T. Hill, private communication, 2005, and P. Minkowski, private communication, 2005).  

In the simpler case of QED, the argument based on Iwasaki's 
theorem is given in Section II of {\sf R24}.  The basic idea is to                                                                                                                
use Iwasaki's result for the vacuum to two photon matrix element of the 
trace of the energy momentum tensor, together with expressions for   
the  electron to electron and the vacuum to two photon 
matrix elements of the ``naive'' trace $m_0\overline \psi \psi$ given by  
application of the Callan--Symanzik equations.  The final result   
for the trace is given by 
\begin{equation}
\theta_{\mu}^{\mu} = [1+\delta(\alpha)] m_0 \overline \psi \psi 
+\frac{1}{4} \beta(\alpha)  N[F_{\lambda \sigma}F^{\lambda \sigma}]
+... ~~~,\nonumber
\end{equation}
with $N[~~~]$ an explicitly defined subtracted operator, with $...$ 
indicating terms that vanish by the equations of motion, and with 
$\delta(\alpha)$ and  $\beta(\alpha)$ the renormalization group 
functions defined by 
$1+\delta(\alpha)=(m/m_0)\partial m_0/\partial m$ and  
$\beta(\alpha)= (m/\alpha) \partial \alpha/\partial m$.  
The first two terms in the power series expansion of the coefficient 
of the $F_{\lambda \sigma}F^{\lambda \sigma}$ 
term in the trace agree with the fourth-order calculation of 
Chanowitz and Ellis.  The trace equation in QCD has a similar structure, 
again with the $\beta$ function appearing as the anomaly coefficient.  The 
fact that the trace anomaly coefficient is given by the appropriate 
$\beta$ function extends to the supersymmetric case, and leads to interesting 
issues that are reviewed in the final section of Adler (2004a).  
The appearance of the $\beta$ function in the anomaly coefficient has  
also played a role in the inference of 
the structure of effective Lagrangians from the form of the 
trace anomaly; see, for example, Pagels and Tomboulis (1978) for an application 
to QCD, and Veneziano and Yankielowicz (1982) for an 
application to supersymmetric Yang--Mills theory.

\section*{References for Chapter 3}

\begin{itemize}

\item Abers, E. S., D. A. Dicus, and V. L. Teplitz (1971). Ward Identities for 
$\eta$ Decay in Perturbation Theory.  
{\it Phys. Rev. D} {\bf 3}, 485-497. 

\item Adler, S. L. (1969) {\sf R16}.  Axial-Vector Vertex in Spinor 
Electrodynamics.  {\it Phys. Rev.} {\bf 177}, 2426-2438. 

\item Adler, S. L. (1970a) {\sf R17}.  $\pi^0$ Decay, in 
{\it High-Energy Physics and 
Nuclear Structure}, S. Devons, ed. (Plenum Press, New York), 
pp. 647-655. 

\item Adler, S. L. (1970b).  Perturbation Theory Anomalies, in {\it Lectures on 
Elementary Particles and Quantum Field Theory}, Vol. 1,  S. Deser, M.   
Grisaru, and H. Pendleton, eds.  (M.I.T. Press, Cambridge, MA), 
pp. 3-164. 

\item Adler, S. L. (1974) {\sf R23}. Anomalies in Ward Identities and 
Current Commutation 
Relations, in {\it Local Currents and Their Applications, Proceedings of 
an Informal Conference}, Princeton, NJ, 8-10 October, 1971, D. H. Sharp 
and A. S. Wightman, eds. (North-Holland, Amsterdam and American Elsevier, 
New York), pp. 142-168. Pages 162-168 are reprinted here. 
 Although overtaken by subsequent events, this 
proceedings was nonetheless seen through to publication by the editors, 
who in their Preface noted that ``Since the speakers chose to survey 
the achievements and prospects of their subjects, they concentrated on 
essentials.  As a result, this collection of their talks seems to 
the organizers to provide a very useful review of the state of the art as 
of 1971.''  The published version was based on a handwritten manuscript  
prepared from my notes by David Sharp, which he sent me with a letter 
dated September 20, 1972, asking me to check it and add references. That 
is why the references include  papers that were circulated in 1972, 
after the conference took place. 

\item Adler, S. L. (2004a). Anomalies to All Orders; arXiv: hep-th/0405040.  Published in 
{\it Fifty Years of Yang-Mills Theory}, G. 't Hooft, ed.  (World Scientific, 
Singapore, 2005). 

\item Adler, S. L. (2004b). Anomalies; arXiv: hep-th/0411038.  To appear in {\it 
Encyclopedia of Mathematical Physics}, to be published by Elsevier in 
early 2006. 

\item Adler, S. L. and W. A. Bardeen (1969) {\sf R19}.  
Absence of Higher-Order Corrections 
in the Anomalous Axial-Vector Divergence Equation.  {\it Phys. Rev.} 
{\bf 182}, 1517-1536. 

\item Adler, S. L. and D. G. Boulware (1969) {\sf R18}. Anomalous 
Commutators and the Triangle 
Diagram.  {\it Phys. Rev.} {\bf 184}, 1740-1744. 
  
\item Adler, S. L., R. W. Brown, T. F. Wong, and B.-L. Young (1971).  Vanishing 
of the Second-Order Correction to the Triangle Anomaly in Landau-Gauge,  
Zero-Fermion-Mass Quantum Electrodynamics.  {\it Phys. Rev. D} {\bf 4}, 
1787-1808. 

\item Adler, S. L., J. C. Collins, and A. Duncan (1977) {\sf R24}. 
Energy-Momentum-Tensor  
Trace Anomaly in Spin-1/2 Quantum Electrodynamics.  {\it Phys. Rev. D}  
{\bf 15}, 1712-1721. 

\item Adler, S. L., B. W. Lee, S. B. Treiman, and A. Zee (1971) {\sf R20}.  
Low Energy 
Theorem for $\gamma + \gamma \to \pi + \pi + \pi$.  {\it Phys. Rev. D} 
{\bf 4}, 3497-3501. 

\item Adler, S. L. and W.-K. Tung (1969) {\sf R21}.  Breakdown of Asymptotic 
Sum Rules in Perturbation Theory.  {\it Phys. Rev. Lett.} {\bf 22},  
978-981. 

\item Adler, S. L. and W.-K. Tung (1970) {\sf R22}.  Bjorken Limit in Perturbation 
Theory. {\it Phys. Rev. D} {\bf 1}, 2846-2859. 

\item Bardeen, W. (1969).  Anomalous Ward Identities in Spinor Field Theories.  
{\it Phys. Rev.} {\bf 184}, 1848-1859. 

\item Bardeen, W. A., H. Fritzsch, and M. Gell-Mann (1972).  Light Cone Current  
Algebra, $\pi^0$ Decay, and $e^+e^-$ Annihilation, CERN preprint TH. 1538, 
21 July, 1972.  Later published in {\it Scale and Conformal Symmetry in 
Hadron Physics}, R. Gatto, ed. (Wiley, New York, 1973), pp. 139-151; 
reissued recently 
as arXiv: hep-ph/0211388.  See also R. J. Oakes, Summary--Second Week, {\it Acta Phys. Austriaca Suppl.} {\bf IX}, 905-909, which summarizes remarks by Gell-Mann at the Schladming Winter School, Feb. 21-Mar. 4, 1972. 

\item B\'eg, M. A. B. (1975). Anomalous Algebras and Neutrino Sum Rules.  {\it 
Phys. Rev. D} {\bf 11}, 1165-1170. 

\item Bell, J. S. and R. Jackiw (1969).  A PCAC Puzzle: $\pi^0 \to \gamma \gamma$ 
in the $\sigma$-Model.  {\it Nuovo Cimento A} {\bf 60}, 47-61. 

\item Bjorken, J. D. (1966).  Applications of the Chiral $U(6)\otimes U(6)$ 
Algebra of Current Densities.  
{\it Phys. Rev.} {\bf 148}, 1467-1478. 

\item Bjorken, J. D. (1967).  Inequality for Backward Electron- and Muon-Nucleon 
Scattering at High Momentum Transfer.  {\it Phys. Rev.} {\bf 163}, 
1767-1769. 

\item Bjorken, J. D. (1969).  Asymptotic Sum Rules at Infinite Momentum. 
{\it Phys. Rev.} {\bf 179}, 1547-1553. 

\item Bjorken, J. D. and S. D. Drell (1965).  {\it Relativistic Quantum Fields} 
(McGraw-Hill, New York). 

\item Bouchiat, C., J. Iliopoulos, and Ph. Meyer (1972). An Anomaly-Free 
Version of  Weinberg's Model.  {\it Phys. Lett. B} 
{\bf 38}, 519-523. 

\item Brandt, R. A. (1969).  Axial-Vector Current in Spinor Electrodynamics.  
{\it Phys. Rev.} {\bf 180}, 1490-1502. 

\item Callan, C. G. (1970). Broken Scale Invariance in Scalar Field Theory.  
{\it Phys. Rev. D} {\bf 2}, 1541-1547. 

\item Callan, C. G. (1972).  Broken Scale Invariance and Asymptotic Behavior.  
{\it Phys. Rev. D} {\bf 5}, 3202-3210. 

\item Callan, C. G. and D. J. Gross (1969).  High-Energy Electroproduction and the 
Constitution of the Electric Current. {\it Phys. Rev. Lett.} {\bf 22}, 
156-159. 

\item Chanowitz, M. S. and J. Ellis (1972). Canonical Anomalies and Broken  
Scale Invariance. {\it Phys. Lett. B} {\bf 40}, 397-400. 

\item Chanowitz, M. S. and J. Ellis (1973).  Canonical Trace Anomalies. {\it Phys. 
Rev. D} {\bf 7}, 2490-2506. 

\item Christ, N., B. Hasslacher, and A. H. Mueller (1972).  Light-Cone Behavior 
of Perturbation Theory.  {\it Phys. Rev. D} {\bf 6}, 3543-3562. 

\item Coleman, S. (1989).  {\it Aspects of Symmetry} (Cambridge University 
Press, Cambridge), pp. 307-327. 

\item Coleman, S. and R. Jackiw (1971).  Why Dilatation Generators  
do not Generate Dilatations. {\it Ann. Phys.} {\bf 67}, 552-598. 

\item Collins, J. C., A. Duncan, and S. D. Joglekar (1977).  Trace and Dilatation 
Anomalies in Gauge Theories.  {\it Phys. Rev. D} {\bf 16}, 438-449. 

\item Crewther, R. J. (1972).  Nonperturbative Evaluation of the Anomalies in 
Low-Energy Theorems.  {\it Phys. Rev. Lett.} {\bf 28}, 1421-1424. 

\item Fritzsch, H. and M. Gell-Mann (1971/1972).  Light Cone Algebra, talk 
at the 1971 
Coral Gables Conference, extended into a preprint a few months later; 
reissued recently as arXiv: hep-ph/0301127. 

\item Fritzsch, H. and M. Gell-Mann (1972). Current Algebra: Quarks and What Else?,  
in {\it Proceedings of the XVI International Conference on High Energy 
Physics}, Chicago--Batavia, IL, J. D. Jackson and A. Roberts, eds.   
(National Accelerator Laboratory, Batavia), Vol. 2, pp. 135-165.  See page 
140 for a discussion of the gauge theory of color octet gluons.  

\item Fukuda, H. and Y. Miyamoto (1949). On the $\gamma$-Decay of Neutral Meson. 
{\it Progr. Theor. Phys.} {\bf 4},  
235 (L)-236 (L)  and 347-363. 

\item Gerstein, I. S. and R. Jackiw (1969).  Anomalies in Ward Identities for 
Three-Point Functions.  {\it Phys. Rev.} {\bf 181}, 1955-1963. 

\item Gilman, F. J. (1969).    Sign of the $\pi^0 \to \gamma \gamma$ Decay  
Amplitude.  {\it Phys. Rev.} {\bf 184}, 1964-1965. 

\item Gross, D. and R. Jackiw (1972). Effect of Anomalies on Quasi-Renormalizable  
Theories. {\it Phys. Rev. D} {\bf 6}, 477-493.

\item Hagen, C. R. (1969).  Derivation of Adler's Divergence Condition from the 
Field Equations.  {\it Phys. Rev.} {\bf 177}, 2622-2623. 

\item Han, M. Y. and Y. Nambu (1965). Three-Triplet Model with Double 
$SU(3)$ Symmetry.  {\it Phys. Rev. } {\bf 139}, B1006-B1010. 

\item Hara, Y. (1964).   Unitary Triplets and the Eightfold Way. {\it Phys. Rev.}
{\bf 134}, B701-B704. 

\item Iwasaki, Y. (1977).  Coupling of the Trace of the Energy-Momentum Tensor 
to Two Photons.  {\it Phys. Rev. D} {\bf 15}, 1172. 

\item Jackiw, R. (1970).  Field Theoretic Investigations in Current Algebra, in 
{\it Lectures on Current Algebra and Its Applications},  S. B. Treiman, 
R. Jackiw, and D. Gross, eds. (Princeton University Press, Princeton, 1972), 
pp. 97-254.  This was extended to the two articles, R. Jackiw,   
Field Theoretic Investigations 
in Current Algebra, and  R. Jackiw, Topological Investigations in 
Quantum Gauge Theories, 
in {\it Current Algebra and Anomalies},  S. Treiman, R. Jackiw, B. Zumino, 
and E. Witten, eds. (Princeton University Press, Princeton and World Scientific, 
Singapore, 1985), pp. 81-210 and pp. 211-359. 

\item Jackiw, R. and K. Johnson (1969). Anomalies of the Axial-Vector Current. 
{\it Phys. Rev.} {\bf 182},  1459-1469. 

\item Jackiw, R. and G. Preparata (1969a).  Probes for the Constituents of the 
Electromagnetic Current and Anomalous Commutators. {\it Phys. Rev. Lett.} 
{\bf 22}, 975-977. 

\item Jackiw, R. and G. Preparata (1969b).  High-Energy Inelastic Scattering of 
Electrons in Perturbation Theory.  {\it Phys. Rev.} {\bf 185}, 1748-1753. 

\item Jackiw, R. and G. Preparata (1969c).  $T$ Products at High Energy and 
Commutators.  {\it Phys. Rev.} {\bf 185}, 1929-1940. 

\item Jauch, J. M. and Rohrlich, F. (1955).   
{\it The Theory of Photons and Electrons} (Addison-Wesley, Cambridge, MA),  
Appendix A5-2, pp. 457-461. 

\item Johnson, K. and F. E. Low (1966). Current Algebras in a Simple Model.  
{\it Progr. Theor. Phys.} Suppl. {\bf 
37-38}, 74-93. 

\item Mack, G. (1968). Partially Conserved Dilatation Current.  
{\it Nucl. Phys. B} {\bf 5}, 499-507. 

\item Maki, A. (1964).  The ``fourth'' Baryon, Sakata Model and Modified 
B-L Symmetry. I  {\it Progr. Theoret. Phys.} {\bf 31}, 331-332. 

\item  Minkowski, P. (1976).  On the Anomalous Divergence of the
Dilatation Current in Gauge Theories.  Univ. Bern preprint, Sept. 1976, 
archived as KEK scanned version. 

 \item Miyamoto, Y. (1965).  Three Kinds of Triplet Model, in {\it Extra Number
Supplement of Progress of Theoretical Physics: Thirtieth Anniversary of 
the Yukawa Meson Theory}, pp. 187-192. 

\item Nambu, Y. (1965)  A Systematics of Hadrons in Subnuclear Physics, in {\it 
Preludes in Theoretical Physics},  A. de-Shalit, H. Feshbach, and L. 
Van Hove, eds. (North-Holland, Amsterdam and John Wiley, New York, 1966), 
pp. 133-142. 

\item Nielsen, N. K. (1977).  The Energy-Momentum Tensor in a Non-Abelian 
Quark Gluon Theory.  {\it Nucl. Phys. B} {\bf 120}, 212-220. 

\item Okubo, S. (1969).  Sign and Model Dependence of $\pi^0 \to 2\gamma$ Matrix 
Element.  {\it Phys. Rev.} {\bf 179}, 1629-1631. 

\item Pagels, H. and E. Tomboulis (1978).  Vacuum of the Quantum Yang-Mills Theory  
and Magnetostatics.  {\it Nucl. Phys. B} {\bf 143}, 485-502. 

\item Polkinghorne, J. C. (1958a). Renormalization of Axial Vector Coupling. 
{\it Nuovo Cimento} {\bf 8}, 179-180. 

\item Polkinghorne, J. C. (1958b). Renormalization of Axial Vector Coupling - II. 
{\it Nuovo Cimento} {\bf 8}, 781. 

\item Rosenberg, L. (1963).  Electromagnetic Interactions of Neutrinos.  
{\it Phys. Rev.} {\bf 129}, 2786-2788. For an analytic evaluation of the 
Feynman integrals, see N. N. Achasov, Once More About the Axial Anomaly 
Pole, {\it Phys. Lett. B} {\bf 287}, 213-215 (1992).   

\item Schwinger, J. (1951).  On Gauge Invariance and Vacuum Polarization.  
{\it Phys. Rev.} {\bf 82}, 664-679. 

\item Schwinger, J. ed. (1958).  {\it Selected Papers on Quantum 
Electrodynamics}
(Dover Publications, New York). 

\item Sen, S. (1970).  University of Maryland Report No. 70-063. [38]

\item Sirlin, A. (1978).  Current Algebra Formulation of Radiative Corrections in 
Gauge Theories and the Universality of Weak Interactions.  {\it Rev. Mod. 
Phys.} {\bf 50}, 573-605.   See Appendix C, p. 600. 

\item Steinberger, J. (1949). On the Use of Subtraction Fields and the Lifetimes  
of Some Types of Meson Decay.  {\it Phys. Rev.} {\bf 76}, 1180-1186. 

\item Sutherland, D. G. (1967).  Current Algebra and Some Non-Strong Mesonic 
Decays.  {\it Nucl. Phys. B} {\bf 2}, 433-440. 

\item Symanzik, K. (1970). Small Distance Behavior in Field Theory and Power   
Counting.  {\it Commun. Math. Phys.} {\bf 18}, 227-246. 

\item Tavkhelidze, A. (1965).  Higher Symmetries and Composite Models of Elementary 
Particles, in {\it High-Energy Physics and Elementary Particles} 
(International Atomic Energy Agency, Vienna), pp. 753-762. 

\item Terent'ev, M. V. (1971).  Possible Connection between the Amplitudes of the 
Processes $e^+e^- \to 3 \pi$, $\gamma \gamma \to 2 \pi$, and $\pi^0 \to 
2 \gamma$.  {\it ZhETF Pis. Red.} {\bf 14}, 140-143 (English translation:
{\it Sov. Phys. JETP Lett.} {\bf 14}, 94-96). 

\item 't Hooft, G. (1976).  Computation of the Quantum Effects due to a 
Four-Dimensional Pseudoparticle.  {\it Phys. Rev. D} {\bf 14}, 3432-3450. 

\item Tung, W.-K. (1969).  Equal-Time Commutators and Equations of Motion for 
Current Densities in a Renormalizable Field-Theory Model.  {\it Phys. Rev.}
{\bf 188}, 2404-2415. 

\item Vainshtein, A. I. and B. L. Ioffe (1967).  Test of Bjorken's Asymptotic  
Formula in Perturbation Theory.  
{\it ZhETF Pis. Red.} {\bf 6}, 917-19
(English translation:  {\it Sov. Phys. JETP Lett.} {\bf 6}, 341-343). 

\item Veltman, M. (1967).  Theoretical Aspects of High Energy 
Neutrino Interactions.  {\it Proc. Roy. Soc. A} {\bf 301}, 107-112. 

\item Veneziano, G. and Y. Yankielowicz (1982).  An Effective Lagrangian for the 
Pure $N=1$ Supersymmetric Yang-Mills Theory.  {\it Phys. Lett. B} {\bf 113}, 
231-235. 

\item Weinberg, S. (1973).  Theory of Weak and Electromagnetic Interactions,  
in {\it Fundamental Interactions in Physics 
and Astrophysics}, G. Iverson et al., eds.  (Plenum Press, New York), 
pp. 157-186. 

\item Wess, J. and B. Zumino (1971).  Consequences of Anomalous Ward Identities. 
{\it Phys. Lett. B} {\bf 37}, 95-97. 

\item Wilson, K. G. (1969).  Non-Lagrangian Models of Current Algebra.  {\it Phys. 
Rev.} {\bf 179}, 1499-1512. 

\item Witten, E. (1983).  Global Aspects of Current Algebra.  {\it Nucl. Phys. B} 
{\bf 223}, 422-432. 

\item Young, B.-L., T. F. Wong, G. Gounaris, and R. W. Brown (1971). Absence of  
Second-Order Corrections to the Triangle Anomaly in Quantum Electrodynamics.  
{\it Phys. Rev. D} {\bf 4}, 348-353. 

\item Zumino, B. (1969).  Anomalous Properties of the Axial Vector Current, in 
{\it Topical Conference on Weak Interactions Proceedings} CERN 69-7,  J. S. Bell, ed.  (CERN, Geneva), pp. 361-370. 

\end{itemize}

\chapter*{4. Quantum Electrodynamics}

\markboth{Adventures in Theoretical Physics}{Quantum Electrodynamics}

\section*{Introduction}
 
My interest in a detailed study of quantum electrodynamics (QED) began during 
my visit to Cambridge, U.K. in the spring of 1968, when I found the anomalous 
properties of the axial-vector triangle diagram discussed in Chapter 3.  
This started me thinking more generally about the properties of fermion 
loop diagrams, and in particular I wondered whether such diagrams in 
quantum electrodynamics could lead to an eigenvalue condition for the 
electric charge, possibly giving an explanation of why the charges 
of different  
particle species (such as the electron and proton) are the same in magnitude.  
This speculation ultimately proved to be wrong, and I look back on the 
investigations that it inspired with mixed feelings, as being somewhat 
of a {\it mis}adventure.  On the one hand, my work on aspects of quantum  
electrodynamics led to a number of important papers with useful results, 
but on the other hand, my preoccupation with this program kept me from 
jumping into the emerging area of Yang--Mills unification at the point when 
much of the interesting theoretical work on non-Abelian theories 
was being done.  

My work on QED divided into three distinct phases, described in the following 
sections.  The first part dealt with a calculation of the process of photon 
splitting in strong magnetic fields, which served as a 
warm-up for getting into   
the study of fermion loop diagrams.  After this calculation was completed, 
I turned to an investigation of the renormalization group properties of 
QED, using as a tool the newly discovered Callan--Symanzik equations.    
Finally, in an attempt to get a better formalism for calculating the 
renormalization group $\beta$ function contribution from closed loop 
diagrams, I worked out a compactification of massless QED on the 4-sphere, 
and applied this formalism to a number of theoretical issues. By the end  
of this phase, it was clear that developments in non-Abelian gauge theories 
were the future of the field of particle physics and, through grand 
unification, offered a compelling way to understand charge quantization,    
which had been the starting motivation for my interest in electrodynamics.  
So at this point I set my QED work aside and moved on to some of the 
phenomenological investigations described in Chapter 5.  
\bigskip

\section*{Strong Magnetic Field Electrodynamics:\hfill\break
Photon Splitting and Vacuum Dielectric Constant}

The discovery of pulsars with ultra-strong trapped magnetic fields led to 
a surge of interest in strong field QED processes, that are unobservably 
small for attainable laboratory magnetic fields.   One of the processes 
of interest is photon splitting in a constant magnetic field, which is 
described by a closed electron loop Feynman diagram.  When conversations 
at the Institute turned to whether this reaction could be of relevance in 
the dynamics of pulsar magnetospheres, my interest in getting into a 
general study of fermion loop processes in QED made it natural for me to 
get involved.  The initial phase of this study led to a paper (Adler, 
Bahcall, Callan, and Rosenbluth, 1970, {\sf R25}), that surveyed the basic 
features of the photon splitting process.  Briefly, the lowest order 
box diagram makes a vanishing contribution, by an argument 
using Lorentz 
invariance and gauge invariance, and so the leading contribution comes 
from the hexagon diagram, with three insertions of the external magnetic 
field. (Earlier calculations had overlooked this fact, and so led to  
the wrong dependence on magnetic field strength.)  
Using the Heisenberg--Euler effective Lagrangian, we calculated the 
photon splitting absorption coefficients for the various photon polarization 
states relative to the magnetic field vector, to leading order in the   
external magnetic field, for photon energies small relative to the 
electron mass.  We also gave the selection 
rules that result from the fact that the dielectric constant for the 
vacuum permeated by a strong magnetic field is different for the different 
photon polarizations (this was where Marshall Rosenbluth's expertise as 
a plasma physicist entered in), and made numerical estimates.    
Some of our results 
were independently obtained around the same time by Bialynicka-Birula and 
Bialynicki-Birula (1970).  

Again with the aim of getting more experience with QED calculations, I 
decided to embark on an exact calculation of photon splitting, for arbitrary 
magnetic fields and for arbitrary photon energies below the pair production 
threshold.  This involved a very lengthy calculation using the   
proper time method, that Schwinger had first used  (Schwinger, 1951) to 
give an elegant 
rederivation of the Heisenberg--Euler effective Lagrangian.  I derived 
general formulas for both the photon splitting amplitudes, and  
the refractive indices needed for the selection rules \big(in the 
latter case correcting an earlier result of Minguzzi 
(1956,1958a,1958b)\big).  
I wrote a computer program to numerically evaluate the photon 
splitting absorption rates, and computed sample results, as well as giving  
a detailed discussion of possible plasma physics corrections to the selection 
rules.  These results were all reported in a comprehensive article 
(Adler, 1971, {\sf R26}) on photon splitting and dispersion in a strong 
magnetic field.  

My overall conclusion was that the leading order 
calculation from the hexagon diagram gives good order of magnitude 
estimates, as graphed in Fig. 8  of {\sf R26}, which plots
the ratio of the exact photon splitting absorption coefficient to the 
hexagon diagram prediction, versus magnetic field, for photon frequencies 
equal to zero and equal to the electron mass $m$.  This plot, incidentally,   
gives a check both on my exact analytic calculation and the numerical work, 
since the 
ratio approaches unity for small field strengths, where the hexagon dominates. 
For magnetic fields 
of order  the ``critical field'' $B_{CR}=m^{2}/e \sim  4.41 \times 10^{9}$ 
Tesla ($4.41 \times 10^{13}$ Gauss), and photon frequencies of
order the electron mass $m$, the photon 
splitting mean free path is  much shorter than characteristic pulsar  
magnetosphere depths.  However, since the absorption coefficients scale as 
$B^{6}$ for small fields, and since the pulsars known in 1971 tended to have 
fields of up to a few tenths of $B_{CR}$, the photon splitting 
process at that 
time seemed to be not of great astrophysical importance.   Stoneham 
(1979) published an analytic recalculation of photon splitting by a 
different method (without numerical evaluation), which as we shall 
see agreed with my calculation.  In an Appendix to his paper, he also 
improved on my estimate of the very small 
corrections that arise from the box diagram,  when finite opening 
angles resulting from photon dispersion are taken into account, and we   
exchanged letters on this aspect of his work.  However, 
after Stoneham's paper, interest in photon splitting waned 
for quite a number of years. 

In the mid 1990's, the discovery of ``magnetars'', pulsars with fields much 
higher than the critical field, revived interest in photon splitting.  
Around April, 1995, John Bahcall told me that  
recent papers by Mentzel, Berg, and Wunner (1994) and Wunner, Sang, and 
Berg (1995) claimed that the photon splitting absorption coefficients 
for energetic photons in strong fields were a factor of $10^{4}$ higher than 
given in my 1971 paper.  If true, this would have had important astrophysical 
ramifications, so I looked back at my own work, and at the papers of the 
Wunner group.  I was struck by the fact that the Wunner group had not  
checked to see whether their calculation 
reproduced the known $B^{6}$ dependence of 
photon splitting for weak fields and low energy photons, a consistency 
test that, as noted above, I had incorporated into my analytic and numerical 
work.  So I strongly suspected that they had made an error, 
possibly through a 
lack of gauge invariance, and wrote a 
letter to this effect to the Wunner group, while John simultaneously wrote   
to {\it Astrophysical Journal Letters}, where their second paper was being 
considered for publication.  Neither of these letters had any effect, and 
the Wunner, Sang, and Berg paper was published in December, 1995.  
John Bahcall and 
Bohdan Paczynski then urged me to make my private misgivings known more 
publicly.  In response, I wrote a short IAS Astrophysics Preprint 
Series article in January, 1996 (Adler, 1996), expanding 
on my letter to the Wunner group, 
and concluding ``it is important that their calculation and mine be rechecked 
by a third party, with the aim of understanding where the discrepancy arises 
and determining who is right.''  I submitted this note to the {\it Astrophysical 
Journal}, which rejected it.  

Although this short note was never published, it had the intended effect 
as a result of its internal circulation within the IAS.  Not 
long afterwards Christian Schubert, an IAS visitor at the time, 
came to my office and said that with new ``stringy'' Feynman rules with 
which he was expert, he thought he could repeat in a few days 
the calculation that had taken me a couple of months by the proper time 
method.  I replied that if he could do that, I would deal with the numerical 
aspects.  A week or two later Christian gave me two equivalent formulas 
for the photon splitting amplitude obtained by his methods; in the meantime, 
the Russian group of Baier, Milstein, and Shaisultanov (1996) 
had produced yet 
another calculation, which agreed numerically with my 1971 paper.  During 
a short visit to the Institute for Theoretical Physics in Santa Barbara, 
I wrote programs to directly compare Schubert's two expressions, my 1971 
result, Stoneham's 1979 formula, and the analytic formula of the 
Russian group, all as applied to the allowed polarization case.  (The 
reason for doing this numerically is that an analytic conversion between 
inequivalent Feynman parameterizations is very difficult, because zero can 
be written as a multidimensional integral in complicated ways.)   The 
programs showed that the five calculations gave precisely identical 
amplitudes.  This was reported in the paper that I drafted with Schubert 
on my return to the IAS (Adler and Schubert, 1996, {\sf R27}).  We 
also posted my computer programs on my web site, and advertised this posting 
in the paper, so that the community at large could verify what we had done. 
About a month later, I received an email from Wunner retracting the earlier 
numerical results of his group, which turned out to result from a 
single sign error in their computer programs.  When this sign 
error was corrected, the analytic results of Mentzel, Berg and Wunner gave 
answers that agreed with everyone else, as discussed in 
Wilke and Wunner (1997).
Thus the photon splitting controversy was finally resolved.  
Subsequently, John Bahcall had me assemble a file of all the relevant 
papers and correspondence for a post-mortem meeting that he held 
with the editors of the {\it Astrophysical Journal}, to analyze 
and improve the process that that had allowed  an incorrect 
paper to get into print, despite several advance warnings 
that the results were suspect.                                         
\bigskip

\section*{The ``Finite QED'' Program via the Callan--Symanzik Equations}

My comprehensive article on photon splitting was finished in early 1971, and 
the following summer I returned to my long-standing interest in a study 
of unresolved issues in the theory of quantum electrodynamics.  Johnson, 
Baker, and Willey (1964), Johnson, Willey, and Baker (1967), 
and Baker and Johnson (1969, 1971a,b)
had written an important series of papers (referred to below as JBW) in which 
they argued that if QED has a Gell-Mann--Low eigenvalue, then the asymptotic 
behavior of both the electron and photon propagators would drastically 
simplify, with the mass term in the electron propagator having power law 
scaling behavior, and the asymptotic photon propagator behaving, after   
charge renormalization, as if it had no photon self-energy part.  
Bill Bardeen and I were both in Aspen for part of the 
summer of 1971, and we embarked on a study of QED using the then very 
new Callan (1970)--Symanzik (1970) equations.  Rather than 
addressing the issue of a 
possible eigenvalue in QED, we studied the simplified model suggested by  
the presence of such an eigenvalue, in which the photon propagator 
is taken as a free propagator with no self-energy part.  In this case 
the $\beta$ function term, which has a coupling constant derivative, is 
not present in the Callan--Symanzik equations, and these equations then can 
be explicitly integrated to give the simple form for the electron propagator 
found by JBW.  These results were described in 
the paper Adler and Bardeen (1971), {\sf R28}. In addition to giving 
results of interest for QED, this paper was one of the first applications of 
the Callan--Symanzik equations, and  
was also a motivation for my remarks at the Princeton  
conference later in 1971 (see {\sf R23}), in which I suggested a possible 
connection between an eigenvalue condition in the strong interactions and 
Bjorken scaling.  

After finishing the paper with Bardeen, I turned to a detailed study of 
the full theory of QED, with photon 
self-energy parts retained, on which I wrote a comprehensive paper 
Adler (1972a), {\sf R29}.  This paper had a number of new results.  I  
began with a review of the original Gell-Mann--Low formulation of the
renormalization group in QED, and then redid their analysis in terms of 
the more modern Callan--Symanzik approach, ending up in Eq.~(53) with the 
explicit map between the Callan--Symanzik $\beta(\alpha)$ function and 
the functions 
$\psi(\alpha)$ and $q(\alpha)$ that enter into the 
Gell-Mann--Low formulation.  \big(An implicit form of this map had 
appeared in Sec. II.3 of Symanzik (1970).\big)
After reviewing the JBW program and the results obtained 
with Bardeen in {\sf R28}, I showed by an argument based on 
the Federbush--Johnson (1960) theorem that {\it if} there is an eigenvalue in QED, then in the 
massless limit all $2n$-point current correlation functions must vanish 
at the eigenvalue.  I then went on to show, in an argument that benefited 
from a conversation with Roger Dashen,  that the vanishing of higher 
correlation functions also implied the vanishing of all coupling constant 
derivatives of the photon proper self-energy part at the eigenvalue; hence 
the eigenvalue, if it existed, must be an {\it infinite order} zero of the 
one-loop $\beta$ function.  These were all correct results that give the 
paper an enduring value. 
                                    
I concluded the paper by proposing that 
in addition to the standard renormalization group result, in which 
the eigenvalue plays the role (through running of the coupling) of the 
unrenormalized fine structure constant $\alpha_0$, there could be an 
additional solution, resulting from a fermion-loopwise summation of the 
theory, in which the eigenvalue plays the role of the physical coupling 
$\alpha$.     
A motivation for this proposal was that the formal power series argument,  
which shows the equivalence of loopwise summation to the usual renormalization 
group analysis, could break down in the presence of an essential singularity  
in the coupling.  I then went on to conjecture that loopwise summation with
an eigenvalue for $\alpha$ was the mechanism 
fixing the physical fine structure constant in a uniform manner for all 
fermion species.  As I have noted in the Introduction to this Chapter, this 
conjecture turned out to be wrong, and in retrospect my excessive emphasis   
on it in writing {\sf R29} distorted the presentation of an otherwise 
good paper.  At the time key people working on the renormalization group,    
in particular Gell-Mann, Low, and Wilson, were all very skeptical.  Wilson,  
in particular, remarked at a Princeton seminar that my demonstration of 
an infinite order zero showed there could be no eigenvalue in QED, 
and although I was privately annoyed at the time, it is now clear that this  
was the correct conclusion.  

Finally, in an Appendix to my paper, I returned to the electron propagator 
analysis carried out in {\sf R28}, this time in a general covariant gauge. 
This investigation was later reanalyzed in more detail, and improved, in 
a comprehensive study by Lautrup (1976).  

The final paper in this section, Adler, Callan, Gross, and Jackiw 
(1972), {\sf R30}, studied the combined implications of the BJL limit, 
the nonrenormalization of anomalies, and the possible presence of an 
eigenvalue in QED. This paper, which was initially drafted by Roman 
Jackiw, grew out of discussions among the authors  at Princeton and 
at the National Accelerator Laboratory.   It shows that the following three phenomena are,  
when taken in combination, incompatible: (1) nonrenormalization of the 
axial-vector anomaly, (2) the existence of an eigenvalue in QED, 
(3) validity of naive scale invariant short-distance expansions involving 
the axial-vector current at the eigenvalue.  Since the finite QED program 
was intended to eliminate the pathologies of QED, through presence of an 
eigenvalue, this showed that its aims could not be attained, and again 
cast strong doubt on the existence of an eigenvalue in QED.  For later work 
coming to the same conclusion, and references to more recent literature, 
see Baker and Johnson (1979), and Acharya and Narayana Swamy (1997).  
On rereading 
{\sf R30} now, it occurs to me that the argument establishing a 
relation between the axial-vector anomaly and the Schwinger term given 
in Section III may be extendable to show that the vanishing 
of anomalies in axial-vector loop diagrams coupling to four or more photons 
in QED implies, through similar use of a BJL limit, that the Schwinger term 
in the two-point function is a $c$-number.  As noted in Chapter 3,  
this is a result that I was unable to prove, before the advent of the  
theory of anomalies, in 1966. For another approach to constraining the 
structure of the Schwinger term, see Jackiw, Van Royen, and West (1970). 
 
\bigskip

\section*{Compactification of Massless QED and Applications}

The fact that the eigenvalue condition for QED can be studied in the 
conformally invariant, massless electron theory, led me to  study remappings 
of the Feynman rules for QED that make use of conformal invariance.  
In Adler (1972b), {\sf R31}, I showed that the equations of motion and 
Feynman rules for massless Euclidean QED can be written in terms of equivalent 
equations of motion and Feynman rules expressed 
in terms of coordinates that are confined 
to the surface of a unit hypersphere in 5-dimensional space (a four-sphere 
in mathematical terms).  For example, letting $\eta^a$ be the coordinate 
on the sphere \big(where $a$ runs from 1 to 5, and ($\eta^{a})^{2}=1$\big), the 
usual four-vector potential is replaced 
by a five-vector $A^a$ obeying the constraint (with repeated indices summed) 
$\eta_a A^a=0$, and the electromagnetic field strength is replaced by a 
three-index tensor $F_{abc}=L_{ab}A_c + L_{bc}A_a + L_{ca}A_b$, with 
$L_{ab}$ the 5-space rotation generators.  This tensor has a two-index 
dual $\hat F_{ab}$, and the Maxwell equations become  
$L_{ab}F_{abc}=2eJ_c~,~~L_{ab}\hat F_{bc}=\hat F_{ac}$.  The 
corresponding $O(5)$ covariant 
Feynman rules are given in Table I of {\sf R31}.  The result of this 
transformation of the theory is an explicit demonstration that massless 
QED can be compactified, so that there are only ultraviolet divergences 
(corresponding to points approaching each other on the surface of the 
sphere, where it becomes tangent to Euclidean 4-space), but no infrared 
divergences.  The $O(5)$ rules, however, are not manifestly conformal 
invariant; in a later section of the paper I showed that they are related, by 
a projective transformation, to a manifestly conformal invariant (but 
non-compact) $O(5,1)$ formalism that was introduced earlier by Dirac (1936).  

In two subsequent papers I further developed and applied 
the $O(5)$ covariant formalism. 
In Adler (1973) I showed that the usual Feynman path integral takes the 
form of an amplitude integral, constructed as  
an infinite product of individually 
well-defined ordinary integrals over coefficients appearing in the 
hyperspherical harmonic expansion of the electromagnetic potential $A_a$. 
In the paper Adler (1974), {\sf R32}, I used the amplitude integral 
formalism to study  a simple model, in which only a single photon mode 
of the form $A_a \propto v_{1a} \eta \cdot v_2 - v_{2a} \eta \cdot v_1$,  
with $v_{1,2}$ orthogonal unit vectors, is retained.  The external field 
Fredholm determinant or vacuum persistence amplitude 
$\Delta(eA)= \det(i\gamma \cdot \partial+e \gamma \cdot A)$ could 
then be studied by exploiting the $O(3) \times O(2)$ residual symmetry   
of this model, which permits the external field problem to be reduced 
to a set of two coupled first order ordinary differential equations, with 
a Wronskian equal to the Fredholm determinant.  A significant result coming 
out of the analysis of this model was that the renormalized 
Fredholm determinant is  
an entire function of order four as $eA$ becomes infinite in a general 
complex direction.  This played a role in a subsequent discussion of  
asymptotic estimates in perturbative QED, as discussed in the paper 
of Balian, Itzykson, Zuber, and Parisi (1978), which followed up on   
an earlier paper of 
Itzykson, Parisi, and Zuber (1977).  Whereas  extrapolation from the   
solvable case of a constant field strength $F_{\mu\nu}$ suggested
 that the order of the Fredholm determinant is two, my solvable 
example showed that two cannot be the correct answer for general vector  
potentials.  Balian et al. noted this and then went on to present 
further arguments for 
the determinant being of order four in four-dimensional spacetime, or 
more generally of order $D$ in $D$-dimensional spacetime.  This in turn had 
important implications for their study of asymptotic behavior of the 
perturbation series in QED.  The subject of the order of the Fredholm 
determinant was further developed by Bogomolny.  In an initial paper 
by Bogomolny and Fateyev (1978), the case of 
fields with an $O(3)\times O(2)$ 
symmetry group that I had initiated in {\sf R32} was taken up again, and 
an asymptotic formula for the Fredholm determinant was obtained.  In 
a subsequent paper, Bogomolny (1979) showed that this asymptotic formula, 
and a similar formula obtained 
by Balian et al. for another special case, could be extended to the 
general result $\lim_{e \to i\infty} \Delta(eA) =(e^4/12 \pi^2) 
\int d^4x \big((A_{\mu})^2\big)^2$, provided $A_{\mu}$ is 
chosen to obey the nonlinear 
gauge condition $\partial_{\mu}(A_{\mu}A^2)=0$.  Thus the order four result 
that I found in my ``one-mode'' model in fact gave the correct general answer 
for QED.  

A further application of the $O(5)$ formalism for QED emerged after the 
discovery of the instanton solution to the Yang--Mills field equations.  
Jackiw and Rebbi (1976) showed that the one-instanton solution is invariant 
under an $O(5)$ subgroup of the full conformal group, and hence can be 
rewritten in an elegant way in terms of the $O(5)$ formulation of 
electrodynamics, as extended to non-Abelian gauge fields.  
Letting $\alpha_a$ be 
the $O(5)$ equivalent of the Dirac $\gamma$ matrices, and $\gamma_{ab}=
(i/4)[\alpha_a,\alpha_b]$, a matrix-valued 
vector potential $A_a$ obeying the 
constraint $\eta \cdot A=0$ can be immediately constructed as $A_a =C 
\eta_b \gamma_{ab}$.  Jackiw and Rebbi showed that when this vector potential 
is substituted into the Yang--Mills field equation as expressed in the 
non-Abelian extension of the $O(5)$ formalism, one gets a cubic equation 
for the coefficient $C$, two roots of which give pure gauge potentials with 
vanishing field strengths, but the third root of which gives the instanton!   
Thus, I had missed a significant opportunity in not pursuing the question, 
raised at least once when I gave seminars, of what  the non-Abelian 
generalization of the $O(5)$ formalism was like.      
A variant of the non-Abelian $O(5)$  formalism was subsequently applied by 
Belavin and Polyakov (1977),  
with corrections by Ore (1977), to give a recalculation  of the Fredholm 
determinant in an instanton background that was first computed by 
't Hooft (1976).

\section*{References for Chapter 4}

\begin{itemize}

\item Acharya, R. and P. Narayana Swamy (1997). No Eigenvalue in Finite Quantum Electrodynamics. {\it Int. J. Mod. Phys. A} {\bf 12}, 3799-3809. 

\item Adler, S. L. (1971) {\sf R26}.  Photon Splitting and Photon Dispersion in a Strong Magnetic Field. {\it Ann. Phys.} {\bf 67}, 599-647.  
Pages 599-601, 609-613, 621-622, and 634-644 are reprinted here. 

\item Adler, S. L. (1972a) {\sf R29}.  Short-Distance Behavior of Quantum Electrodynamics and an Eigenvalue Condition for $\alpha$.  {\it Phys. Rev. D} {\bf 5}, 3021-3047. 

\item Adler, S. L. (1972b) {\sf R31}.  Massless, Euclidean Quantum Electrodynamics on the 5-Dimensional Unit Hypersphere.  {\it Phys. Rev. D} {\bf 6}, 3445-3461. 

\item Adler, S. L. (1973).  Massless Electrodynamics on the 5-Dimensional Unit Hypersphere:  An Amplitude-Integral Formulation.  {\it Phys. Rev. D} {\bf 8}, 2400-2418. 

\item Adler, S. L. (1974) {\sf R32}.  Massless Electrodynamics in the One-Photon-Mode Approximation.  {\it Phys. Rev. D} {\bf 10}, 2399-2421. 

\item Adler, S. L. (1996).  Comment on ``Photon Splitting in Strongly Magnetized Objects Revisited''.  IASSNS-AST 96/4 (unpublished). 

\item Adler, S. L., J. N. Bahcall, C. G. Callan, and M. N. Rosenbluth (1970) {\sf R25}. Photon Splitting in a Strong Magnetic Field.  {\it Phys. Rev. Lett.} {\bf 25}, 1061-1065. 

\item Adler, S. L. and W. A. Bardeen (1971) {\sf R28}.  Quantum Electrodynamics without Photon Self-Energy Parts:  An Application of the Callan-Symanzik Scaling Equations.  {\it Phys. Rev. D} {\bf 4}, 3045-3054. 
  
\item Adler, S. L., C. G. Callan, D. J.  Gross, and R. Jackiw (1972) {\sf R30}. Constraints on Anomalies.  {\it Phys. Rev. D} {\bf 6}, 2982-2988. 

\item Adler, S. L. and C. Schubert (1996) {\sf R27}.  Photon Splitting in a Strong Magnetic Field: Recalculation and Comparison with Previous Calculations. {\it Phys. Rev. Lett.} {\bf 77}, 1695-1698. 

\item Baier, V. N., A. I. Milstein, and R. Zh. Shaisultanov (1996). Photon Splitting in a Very Strong Magnetic Field.  {\it Phys. Rev. Lett.} {\bf 77}, 1691-1694. 

\item Baker, M. and K. Johnson (1969).  Quantum Electrodynamics at Small Distances. {\it Phys. Rev.} {\bf 183}, 1292-1299. 

\item Baker, M. and K. Johnson (1971a).  Asymptotic Form of the Electron Propagator and the Self-Mass of the Electron.  {\it Phys. Rev. D} {\bf 3}, 2516-2526. 

\item Baker, M. and K. Johnson (1971b).  Simplified Equation for the Bare Charge in Renormalized Quantum Electrodynamics.  {\it Phys. Rev. D} {\bf 3}, 2541-2542.  

\item Baker, M. and K. Johnson (1979).  Applications of Conformal Symmetry in Quantum Electrodynamics.  {\it Physica} {\it A} {\bf 96}, 
120-130. 

\item Balian, R., C. Itzykson, J. B. Zuber, and G. Parisi (1978).  Asymptotic Estimates in Quantum Electrodynamics. II.  {\it Phys. Rev. D} {\bf 17}, 1041-1052. 

\item Belavin, A. A. and A. M. Polyakov (1977).  Quantum Fluctuations of 
Pseudoparticles.  {\it Nucl. Phys. B} {\bf 123}, 429-444. 

\item Bialynicka-Birula, Z. and I. Bialynicki-Birula (1970).  Nonlinear Effects in Quantum Electrodynamics.  Photon Propagation and Photon Splitting in an External Field. {\it Phys. Rev. D} {\bf 2}, 2341-2345. 

\item Bogomolny, E. B. (1979).  Large-Complex-Charge Behavior of the Dirac 
Determinant.  {\it Phys. Lett.} {\it B} {\bf 86}, 199-202. 

\item Bogomolny, E. B. and V. A. Fateyev (1978). The Dyson Instability and 
Asymptotics of the Perturbation Series in QED. {\it Phys. Lett.} 
{\it B} {\bf 76}, 210-212. 

\item Callan, C. G. (1970).  Broken Scale Invariance in Scalar Field Theory. {\it Phys. Rev. D} {\bf 2}, 1541-1547. 

\item  Dirac, P. A. M. (1936). Wave Equations in Conformal Space. {\it Ann. Math.} {\bf 37}, 429-442. 

\item Federbush, P. G. and Johnson, K. A. (1960).  Uniqueness Property of the 
Twofold Vacuum Expectation Value.  {\it Phys. Rev.} {\bf 120}, 1926. 

\item Itzykson, C., G. Parisi, and J-B. Zuber (1977).  Asymptotic Estimates in Quantum Electrodynamics.  {\it Phys. Rev. D} {\bf 16}, 
996-1013. 

\item Jackiw, R. and C. Rebbi (1976).  Conformal Properties of a Yang-Mills Pseudoparticle.  {\it Phys. Rev. D} {\bf 14}, 517-523. 

\item Jackiw, R., R. Van Royen, and G. B. West (1970). Measuring 
Light-Cone Singularities. {\it Phys. Rev. D} {\bf 2}, 2473-2485.  See 
especially Sec. II B. 

\item Johnson, K., M. Baker, and R. Willey (1964).  Self-Energy of the Electron. {\it Phys. Rev.} {\bf 136}, B1111-B1119. 

\item Johnson, K., R. Willey, and M. Baker (1967).  Vacuum Polarization in 
Quantum Electrodynamics.  {\it Phys. Rev.} {\bf 163}, 1699-1715. 

\item Lautrup, B. (1976).  Renormalization Constants and Asymptotic Behavior in Quantum Electrodynamics.  {\it Nucl. Phys. B} {\bf 105}, 
23-44. 

\item Mentzel, M., D. Berg, and G. Wunner (1994).  Photon Splitting in Strong Magnetic Fields.  {\it Phys. Rev. D} {\bf 50}, 1125-1139. 

\item Minguzzi, A. (1956). Non-Linear Effects in the Vacuum Polarization.  {\it Nuovo Cimento} {\bf 4}, 476-486. 

\item Minguzzi, A. (1958a).  Non Linear Effects in the Vacuum Polarization (II). {\it Nuovo Cimento} {\bf 6}, 501-511. 

\item Minguzzi, A. (1958b).  Causality and Vacuum Polarization due to a Constant and a Radiation Field. {\it Nuovo Cimento} {\bf 9}, 145-153. 


\item Ore, F. R. (1977). How to Compute Determinants Compactly.  {\it Phys. Rev. D} {\bf 16}, 2577-2580. 

\item Schwinger, J. (1951).  On Gauge Invariance and Vacuum Polarization.  
{\it Phys. Rev.} {\bf 82}, 664-679. 

\item Stoneham,  R. J. (1979).  Photon Splitting in the Magnetized Vacuum. 
{\it J. Phys. A: Math. Gen.} {\bf 12}, 2187-2203. 

\item Symanzik, K. (1970).  Small Distance Behavior in Field Theory and Power Counting. {\it Commun. Math. Phys.} {\bf 18}, 227-246.

\item 't Hooft, G. (1976).  Computation of the Quantum Effects due to a 
Four-Dimensional Pseudoparticle.  {\it Phys. Rev. D} {\bf 14}, 
3432-3450. 

\item Wilke, C. and Wunner, G. (1997).  Photon Splitting in Strong Magnetic Fields: Asymptotic Approximation Formulas versus Accurate Numerical Results. {\it Phys. Rev. D} {\bf 55}, 997-1000. 

\item Wunner, G., R. Sang, and D. Berg (1995). Photon Splitting in Strongly Magnetized Cosmic Objects  -- Revisited.   {\it Astrophys. J.} {\bf 455}, L51-L53. 

\end{itemize}

\chapter*{5. Particle Phenomenology and Neutral Currents}

\markboth{Adventures in Theoretical Physics}{Particle Phenomenology and Neutral Currents}

\section*{Introduction} 
Much of the work described in Chapters 2 and 3 on soft pion theorems, 
sum rules, anomalies, 
and neutrino reactions falls in the category of phenomenology, 
but both the interrelations between different aspects of this research, 
and the chronology, suggested that it be discussed earlier.   Even before 
this work was done, I wrote my first particle phenomenology paper in 
collaboration with my first year Princeton graduate school roommate,    
and former Harvard classmate, Alfred Goldhaber 
(Adler and Goldhaber, 1963).  In this paper we analyzed the possibility of 
using the deuteron to provide a polarized proton target, by determining the 
polarization of the recoiling spectator neutron through its scattering   
on ${\rm He}^4$.  Although perhaps 
feasible, this proposal was never implemented, and much better methods for 
directly obtaining polarized targets are now available.  After I completed 
the work on quantum electrodynamics described in Chapter 4, I returned 
to phenomenology in a number of papers written, or conceived, during 
visits to the National Accelerator Laboratory (subsequently renamed the 
Fermi National Accelerator Laboratory, or Fermilab), and continued with   
related work in a number of papers written at the IAS.   
I discuss the earlier work done at Fermilab in the first section that 
follows, and then in the second section take up work at both Fermilab 
and the IAS relating to neutral currents.  
\bigskip
\section*{Visits to Fermilab}

When the National Accelerator Laboratory was inaugurated, my former thesis 
advisor Sam Treiman was brought in, on a succession of leaves from 
Princeton starting in 1970,  to serve as temporary head of the Theory 
Group, with the charge of setting it up and recruiting a permanent head.   Subsequently, Ben Lee 
was hired to be the permanent head of the Theory Group.  During this period 
many theorists from outside institutions were invited to be term time 
and/or summer visitors, and as part of this program I made a series of visits 
to Fermilab, and wrote a number of phenomenological papers growing out of 
discussions with people there.  

As already noted in Chapter 3, during a 1971 visit to Fermilab I 
collaborated with Lee, Treiman, and Tony Zee to study the 
anomaly-based prediction  
for the process $\gamma \gamma \to 3 \pi$, described in the paper {\sf R20}.   
This was applied in a subsequent paper that I wrote with Glennys Farrar and 
Treiman (Adler, Farrar, and Treiman, 1972, 
{\sf R33}) to an analysis of the contribution of three pion intermediate 
states to the rare kaon decay $K_L \to \mu^+\mu^-$.  The background for this 
study was what was then called the ``$K_L \to \mu^+\mu^-$ puzzle'', the 
fact that experiment had not detected this kaon decay mode at a level 
considerably below that given by a unitarity bound based on the  
assumed dominance of a two photon intermediate state in the absorptive part   
of the decay amplitude.  There were thus two possibilities, either an 
experimental problem, or destructive interference with another intermediate 
state, for which the three pion intermediate state was a prime candidate.   
Aviv and Sawyer (1971) had proposed to use soft pion methods to estimate 
the three pion contribution, and had concluded that the contribution was 
much too small to be relevant.  However, the Aviv--Sawyer analysis 
used an expression 
for the $3 \pi \to \gamma \gamma$ amplitude  which had been shown in {\sf R20} to be 
incorrect.  In {\sf R33}, we estimated the three pion contribution by 
using the corrected 
$3 \pi \to \gamma \gamma$ amplitude calculated in {\sf R20}, but still found  
that it gave much too small a contribution to explain the lack of 
observed $K_L \to \mu^+\mu^-$ events. Similar conclusions, again using the 
results of {\sf R20}, were reached independently by  Pratap, Smith, and Uy 
(1972).  Ultimately, the origin of the ``$K_L \to \mu^+\mu^-$ puzzle'' 
turned out to be experimental, and this decay mode has now been seen 
in a number of experiments, with the Particle Data Group giving an 
average value for $\Gamma(\mu^+\mu^-)/\Gamma_{\rm TOT}$ of $\sim 7.2 \times 
10^{-9}$, as compared with the theoretical unitarity lower 
bound of $7.0 \times 10^{-9}$ based on the 
current $K_L \to \gamma \gamma$ branching ratio.    

During the years 1973-1974, my Fermilab visits led to papers in two 
separate areas, searches for neutral currents in weak pion production, and 
the analysis of what was then a discrepancy between theory and experiment    
in $\mu$-mesic atom x-ray spectra.  I will take up this second area first,  
because the neutral current work leads directly 
into the papers discussed in the next section. 
My interest in the $\mu$-mesic atom discrepancy was stimulated by my earlier 
work on quantum electrodynamics, since an eigenvalue in QED could show up 
as deviations from the standard perturbation theory predictions for vacuum 
polarization effects.  Thinking about tests for vacuum polarization
discrepancies in QED led me to think more generally about other aspects 
of vacuum polarization, in particular the predictions for the ratio 
$R(s)=\sigma(e^+e^- \to {\rm hadrons}; s)/\sigma(e^+e^- \to \mu^+\mu^-; s)$ 
in various models for quark structure of hadrons.  This offshoot of the 
QED work led to results that are still used today, introduced in the paper 
Adler (1974a), {\sf R34}, dealing with ``Some simple vacuum-polarization   
phenomenology...''.  My basic observation was that whereas $R$ is measured 
in the timelike region, the natural place to compare experiment with 
scaling predictions of various theories is in the spacelike region, where 
(since there are no threshold effects) one might expect an early or 
``precocious'' onset of scaling.   Rather than directly using the 
dispersion relation for the vacuum polarization part to calculate the 
spacelike continuation, I proposed using its first derivative, and so 
defined a function 
\begin{equation}
T(-s)=\int_{4m_{\pi}^2}^{\infty}\frac {du  R(u)}{(s+u)^2}~~~.\nonumber
\end{equation}
This function is the one for which parton models and QCD most directly make predictions, 
and since it is positive definite and 
involves a strongly convergent integral (for $R$ approaching a constant), 
the experimentally inaccessible high energy tail has a known sign and a  
magnitude that can be bounded.   For a parton model in which 
$R$ asymptotically approaches a constant $C$, one has $T(-s) \sim C/s $ 
as $s \to \infty$, and a similar formula holds in QCD with a known 
logarithmic correction. The paper {\sf R34} used the function $T(-s)$  
to propose a test of the colored quark hypothesis.  Subsequently, 
De R\'ujula and Georgi (1976) used a modified version of this idea, 
defining $D(s)= s T(-s) $, to analyze the new SPEAR data. They found that 
the original colored quark model was excluded, and among various viable 
possibilities, noted that ``the standard model
with charm is acceptable if heavy leptons are produced,'' a conclusion that  
was borne out by experiment with the subsequent discovery of the   
$\tau$ lepton.  Shortly afterwards, Poggio, Quinn, and  
Weinberg (1976) proposed a generalized method in which the derivative of 
the hadronic vacuum polarization that I had used is replaced by a 
finite difference between the hadronic vacuum polarization values at points 
a distance $\pm i\Delta$ from the timelike real axis, leading to a ``smeared'' 
average of $R(s)$ that retains sensitivity to threshold effects.  
Recently, my original method, generally in the form $D(s)$ used by 
De R\'ujula and Georgi, has been revived under the name of the  ``Adler 
function'', in a number of papers; see, for example, Broadhurst and 
Kataev (1993),  Kataev (1996); Peris, 
Perrottet, and de Rafael (1998); Beneke (1999);  
Eidelman, Jegerlehner, Kataev, and Veretin (1999);  
Kataev (1999); Cveti\v c, Lee, and Schmidt (2001); Cveti\v c, Dib, Lee, and 
Schmidt (2001); Milton, Solovtsov, and Solovtsova (2001); and Dorokhov (2004).  

In the second part of {\sf R34} I examined what was then a discrepancy 
between theory and experiment in $\mu$-mesic atom x-ray transition energies, 
under the assumption that (if real) the discrepancy arose from a 
nonperturbative 
correction $\delta \rho$ to the vacuum polarization absorptive part.  
Assuming that $\delta \rho$  is positive, or positive and monotonic, 
I derived lower bounds on the corresponding deviation that would be  
expected in $a_{\mu}=\frac{1}{2}(g_{\mu}-2)$.  For instance, if 
$\delta \rho$ is assumed positive and monotonic, comparison of the kernels 
that weight $\rho$ in the formulas for the x-ray transition energies and 
for $a_{\mu}$ gives the bound $\delta a_{\mu} \leq -(0.98\pm 0.18) \times 
10^{-7}$.  In a follow-up paper with Roger Dashen and Sam Treiman (Adler,
Dashen, and Treiman, 1974) we discussed other tests for a nonperturbative
vacuum polarization contribution, and also placed bounds on the mass of 
a light scalar meson that could be invoked to explain the x-ray discrepancy.
A few months later, Barbieri (1975) extended the method of {\sf R34}
to show that 
precision measurements of the $(\mu\, {}^4{\rm He})^+$ system were already 
at variance, within the vacuum polarization deviation or scalar meson 
exchange hypotheses, with the supposed x-ray discrepancy. A later paper  
of Barbieri and Ericson (1975) gave additional evidence against the 
scalar meson explanation for the x-ray discrepancy.  In the meantime, during  
1975 and the few years following, 
there were a number of experimental developments, 
reviewed in detail in Borie and Rinker (1982), as a result of which 
the muonic x-ray discrepancy was eliminated.   
Incidentally, the current theoretical and experimental 
values of $a_{\mu}$ differ by a few 
parts in $10^{-9}$, well below the lower bounds on $\delta a_{\mu}$ 
inferred in {\sf R34} from  
the $\mu$-mesic atom x-ray data at that time, giving an additional 
indication that that 
the purported x-ray discrepancy was an experimental artifact.    
\bigskip
\section*{Neutral Currents}

The existence of weak neutral currents is a principal prediction of the 
Glashow--Weinberg--Salam electroweak theory, and commanded much attention 
in the 1970s.  Failure to find weak neutral currents would have falsified 
the electroweak theory, and on the other hand, detection of weak neutral 
currents would give a value for the electroweak mixing angle $\theta_W$, 
which in turn determines the masses of the heavy intermediate bosons of the 
theory.  As a result of my thesis work on weak pion production, it was 
natural for me to get interested in theoretical estimates of the neutral 
current weak pion production channels $\nu + N_i \to \nu + \pi +N_f$, with 
$N_{i,f}$ a nucleon  (either a neutron or proton) and with $\pi$ a pion 
of appropriate charge.  In July 1972,  a collaboration 
with Wonyong Lee as spokesman proposed 
a study of weak neutral currents in both the purely leptonic and the 
pion production channels at the Brookhaven AGS accelerator, and a copy  
of their proposal is in my files. Through this, 
and through related correspondence of Ben Lee with Sam Treiman, I got 
interested in doing detailed calculations for this process, and over the 
next few years was in frequent touch with the experimental group for 
which Wonyong Lee was spokesman.  

My initial papers were motivated by the fact that preliminary estimates 
of neutral current weak pion production by Ben Lee (1972) appeared 
to conflict with experiments in complex nuclei reported by Wonyong Lee (1972),
subject to two caveats.  The first caveat
was that Ben Lee's static model estimates 
didn't include $I=1/2$ contributions to weak 
pion production, and the second caveat was that nuclear charge 
exchange corrections
could be important, as noted by Perkins (1972).     
The first of these issues was dealt with in a short 
paper Adler (1974b), {\sf R35}, where I used my model of {\sf R15}, as 
adapted to the neutral current case, to estimate the effects of including 
the nonresonant isospin $1/2$ channels, and concluded that they had little 
effect on Ben Lee's estimate from the dominant isospin $3/2$ channel.  The 
second issue was dealt with in a paper on nuclear charge exchange corrections 
to pion production in the $\Delta(1232)$ region, that I wrote 
in collaboration 
with Shmuel Nussinov and Emmanuel Paschos (Adler, Nussinov, and Paschos, 
1974, {\sf R36}).  In this paper, we estimated the effects of multiple 
charge exchange scattering on pion production in nuclear targets, using an  
extension of techniques used by Fermi and others to calculate multiple 
neutron scattering in the early days of neutron physics.  A considerable 
part of the fun of writing this paper was learning about this older work on 
neutron physics, and feeling a sense of continuity between current concerns 
of weak interaction physics and the quite differently motivated work of an 
earlier generation.  In addition to giving analytic formulas, we tabulated 
various results for the 
case of a ${}_{13}{\rm Al}^{27}$ target, as appropriate to 
experiments with aluminum spark chamber plates.  In {\sf R36}, we made 
the simplifying assumption of an isotopically neutral target (that is, 
equal numbers of neutrons and protons), which is exact 
for  ${}_6 {\rm C}^{12}$, 
and a good approximation for aluminum.  In a follow up paper (Adler, 1974c), 
I extended the model to nuclear targets with a neutron excess.  As can be 
seen from Table II of {\sf R36}, charge exchange corrections are sizable, 
and in our model typically reduce the ratio of neutral current 
to charged current $\pi^0$ production by about 40\%.  

My next paper on neutral currents was motivated by the fact that preliminary 
results of an experiment  
on weak pion production in hydrogen at Argonne National Laboratory  
showed a cluster of neutral current events just above threshold.  
In this kinematic regime soft pion methods should apply, allowing one 
to relate threshold neutral current weak pion production in the standard 
electroweak theory to the 
elastic neutral current cross section for $\nu +p \to \nu + p$.  Using   
this relation, I showed in Adler (1974d), {\sf R37} that one could place   
bounds on the expected number of neutral current pion production 
events in the threshold region, with the Argonne results exceeding these 
bounds.  Thus, there seemed to be stronger neutral current 
weak pion production than suggested by the $SU(2) \times U(1)$  
electroweak theory.  

Subsequent events then proceeded on several parallel tracks. 
In a follow-up paper to {\sf R37}, published as Adler (1975a), {\sf R38}, 
I used the full apparatus of my weak pion production 
calculation of {\sf R15} to 
extend the neutral current calculation above the threshold region to 
include the regime where $\Delta(1232)$ production dominates. This analysis    
reinforced the conclusions about the preliminary Argonne data 
already reached in {\sf R37}.  Simultaneously, with a large  
group of postdocs at the Institute, I embarked on a study of  weak pion 
production in alternative models of neutral currents with scalar,
pseudoscalar, and tensor currents, and also with so-called ``second class'' 
(abnormal $G$-parity) currents. Additionally, in Adler, Karliner, Lieberman,  
Ng, and Tsao (1976), we did a detailed study of isospin-$1/2$ resonance 
production by $V,A$ neutral currents.  Perhaps 
the one part of the group effort on alternative current structures 
to have lasting value was a calculation 
of nucleon to nucleon and pion to pion matrix elements of scalar, 
pseudoscalar, and tensor current densities, using all the theoretical 
tools then at our disposal: flavor  $SU_3$ and chiral $SU_3 \times SU_3$ 
symmetries, the quark model, and the MIT ``bag'' model.  The results of 
these calculations were checked by several of us, and tabulated in 
Adler, Colglazier, Healy, Karliner, Lieberman, Ng, and Tsao (1975),
{\sf R39}; they were subsequently relevant for estimates of the coupling 
to nucleons of hypothetical scalar and pseudoscalar 
particles, such as axions.  The main part of the group effort was a  
current algebra soft pion production calculation for the alternative current 
case, which involved extensive algebra and computer work.   
{}From this, we found that 
one could explain roughly half of the reported Argonne threshold events  
with currents of scalar, pseudoscalar, and tensor type, by allowing some 
deviations from the matrix element 
estimates of {\sf R39}, as I reported at the January, 1975 Coral Gables    
Conference (Adler, 1975b).  In the meantime, the Argonne 
group reexamined possible background 
problems affecting their preliminary results, with the 
result that they ultimately discounted the cluster of pion production 
events near threshold. So by September of 1975, 
when I reviewed the subject of gauge theories and neutrino interactions 
at a conference at 
Northeastern University (Adler, 1976a), the electroweak theory 
predictions for neutral current weak pion production, following from purely  
$V$ and $A$ currents, were no longer in conflict with experiment.  This   
conclusion was reinforced by a subsequent detailed analysis by Monsay (1978) 
of neutral  current weak pion production, using my model together with  
the charge exchange corrections of {\sf R36}.    

In the summer of 1975 I lectured on neutrino interactions and neutral 
currents at the Sixth Hawaii Topical Conference on Particle Physics, and 
gave a comprehensive survey of neutral current phenomenology based on 
parton model methods, soft pion theorems, and quark model calculations of 
baryon static properties.    This appeared both in the conference 
proceedings (Adler, 1976b) and again in a tenth year anniversary volume 
selecting highlights from the preceding summer schools (Pakvasa and 
Tuan, 1982).  My hope in preparing the 1975 lectures was that surveying 
all available tools would hasten the day when one could determine 
electroweak parameters based on using all available data for a global fit, 
instead of doing piecemeal fits channel-by-channel.  Such a global 
fit was carried out a few years later by Abbott and 
Barnett (1978a,b), who included four types of data:  deep inelastic neutrino 
scattering $\nu N \to \nu  X$,  elastic neutrino-proton scattering 
$\nu p \to \nu  p$, neutrino induced inclusive pion production 
$\nu  N \to \nu \pi X$, and neutrino induced exclusive pion production 
$\nu N \to \nu \pi N$.  For the exclusive pion process, they employed my 
weak pion 
production calculation of {\sf R15} as extended 
to neutral currents in {\sf R38}, 
using test data that I ran for them from my programs as benchmarks 
to help debug their programming.  Their 
results were, in the words of their letter Abstract, ``for the 
first time, a unique determination of the weak neutral-current couplings 
of $u$ and $d$ quarks.  Data for exclusive pion production are a crucial 
new input in this analysis.''  Their multi-channel fit  
gave the first full confirmation that 
the Glashow--Weinberg--Salam model, with $\sin^2\theta_W$ between 0.22 and 
0.30, was in agreement with the experimental up and down quark  
neutral current coupling parameters.  To me, the Abbott--Barnett analysis  
was valued recompense for the several years of hard calculation and scholarly 
attention to detail that I had put into the subject of weak pion production.  

\bigskip

\section*{References for Chapter 5}

\begin{itemize}

\item Abbott, L. F. and R. M. Barnett (1978a).  Determination of the Weak 
Neutral-Current Couplings. {\it Phys. Rev. Lett.} {\bf 40}, 1303-1306. 

\item Abbott, L. F. and R. M. Barnett (1978b).  Quark and Lepton Couplings 
in the Weak Interactions.  {\it Phys. Rev. D} {\bf 18}, 3214-3229. 

\item Adler, S. L. (1974a) {\sf R34}.  Some Simple Vacuum-Polarization 
Phenomenology:  $e^+e^- \to {\rm Hadrons}$; The Muonic-Atom X-Ray 
Discrepancy and $g_{\mu}-2$.  {\it Phys. Rev. D} {\bf 10}, 3714-3728. 

\item Adler, S. L. (1974b) {\sf R35}.  $I=\frac{1}{2}$ Contributions to $\nu_{\mu} 
+ N \to \nu_{\mu} + N + \pi^0$ in the Weinberg Weak-Interaction Model.  
{\it Phys. Rev. D} {\bf 9},   229-230. 

\item Adler, S. L. (1974c).  Pion Charge-Exchange Scattering in the (3,3)-Resonance 
Region in Nuclei with a Neutron Excess.  {\it Phys. Rev. D} 
{\bf 9}, 2144-2150. 

\item Adler, S. L. (1974d) {\sf R37}. Application of Current Algebra Techniques 
to Neutral-Current--Induced Threshold Pion Production. {\it Phys. Rev. Lett.} 
{\bf 33}, 1511-1514. 

\item Adler, S. L. (1975a) {\sf R38}.  Application of Current-Algebra Techniques 
to Soft-Pion Production by the Weak Neutral Current: $V,\,A$ Case. 
{\it Phys. Rev. D} {\bf 12}, 2644-2655. 

\item Adler, S. L. (1975b).  Theoretical Interpretation of Recent Neutral 
Current Results, in {\it Theories and Experiments in High-Energy Physics}, 
Proceedings of Orbis Scientiae, the second Coral Gables Conference 
at the University of Miami, January 20-24, 1975,  B. Kursunoglu, chair, 
 A. Perlmutter and S. M. Widmayer, eds. (Plenum Press, New York), pp. 297-327. 

\item Adler, S. L. (1976a).  Gauge Theories and Neutrino Interactions, in 
{\it Gauge Theories and Modern Field Theory}, R. Arnowitt and P. Nath, eds.  
(MIT Press, Cambridge, MA), pp. 127-160. 
  
\item Adler, S. L. (1976b).  Neutrino Interaction Phenomenology and Neutral 
Currents, in {\it Proceedings of the Sixth Hawaii Topical Conference in  
Particle Physics (1975)},  P. N. Dobson, S. Pakvasa, V. Z. Peterson, and 
S. F. Tuan, eds. (University of Hawaii, Manoa/Honolulu), pp. 1-207. 

\item Adler, S. L., E. W. Colglazier, J. B. Healy, I. Karliner, J. Lieberman, 
Y. J. Ng, and H.-S. Tsao (1975) {\sf R39}. Renormalization Constants 
for Scalar, Pseudoscalar, and Tensor Currents.  {\it Phys. Rev. D} 
{\bf 11}, 3309-3318. 

\item Adler, S. L., R. F. Dashen, and S. B. Treiman (1974).  Comments on Proposed 
Explanations for the Muonic-Atom X-Ray Discrepancy.  {\it Phys. Rev. D} 
{\bf 10}, 3728-3735. 

\item Adler, S. L., G. R. Farrar, and S. B. Treiman (1972) {\sf R33}.  
Three-Pion States in the $K_L \to \mu^+\mu^-$ Puzzle.  {\it Phys. Rev. D} 
{\bf 5}, 770-772. 

\item Adler, S. L. and A. S. Goldhaber (1963).  Use of the Deuteron to Provide 
a Polarized Proton Target.  {\it Phys. Rev. Lett.} {\bf 10}, 448-450. 

\item Adler, S. L., I. Karliner, J. Lieberman, Y. J. Ng, and H.-S. Tsao (1976).  
Isospin-$\frac{1}{2}$ Nucleon-Resonance Production by a $V,\,A$ Weak Neutral 
Current. {\it Phys. Rev. D} {\bf 13}, 1216-1233. 

\item Adler, S. L., S. Nussinov, and E. A. Paschos (1974) {\sf R36}.  Nuclear 
Charge-Exchange Corrections to Leptonic Pion Production in the 
(3,3)-Resonance  Region.  {\it Phys. Rev. D} {\bf 9}, 2125-2143. 

\item Aviv, R. and R. F. Sawyer (1971).  Three-Pion Intermediate State and the 
$K_L^0 \to \mu^+ \mu^-$ Puzzle.  {\it Phys. Rev. D} {\bf 4}, 2740-2742. 

\item Barbieri, R. (1975).  Vacuum Polarization Phenomenology:  The $\mu$ Mesic 
Atom X-Ray Discrepancy and the $2P_\frac{3}{2}-2S_\frac{1}{2}$ Separation 
in the $(\mu \, {}^4{\rm He})^+$ System.  {\it Phys. Lett. B} {\bf 56},  
266-270. 

\item Barbieri, R. and T. E. O. Ericson (1975).  Evidence against the Existence 
of a Low Mass Scalar Boson from Neutron-Nucleus Scattering.  
{\it Phys. Lett. B}  {\bf 57}, 270-272. 

\item Beneke, M. (1999). Renormalons.  {\it Physics Reports} {\bf 3}, 1-142, 
Sec. 2.2.  Beneke uses the notation $D(Q^2)$ but does not include an 
additional factor of $Q^2$. 

\item Borie, E. and G. A. Rinker (1982).  The Energy Levels of Muonic Atoms. 
{\it Rev. Mod. Phys.} {\bf 54}, 67-118.  See particularly pp. 105-108. 

\item Broadhurst, D. J. and A. L. Kataev (1993).  Connections 
between Deep-Inelastic and Annihilation Processes at 
Next-to-Next-to-Leading Order and Beyond.  {\it Phys. Lett. B} {\bf 315}, 179-187.  

\item Cveti\v c, G., C. Dib, T. Lee, and I. Schmidt (2001).  
Resummation of the Hadronic Tau
Decay Width with the Modified Borel Transform Method.  {\it Phys. Rev. D} 
{\bf 64}, 093016. 

\item Cveti\v c, G., T. Lee, and I. Schmidt (2001).  Resummations 
with Renormalon 
Effects for the Leading Hadronic Contribution to the Muon $g_{\mu}-2$.  
{\it Phys. Lett. B} {\bf 520}, 222-232. 

\item De R\'ujula, A. and H. Georgi (1976).  Counting Quarks in $e^+e^-$ Annihilation.
{\it Phys. Rev. D} {\bf 13}, 1296-1301. 

\item Dorokhov, A. (2004). Adler Function and Hadronic Contribution to the Muon 
$g-2$ in a Nonlocal Chiral Quark Model.  {\it Phys. Rev. D} {\bf 70}, 
094011. 

\item Eidelman, S., F. Jegerlehner, A. L. Kataev, and O. Veretin (1999).  
Testing Non-Perturbative Strong Interaction Effects via the Adler Function. 
{\it Phys. Lett. B} {\bf 454} 369-380. 

\item Kataev, A. L. (1996).  The Generalized Crewther Relation: The 
Peculiar 
Aspects of the Analytical Perturbative QCD Calculations; arXiv: hep-ph/9607426. 

\item Kataev, A. L. (1999).  Adler Function from $R^{e^+e^-}(s)$ Measurements: 
Experiments vs QCD Theory, in {\it Moscow 1999, Particle Physics at the 
Start of the New Millennium}, pp. 43-52; arXiv: hep-ph/9906534. 

\item Lee, B. W. (1972).  The Process $\nu_{\mu} + p \to \nu_{\mu} + p + \pi^0$ 
in Weinberg's Model of Weak Interactions.  {\it Phys. Lett. B} {\bf 40}, 
420-422. 

\item Lee, W. (1972) Experimental Limit on the Neutral Current in the 
Semileptonic Processes.  {\it Phys. Lett. B} {\bf 40}, 423-425. 

\item Milton, K. A., I. L. Solovtsov, and O. P. Solovtsova (2001). Adler 
Function for Light Quarks in Analytic Perturbation Theory.  {\it Phys. 
Rev. D} {\bf 64}, 016005. 

\item Monsay, E. H. (1978).  Single-Pion Production by the Weak Neutral Current. 
{\it Phys. Rev. D} {\bf 18}, 2277-2289.

\item Pakvasa, S. and S. F. Tuan, eds. (1982).   {\it Selected Lectures, Hawaii  
Topical Conference in  Particle Physics} (World Scientific, Singapore).  
My 1975 lectures Adler (1976b) are reprinted in Vol. 2 of 
this compilation, 
pp. 499-705. 

\item Peris, S., M. Perrottet, and E. de Rafael (1998).  Matching Long and Short 
Distances in Large-$N_c$ QCD.  {\it JHEP} {\bf 05}, 011. 

\item Perkins, D. H. (1972)  Neutrino Interactions, in 
{\it Proceedings of the XVI International 
Conference on High Energy Physics}, Chicago--Batavia, IL, 
 J. D. Jackson and A. Roberts, eds. (National Accelerator Laboratory, Batavia), 
Vol. 4, pp. 189-247.  The remarks on charge-exchange corrections are 
on pp. 205-206.

\item Poggio, E. C., H. R. Quinn, and S. Weinberg (1976). 
 Smearing Method in 
the Quark Model.  {\it Phys. Rev. D} {\bf 13}, 1958-1968.

\item Pratap, M., J. Smith, and Z. E. S. Uy (1972).  Total Cross Sections for 
the Reactions $2\gamma \to 3\pi$ and $e^+e^- \to e^+e^-3\pi$.  
{\it Phys. Rev. D} {\bf 5}, 269-270. 

\end{itemize}

\chapter*{6. Gravitation}

\markboth{Adventures in Theoretical Physics}{Gravitation}
\bigskip
\section*{Introduction} 
During the first half of the 1970's, I started to get interested in learning 
more about gravitational physics.  When I was a graduate student at Princeton 
in the early 1960's, particle physics and gravitational physics were 
quite separate subjects, with the former the domain of Goldberger and Treiman, 
and the latter the domain of Wheeler and Dicke, to mention just a few key   
faculty members.  
Under the unstructured system at Princeton, I 
never took a course in gravitation, and for my general exam got by with the 
introduction to  general relativity that I obtained by reading the 
text of Peter Bergmann (1942), as well as reading some of the 
original Einstein 
papers reprinted 
in a Dover edition. (Working through the Dover volume was a project 
of an informal reading and discussion group during my senior year at Harvard,  
organized by Norval Fortson, an experimental physics graduate 
student affiliated with the residential house where I lived then.)  
However, in the 1970's it  
became clear both that many new results had been obtained in general 
relativity, so that my undergraduate knowledge was out-of-date, and  that 
general relativity was becoming part of the essential tool kit of people 
working in quantum field theory.  Among the things that convinced me of 
this were reading the thesis of Stephen Fulling (1972) on scalar quantum 
field theory in de Sitter space, while I was working on the $O(5)$ formulation 
of QED, the work of 't Hooft and Veltman (1974) and Deser (1975)  
on one-loop divergences of 
quantum gravity, and the availability of the new books on gravitation of 
Weinberg (1972), Misner, Thorne, and Wheeler (1973), and Hawking and Ellis 
(1973).

My intention in writing my comprehensive Hawaii lectures in the summer of 
1975 was to wind up my involvement with neutrino physics, so that I could 
turn to something new.  
Since in 1976 I was due for a sabbatical, and my family did not want to 
travel away from Princeton,  I decided that to learn relativity I would  
take a ``reverse sabbatical'', by  going to Princeton University to teach 
the relativity course for a year.  So I spent my evenings during the 
1975-1976 academic year reading the texts of Weinberg and of Misner, Thorne, 
and Wheeler, and then took my sabbatical during the 1976-1977 academic year, 
teaching both the fall term course in Special Relativity and the spring 
term continuation course in General Relativity.  I also was the faculty
advisor for John David Crawford, who did a senior thesis  on experimental tests 
for curvature squared additions to the gravitational action.  With this 
reading and teaching as background, I embarked on a 
number of relativity-related research 
projects, described in the next two sections. 
\bigskip
\section*{First Papers}

My first papers on gravity were the working out of a very speculative 
idea, that gravitation might be a composite phenomenon, with the 
gravitational fields arising as composite ``pairing'' amplitudes of photons 
in analogy with the energy gap order parameter for superconductivity. 
In the paper Adler, Lieberman, Ng, and Tsao (1976), we looked for weak 
coupling singularities in the electromagnetic 
photon-photon ladder graph sum in a conformally 
flat spacetime, and found some resemblances to the helicity structure of 
graviton exchange amplitudes.  In a follow-up paper (Adler, 1976) I gave 
a linearized Hartree formulation for the photon pairing problem in a 
general background metric.  I was never able to establish a detailed  
connection between photon pairing amplitudes and graviton couplings in the 
general case, and the fact that no weak coupling singularities occurred in  
flat spacetime meant that one could not establish a connection with the 
standard results of linearized general relativity.  In retrospect, the 
absence of pairing effects in flat spacetime could have been expected  
from a subsequent theorem of Weinberg and Witten (1980), 
that ruled out spin-2 composites under quite 
general assumptions, and effectively doomed the program as set up in the   
1976 papers.  However, a useful outcome of writing these papers was that 
it started me thinking more generally about the idea of gravitation as an 
effective theory, and in particular about Sakharov's ideas on gravitation, 
which I briefly discussed in the paper Adler (1976); following up this 
direction later on led to my work on the Einstein action as a 
symmetry-breaking effect, discussed in the  
next section. 

A second topic that I worked on in 1976 was the regularization of the 
stress-energy tensor for particles propagating in a general background 
metric.  In the paper Adler, Lieberman, and Ng (1977), we applied covariant  
point-splitting techniques to the Hadamard series for the Green's functions, 
which we used to define a regularized stress-energy tensor for vector and 
scalar particles.  This was a very technical computation, and contained 
useful formulas among its results, but also produced an embarrassment: by 
our method of regularization, we did not find the trace anomaly that had 
been found by others using different methods.  We rechecked our calculation 
carefully, but could not find the source of the discrepancy.  The problem 
was finally resolved by Wald (1978) (in time to be described in a note 
added in proof to our 1977 paper).  Wald had earlier (Wald, 1977)  
set up a general axiomatization for the stress-energy tensor, and 
in  Wald (1978) had shown that it leads to an essentially unique result.  
Applying a point-separation method similar to ours, he had also found 
no trace anomaly, but then went on to note that there was a subtle error in 
our analysis.  We had assumed that the  
local and boundary-condition-dependent parts of the Hadamard solution are 
separately symmetric in their arguments, but this is in fact not the case; 
only their sum is symmetric.  Wald (1978) showed in the scalar case that 
when the analysis is repeated without the incorrect assumption, one gets 
the standard trace anomaly.  Judy Lieberman and I then did the corresponding 
calculation in the vector particle case 
(Adler and Lieberman, 1978, {\sf R40}), again finding that when the asymmetry 
of the two pieces of the Hadamard solution is taken into account, one gets 
the correct trace anomaly.  

In a lunchtime conversation at some point during the 1977-1978 academic 
year, Robert Pearson asked whether the ``no-hair'' theorems of general 
relativity applied to the case of spontaneous symmetry breaking.  I thought 
this was interesting and looked into it, finding no relevant papers in the 
literature.  This became the subject of a joint paper (Adler and Pearson, 
1978, {\sf R41}), which showed that the standard ``no-hair'' theorems 
generalize to the vector field in the Abelian Higgs model, and to the 
non-conformally invariant Goldstone scalar field model. In our paper, we  
restricted ourselves to static, spherically symmetric black holes,  
and made the 
physically motivated assumption that any ``hair'' would also be static  
and spherically symmetric.  This permits a simplifying choice of gauge 
for the Abelian Higgs model 
introduced by Bekenstein (1972).  He observes that static electric charge 
distributions must give rise to static electric fields and vanishing magnetic 
fields.  Thus one can find a special gauge in which the potentials $A_{\mu}$ 
obey $\vec A=0,~dA_0/dt=0$.  Since the gauge-independent 
source current $j_{\mu}$ obeys similar 
conditions $\vec j=0,~dj_0/dt=0$, and since the gauge-independent magnitude of the Higgs scalar field is static, 
one finds that the residual phase of 
the Higgs scalar field in the special gauge is a space-independent, linear 
function of time, which can be eliminated by a further gauge transformation 
that preserves the gauge conditions $\vec A=0,~dA_0/dt=0$.  Thus one can 
do the analysis of possible ``hair'' taking the vector potential to be 
zero, and the Abelian Higgs field to be real. I have described Bekenstein's   
argument here in some detail because the choice of gauge in {\sf R41}
is the basis of rather loosely worded objections to our 
paper in lectures of Gibbons (1990); his assertion (and that of authors  
who have quoted his lectures) that the gauge choice is problematic is not   
correct, as working through the Bekenstein argument given above makes clear. 
Also, I have rechecked the proof given in {\sf R41}, and apart from 
the minor problem found by Ray, as 
discussed below, I find that the proof is correct, in disagreement with 
further statements in Gibbons' lecture. However, in response to Gibbons' 
comments 
about our choice of gauge, proofs of the ``no-hair'' theorem for 
the Abelian Higgs model that do not use a special gauge choice have 
since been given by Lahiri (1993) and by Ay\'on-Beato (2000).  

Our argument starting from Eq.~(24) of {\sf R41}  was subsequently 
considerably 
simplified, and in the case when $d\theta/d\lambda|_H=0$  corrected, in a 
paper of Ray (1979). (The subscript $H$ here refers to evaluation at the 
horizon; see {\sf R41} for details of this and other notation used in the 
following discussion of Ray's paper.)  The minor problem noted by Ray resulted from our 
not dropping the 
subdominant term $d\theta/d\lambda$ on the right-hand side of Eq.~(31) 
when integrating this equation to get Eq.~(33), so as to be consistent 
with our dropping this term elsewhere, such as in Eq.~(32).  
When this term is dropped, the $\theta^{-1/2}$ factor in Eq.~(33) is replaced 
by a constant,  and the approximate solution of Eq.~(33) agrees with the 
exact solution of Eq.~(24) given by Ray.  As Ray points out, 
with this correction one still finds 
that $q^{-1}\phi^2$ is infinite at the 
horizon unless $K=0$, which is what is needed to complete the proof.  

Finally, I note that the subject of black hole ``hair'' in gauge theories has  
taken on new 
interest recently with the discovery that topological charges on a 
black hole can give nonzero effects outside the horizon; see, for 
example, Coleman, Preskill, and Wilczek (1992) and the related lectures 
of Wilczek (1998).

\bigskip
\section*{Einstein Gravity as a Symmetry Breaking Effect}  

In late January of 1978  
I organized a small conference on ``Geometry, Gravity and Field Theory'' 
 for the 
EST Foundation in San Francisco; this was a memorable event that was attended 
by a large fraction of the leading people with interests 
in quantum gravity.  During my plane  
travel for this conference, and afterwards, 
I started to think about the confinement problem in QCD, and 
this became the main focus of my research for the next two years, as 
described in the following chapter.  However, learning about scale breaking 
in QCD also led me back into gravitational physics,  
through considering the role similar mechanisms might play  in 
giving a quantitative form to the suggestion by Sakharov (1968) (see also  
Klein, 1974) 
that Einstein  gravity is the ``metric elasticity'' of spacetime.  I 
did not arrive at the correct formulation immediately; I find in my files 
two unpublished manuscripts, the first positing  monopole boundary 
conditions, and the second positing dimension-2 operators, as a source 
for symmetry breaking, in both cases suggesting connections with the  
Einstein-Hilbert action.  I went as far as submitting  
a manuscript based on the second for publication, 
and also gave a seminar on it at Princeton University, where 
my arguments were 
torn to shreds by David Gross (following which I withdrew the manuscript).  
The criticism proved useful; I went home, learned more about dimensional  
transmutation and the theory of calculability versus renormalizability,  
and came up with the correct formulation given in Adler (1980a), {\sf R42}. 
The basic idea here is that in theories which contain no scalars, so that 
scale invariance is spontaneously broken (QCD is a prime example, but 
``technicolor'' type unification models also fit this description), there 
will be an induced order $R$ term in the action in a curved background, with 
a coefficient that is calculable in terms of the scale mass of the theory. 
Thus, if an underlying unified theory spontaneously breaks scale invariance 
at the Planck scale, one can induce the Einstein gravitational action as 
a scale-symmetry breaking effect, giving an explicit realization of the 
Sakharov--Klein idea.  

I followed up this paper with a second one (Adler, 1980b, {\sf R43}) in which 
I gave an explicit formula for the ``induced gravitational constant'' in 
theories with dynamical breakdown of scale invariance, expressed in   
terms of the vacuum expectation of the autocorrelation function 
of the trace of the renormalized 
stress-energy tensor $\tilde T_{\mu\nu}$, 
\begin{equation}
(16 \pi G_{\rm ind})^{-1}=\frac{i}{96}\int d^4x [(x^0)^2-(\vec x)^2] 
\langle T\big(\tilde T_{\lambda}^{\lambda}(x) \tilde T_{\mu}^{\mu}(0)\big) 
\rangle^{\rm flat\, spacetime}_{0,\,{\rm connected}}~~~.\nonumber
\end{equation}
This formula for the induced Newton constant 
was independently obtained at about the same time by Zee (1981), and    
in the subsequent literature, 
the term ``induced gravity'' has come to be frequently used  
to describe the whole set of ideas involved.   
These papers 
attracted considerable attention in the gravity community, one result of 
which was that Claudio Teitelboim and his colleagues at the University of 
Texas in Austin invited me to give the Schild lectures in April of 1981.  
(My four lectures over a two week period, entitled ``Einstein Gravity as 
a Symmetry-Breaking Effect in Quantum Field Theory'', were the eleventh 
in the Schild series.)   
This proved memorable for an unanticipated reason; shortly before I was to 
go to Texas I contracted a mild case of what was probably type-A hepatitis
(the kind transmitted by shellfish), and so was sick in bed with very little 
energy.  I dragged myself out of bed on alternate days to write lecture notes, 
and then was so tired I had to sleep the entire day following.  At any rate, 
I improved enough so that my doctor gave me permission to go to Texas, 
where Philip Candelas took me into his home and helped me get through my 
scheduled lectures.  Ultimately, I expanded the lectures into a much-cited 
comprehensive 
article that appeared in {\it Reviews of Modern Physics} (Adler, 1982, {\sf R44}). 
A year later, I wrote a briefer synopsis of the program of generating 
the Einstein action as an effective field theory, for a Royal Society 
conference on ``The Constants of Physics'', which was published as 
Adler (1983).  

The explicit formula for the induced gravitational constant raises a 
number of interesting issues.  First of all, if one assumes an unsubtracted 
dispersion relation for the Fourier transform $\psi(q^2)$ of the 
autocorrelation function of the stress-energy 
tensor trace, the induced gravitational constant is negative. However, 
as shown by Khuri (1982a) using analyticity methods, in asymptotically 
free theories there are three possible cases, depending on the distribution 
of zeros of $\psi(q^2)$, and in one of these cases $G_{\rm ind}$ 
has positive sign.  In 
further papers Khuri (1982b,c) showed that in this case one can place 
useful bounds on the induced gravitational constant, expressed 
in terms of the scale mass of the theory.   

The question of whether the 
formula for the induced gravitational constant gives a unique answer has been  
discussed, from the point of view infrared renormalon singularities, 
by David (1984) and in a follow-up paper of David and Strominger (1984). 
These authors argue that renormalons introduce an arbitrariness into  
the calculation of $G_{\rm ind}$, as manifested through the fact that in the 
dimensional regularization of the ultraviolet singular ``comparison function'' 
$\Psi_c(t)$ introduced in Eq.~(5.48) of {\sf R44}, one has to continue 
onto a cut.  In Appendix B, 
Section 3 of {\sf R44}, I used a principal value prescription to deal with 
this, which David argues can be modified by taking complex weightings of 
the upper and lower sides of the branch cut, allowing a free parameter 
multiple of the imaginary part to be introduced into the calculation of the   
integral over the comparison function.  
David argues that this means that the expression for 
$G_{\rm ind}$ has an inherent ambiguity.  I believe that this conclusion 
is suspect; since QCD and similar theories that spontaneously generate a 
mass scale are believed to be consistent field theories, their 
curved spacetime embeddings should, by the equivalence principle, also be 
consistent theories.  This strongly suggests that the coefficient of the 
order $R$ term in a curvature expansion of the vacuum action functional 
should be well defined, and that the ambiguity is an artifact of the 
comparison function procedure.  This view is supported by the review 
article of Beneke (1999) on renormalons, where it is argued that renormalon 
ambiguities are typically canceled by corresponding ambiguities in  
non-perturbative terms \big(such as the integral $\Delta I_{UV}$ with 
integrand $\Psi-\Psi_c(t)$ in Eq.~(5.48)\big), giving total physical 
amplitudes that are unambiguous.  In other words, the renormalon ambiguities 
are an artifact of an attempted separation of QCD physical amplitudes into 
a ``perturbative'' and a ``non-perturbative'' part, and only indicate that 
if a branching prescription (such as a principal value) is needed for 
the perturbative part, then a corresponding branching prescription is 
also needed for the non-perturbative part.  This will make the 
calculation of quantities like $G_{\rm ind}$ difficult, but does not imply that 
the calculation cannot, in principle, give a unique, physical answer. In 
the paper of  David and Strominger (1984), the authors show that $G_{\rm ind}$ 
is unambiguous in {\it finite} supersymmetric theories, giving an existence   
proof that there are theories with a finite induced Newton's constant. 
In the general case,  
they acknowledge that ``there is no {\it proof} that $G_{\rm ind}$ will 
{\it necessarily} be ambiguous'', and I suspect that in fact $G_{\rm ind}$  
will turn out to be well defined in a much wider class of supersymmetric and 
non-supersymmetric theories than only finite ones. 
Clearly, this is a question that merits further study.  

If one thinks more generally about the structure of a fundamental 
theory of gravitation, there are a number of possibilities.  It may 
be that the Planck length is the minimum length scale possible, because 
of an underlying ``graininess'' of spacetime.  Or spacetime may 
be a continuum, as generally assumed, in which case the   
Planck length plays the role of the scale at which   
a classical metric breaks down, with new dynamical principles taking  
over at shorter distances. 
The suggestion that the order $R$ gravitational action is an 
expression of 
scale symmetry breaking in a more fundamental scale-invariant theory    
is clearly  based on a continuum picture of spacetime.  A continuum 
assumption is also made in string theories, which however are 
not scale-invariant; in string theories 
a fundamental length scale (the string tension) appears in the 
action, and this directly sets the scale for the gravitational action.         
Should spacetime turn out to be discrete or grainy, there may be more general 
forms of the induced gravitation idea that are relevant.    
Ultimately, the origin of the spacetime metric, and of the 
Einstein--Hilbert gravitational action that governs its dynamics, will 
not be certain 
until we have a unifying theory that also resolves the cosmological constant 
problem, which is not dealt with in any of the current ideas about 
quantum gravity.

\section*{References for Chapter 6}

\begin{itemize}

\item Adler, S. L. (1976).  Linearized Hartree Formulation of the Photon 
Pairing Problem.  {\it Phys. Rev. D} {\bf 14}, 379-383. 

\item Adler, S. L. (1980a) {\sf R42}.  Order-$R$ Vacuum Action Functional in 
Scalar-Free Unified Theories with Spontaneous Scale Breaking.  
{\it Phys. Rev. Lett.} {\bf 44}, 1567-1569. 

\item Adler, S. L. (1980b) {\sf R43}.  A Formula for the Induced Gravitational 
Constant.  {\it Phys. Lett. B} {\bf 95}, 241-243. 

\item Adler, S. L. (1982) {\sf R44}. Einstein Gravity as a Symmetry-Breaking 
Effect in Quantum Field Theory. {\it Rev. Mod. Phys.} {\bf 54}, 729-766. 

\item Adler, S L. (1983).  Einstein Gravitation as a Long Wavelength 
Effective Field Theory.  {\it Phil. Trans. R. Soc. Lond. A} 
{\bf 310}, 273-278.  This paper also appears in {\it The Constants of Physics}, 
W. H. McCrea and M. J. Rees, eds. (The Royal Society, London, 1983), 
pp. [63]-[68].

\item Adler, S. L. and J. Lieberman (1978) {\sf R40}.  Trace Anomaly of the 
Stress-Energy Tensor for Massless Vector Particles Propagating in a General 
Background Metric.  {\it Ann. Phys.} {\bf 113}, 294-303. 

\item Adler, S. L., J. Lieberman, and Y. J. Ng (1977).  Regularization of the 
Stress-Energy Tensor for Vector and Scalar Particles Propagating in a 
General Background Metric. {\it Ann. Phys.} {\bf 106}, 279-321. 

\item Adler, S. L., J. Lieberman, Y. J. Ng, and H.-S. Tsao (1976).  Photon Pairing 
Instabilities: A Microscopic Origin for Gravitation?  {\it Phys. Rev. D} 
{\bf 14}, 359-378.

\item Adler, S. L. and R. B. Pearson (1978) {\sf R41}.  ``No-Hair'' Theorems for 
the Abelian Higgs and Goldstone Models.  {\it Phys. Rev. D} {\bf 18}, 
2798-2803. 
  
\item Ay\'on-Beato, E. (2000).  ``No-Hair'' Theorem for Spontaneously Broken 
Abelian Models in Static Black Holes.  {\it Phys. Rev. D} {\bf 62}, 
104004. 

\item Bekenstein, J. (1972).  Nonexistence of Baryon Number for Static Black 
Holes.  {\it Phys. Rev. D} {\bf 5}, 1239-1246. 

\item Beneke, M. (1999).  Renormalons. {\it Physics Reports} {\bf 317}, 1-142.
See especially  Sec. 2. 

\item Bergmann, P. G. (1942).  {\it Introduction to the Theory of Relativity} 
(Prentice Hall, Englewood Cliffs). 

\item Coleman, S., J. Preskill, and F. Wilczek (1992).  Quantum Hair on Black 
Holes.  {\it Nucl. Phys. B} {\bf 378}, 175-246. 

\item David, F. (1984).  A Comment on Induced Gravity.  {\it Phys. Lett. B} 
{\bf 138}, 383-385. 

\item David, F. and A. Strominger (1984).  On the Calculability of Newton's 
Constant and the Renormalizability of Scale Invariant Quantum Gravity. 
{\it Phys. Lett. B} {\bf 143}, 125-129. 

\item Deser, S. (1975).  Quantum Gravitation: Trees, Loops and Renormalization, 
in {\it Quantum Gravity, an Oxford Symposium}, C. J. Isham, R. Penrose, 
and D. W. Sciama, eds. (Clarendon Press, Oxford), pp. 136-173. 

\item Fulling, S. (1972).  Scalar Quantum Field Theory in a Closed Universe of 
Constant Curvature.  Princeton University Dissertation. 

\item Gibbons, G. W. (1990). Self-Gravitating Magnetic Monopoles, Global Monopoles 
and Black Holes, in {\it The Physical Universe: The Interface between 
Cosmology, Astrophysics, and Particle Physics, Lecture Notes in Physics 
Vol. 383}, J. D. Barrow, A. B. 
Henriques, M. T. V. T. Lago, and M. S. Longair, eds. (Springer-Verlag, 
Berlin, 1991), pp. 110-133. 

\item Hawking, S. W. and G. F. R. Ellis (1973).  {\it The Large Scale 
Structure of Space-Time} (Cambridge University Press, Cambridge). 

\item Khuri, N. N. (1982a).  Sign of the Induced Gravitational Constant. 
{\it Phys. Rev. D} {\bf 26}, 2664-2670. 

\item Khuri, N. N. (1982b).  Upper Bound for Induced Gravitation.  {\it Phys. Rev. 
Lett.} {\bf 49}, 513-516. 

\item Khuri, N. N. (1982c).  Induced Gravity and Planck Zeros.  {\it Phys. Rev. 
D} {\bf 26}, 2671-2680. 

\item Klein, O. (1974).  Generalization of Einstein's Principle of Equivalence   
so as to Embrace the Field Equations of Gravitation.  
{\it Phys. Scr.} {\bf 9}, 69-72. 

\item Lahiri, A. (1993). The No-Hair Theorem for the Abelian Higgs Model. 
{\it Mod. Phys. Lett. A} {\bf 8}, 1549-1556. 

\item Misner, C. W., K. S. Thorne, and J. A. Wheeler (1973).  {\it Gravitation}   
(W. H. Freeman, San Francisco). 

\item Ray, D. (1979).  Comment on the ``No-Hair'' Theorem for the Abelian-Higgs 
Model. {\it Phys. Rev. D} {\bf 20}, 3431. 

\item Sakharov, A. D. (1968).  Vacuum Quantum Fluctuations in Curved Space and the 
Theory of Gravitation.  {\it Dok. Akad. Nauk. SSSR} {\bf 177}, 70-71 
(English translation:  {\it Soviet Phys.  -- Doklady}
 {\bf 12}, 1040-1041). 

\item 't Hooft, G. and M. Veltman (1974).  One-Loop Divergencies in the Theory of 
Gravitation.  {\it Ann. Inst. Henri Poincar\'e A: Physique th\'eorique} 
{\bf 20}, 69-94. 

\item Wald, R. M. (1977).  The Back Reaction Effect in Particle Creation in 
Curved Spacetime.  {\it Commun. Math. Phys.} {\bf 54}, 1-19. 

\item Wald, R. M. (1978).  Trace Anomaly of a Conformally 
Invariant Quantum Field 
in Curved Spacetime.  {\it Phys. Rev. D} {\bf 17}, 1477-1484. 

\item Weinberg, S. (1972).  {\it Gravitation and Cosmology:  Principles and 
Applications of the General Theory of Relativity} (John Wiley, New York). 

\item Weinberg, S. and E. Witten (1980).  Limits on Massless Particles. {\it Phys. Lett. B} {\bf 96}, 59-62. 

\item Wilczek, F. (1998).  Lectures on Black Hole Quantum Mechanics: 
Lectures 3 \& 4, in {\it The Black Hole 25 Years After}, 
C. Teitelboim and J. Zanelli, eds. (World Scientific, Singapore), 
pp. 229-326. 

\item Zee, A. (1981).  Spontaneously Generated Gravity.  
{\it Phys. Rev. D} {\bf 23}, 858-866. 

\end{itemize}

\chapter*{\centerline{7. Non-Abelian Monopoles, Confinement Models,}
\centerline{and Chiral Symmetry Breaking}}

\markboth{Adventures in Theoretical Physics}{Non-Abelian Monopoles, Confinement Models, and Chiral Symmetry Breaking}
\bigskip
\section*{Introduction} 
The somewhat disparate topics to be discussed in this chapter are all 
connected through my interest during the late 1970's and 
early 1980's in studying 
nonperturbative properties of quantum chromodynamics (QCD), the theory of 
the strong interactions.  I began these investigations by looking for 
a semi-classical model for heavy quark confinement.  My first idea, that 
quarks might be confined in a non-Abelian monopole background field, did 
not work, but led to interesting progress in the theory of monopoles, as   
described in the first section. Most significantly, as discussed in detail, 
the monopole work led indirectly to   
the completion by Clifford Taubes of his multimonopole existence theorem 
during a visit to the IAS in the spring of 1980. 
I then turned to models based on the 
nonlinear dielectric properties of the QCD vacuum, which led to 
the confinement  of quarks in ``bag''-like structures which yield  
good heavy quark static potentials, as discussed in the second section.  
Finally, at the end of this period I worked briefly on the spontaneous 
breaking of chiral symmetry in QCD within the framework of pairing 
models patterned after superconductivity, as discussed in the final section.  
All three of these aspects of my study of QCD involved heavy numerical 
work, which in turn led to my interest in algorithms discussed in the 
next chapter. 

\bigskip                            
\section*{Non-Abelian Monopoles}

My first attempt at the confinement problem, which did 
not succeed but which had useful by-products that I shall describe here,  
was based on the idea of considering the potential between classical 
quark sources in the background of a non-Abelian
 't Hooft (1974a)--Polyakov (1974)--Prasad--Sommerfield (1975)--Bogomol'nyi (1976)  
monopole or its generalizations, which I conjectured in Adler (1978b), 
{\sf R45}  
might act as a quark-confining ``bag''.   
To justify considering classical 
quark sources, I initially resorted to a scheme (Adler, 1978a) that I 
called ``algebraic 
chromodynamics'', which involved looking at the color space spanning the 
direct product of independent color charge matrices. However, I eventually 
dropped this apparatus in my pursuit of the confinement problem, and 
used instead the popular approximation of 
color charge matrices lying in a maximal Abelian subgroup 
of the $SU(3)$ color group of conventional QCD, which gives   
a good first approximation to the full QCD color structure.  Since it is clear that   
source charges in classical Yang--Mills theory are not confined, I looked 
for a simple modification of this theory that 
might lead to a linear potential.  
The first idea I tried was to look at classical Yang--Mills charges in 
the field of a background monopole. This had the obvious   
problem that the monopole scale has no clear relation to the QCD scale 
set by dimensional transmutation, but I simply ignored this difficulty 
and plunged ahead.  

To pursue (and ultimately rule out) the conjecture that a monopole 
background would confine, I did a number of 
calculations of properties of monopole solutions. The first was a calculation 
of the Green's function for a single Prasad--Sommerfield monopole, by using 
the multi-instanton representation of the monopole and a formalism for  
calculating multi-instanton Green's functions given by Brown et al. (1978).  
This calculation was spread over two papers that I wrote; setting up contour  
integral expressions for the Green's function was done in Appendix A 
of Adler (1978b), {\sf R45}, and the final result for the monopole 
propagator, after evaluation 
of the contour integrals and considerable algebraic simplification, was 
given in Appendix A of Adler (1979a), {\sf R46}.  \big(The fact that many 
lengthy expressions for parts of the Green's function collapsed, after 
algebraic rearrangement, into simple formulas, suggested that there should be a more efficient way to find the monopole Green's function.  Not long 
afterwards, Nahm (1980) gave a new representation for the monopole that 
permitted a much simpler calculation of the Green's function given in  
{\sf R46}.\big) To check that the lengthy  
expression that I had obtained for the propagator really satisfied the 
differential equation for the Green's function, I used numerical methods, 
calculating the partial derivatives acting on the propagator 
by finite difference methods on a very fine mesh.  From numerical 
calculations based on the propagator formula, it was clear     
that a single monopole background would not lead to confinement; all that 
happened was that a Coulombic attractive $-1/r$ potential was reversed into a 
repulsive $1/r$ potential for large quark separations, a result that could 
have been anticipated from the large distance structure of the monopole field. 

Not yet ready to give up on the monopole background idea, I then wrote two  
papers speculating that the Prasad--Sommerfield monopole might 
be a member of a larger class of solutions, in which the point at which the 
monopole Higgs field vanishes is extended to a higher-dimensional region,  
and in particular to a ``string''-like configuration with a line segment 
as a zero set. In the first of these papers (Adler 1979b) I studied small 
deformations around the Prasad--Sommerfield monopole and found several series 
of such deformations.  For normalized deformations I recovered the monopole 
zero modes that had already been obtained by Mottola (1978, 1979), 
but I found that 
``if an axially symmetric extension exists, it cannot be reached by 
integration out along a tangent vector defined by a nonvanishing, 
non-singular small-perturbation mode''.  This work was later extended into 
a complete calculation of the perturbations around the Prasad--Sommerfield 
solution by Akhoury, Jun, and Goldhaber (1980), who also found ``no acceptable 
nontrivial zero energy modes.''  In my second paper, Adler (1979c), 
I employed nonperturbative methods and suggested that despite the negative 
perturbative results, there might still be interesting extensions of the 
Prasad--Sommerfield solution with extended Higgs field zero sets.  

At just around the same time, Erick Weinberg  
wrote a paper (Weinberg, 1979b) extending an index theorem of 
Callias (1978) to give a parameter counting theorem for multi-monopole 
solutions.  Weinberg concluded that ``any solution with $n$ units 
of magnetic charge belongs to a $(4n-1)$-parameter family of solutions.  It  
is conjectured that these parameters correspond to the positions and 
relative $U(1)$ orientations of $n$ noninteracting unit monopoles''.  For 
$n=1$, his results agreed with the zero-mode counting implied by Mottola's 
explicit calculation.  Weinberg and I were aware of each other's work, as 
evidenced by correspondence in my file dating from March to June of 1979, 
and references relating to this correspondence in our papers Adler (1979c) 
and Weinberg (1979b).  

My contact with Clifford Taubes was initiated by an April, 1979 letter from 
Arthur Jaffe, after I gave a talk at Harvard while 
Jaffe, as it happened, was visiting Princeton!  In his letter, Jaffe  
noted that I was working on problems similar to those on 
which his students were working, and enclosed a copy of a paper by 
Clifford Taubes.   (This preprint was not filed with Jaffe's letter, so I 
am not sure which of the early Taubes papers listed on the SLAC Spires 
archive that it was.)  
Jaffe's letter initiated 
telephone contacts with Taubes and some correspondence from him.  On Jan. 6, 
1980 Taubes wrote to me that he was making progress in proving the existence 
of multi-monopole Prasad--Sommerfield solutions, and in this letter and a 
second one dated on January 18, 1980 he reported results that were relevant 
to my conjectures on the possibility of deformed monopoles.  His results 
placed significant restrictions on my conjectures; in a letter dated 
Feb. 1, 1980 I wrote to Lochlainn O'Raifeartaigh, who had also 
been interested in axially symmetric monopoles, saying that 
``On thinking some more 
about your paper (O'Raifeartaigh's preprint was unfortunately not 
retained in my files) I 
realized that the enclosed argument by Cliff Taubes is strong evidence 
against $n=2$ monopoles involving a line zero.  What Taubes shows is that 
a finite action solution of the Yang--Mills--Higgs Lagrangian cannot have 
a line zero of arbitrarily great length; hence if $n=2$ monopoles contained 
a line zero joining the monopole centers, the monopole separation would be 
bounded from above.  But this seems unlikely...''.   This correspondence 
and the result of Taubes was mentioned at the end of the published version, 
Houston and O'Raifeartaigh (1980).  

As a result of our overlapping interests, I arranged for Taubes to make 
an informal visit, of two or three months, to the IAS 
 during the spring of 1980.  Clifford had expressed 
interest in this, he noted in a recent email, in part 
because Raoul Bott had suggested 
that he visit the Institute to get acquainted with Karen Uhlenbeck, who 
was visiting the IAS that year.  In the course of his visit he met and  
interacted with Uhlenbeck, who, along with Bott, had a major 
impact on his development as a mathematician.  

Taubes began the visit by looking at my conjecture 
of extended zero sets, but after a while told me that he could not 
find an argument for them.  Partly as a result of his work, I was getting 
disillusioned with my own conjecture, so I asked him what was happening 
with his attempted proof of multi-monopole solutions.  Taubes replied that 
he was stuck on that, and not sure whether they existed. I then mentioned 
to him Erick Weinberg's parameter counting result, which strongly suggested 
a space of moduli much like that in the instanton case, where looking at 
deformations correctly suggests the existence and structure of the 
multi-instanton solutions.  To my surprise, Taubes was not aware of Erick's 
result, and knowing it impelled him into action on his multi-monopole proof. 
Within a week or two he had completed a proof,
and wrote it up on his  return to Harvard.  \big(Thus, there was a parallel  
to what happened a year before with respect to solutions of the first order 
Ginzburg--Landau equations.  In that case, Taubes had heard a lecture at 
Harvard by Erick Weinberg on parameter counting for multi-vortex solutions 
(written up as Weinberg, 1979a) and then went home and came up 
with his existence proof for multi-vortices, Taubes (1980).  The vortex work 
provided the initial impetus for Taubes' turning to the monopole problem.\big)
In his paper Taubes  showed that ``for every 
integer $N\not=0$ there is at least a countably infinite set of solutions 
to the static $SU(2)$ Yang-Mills-Higgs equations in the Prasad-Sommerfield 
limit with monopole number $N$.  The solutions are partially parameterized 
by an infinite sublattice in $S_N(R^3)$, the $N$-fold symmetric product 
of $R^3$ and correspond to noninteracting, distinct monopoles.'' This 
quote is taken from the Abstract of his preprint ``The Existence of 
Multi-Monopole Solutions to the Static, $SU(2)$ Yang--Mills--Higgs Equations 
in the Prasad--Sommerfield Limit'', which was received on the SLAC Spires 
data base in June, 1980, and which carried an acknowledgement on the 
title page noting that ``This work was completed while the author was a guest 
at the Institute for Advanced Studies, Princeton, NJ 08540''.  
His preprint also ended with an Acknowledgment section
noting his conversations with me, with Arthur Jaffe, and 
with Karen Uhlenbeck, as well as the Institute's hospitality.  The proof    
was not published in this form, however, but instead appeared (with  
acknowledgments edited out at some stage) as Chapter IV of the book 
Jaffe and Taubes (1980)
that was completed in August of 1980. The multimonopole 
existence proof was a milestone in Taubes' career; in a recent exchange 
of emails relating to the events described in this section, Taubes 
commented on his visit ``to hang out at the 
IAS during the spring of 1980.  It profoundly affected my subsequent 
career...''.  He went on to further investigations of monopole solutions, that 
lead him to studies of 4-manifold theory which have had a great impact on   
mathematics. 

O'Raifeartaigh, who had been following the monopole work at a distance, 
invited me during the spring of 1980 to come to Dublin that summer to lecture 
on my papers. However, since Taubes had much more interesting results 
I suggested to Lochlainn that he ask Clifford instead, and Taubes did go to  
Dublin to lecture. After Clifford's visit, I redirected my search for   
semiclassical confinement models to a study of nonlinear dielectric models 
by analytic and numerical methods, in collaboration  with Tsvi Piran; these 
models do give 
an interesting class of confining theories, and are described in the  
following section.  Based on the observation that  
the Yang--Mills action is multiquadratic (that is, at most quadratic in each 
individual potential component), Piran and I also applied 
the same numerical 
relaxation methods to give an efficient method for the 
computation of axially symmetric multimonopole 
solutions. \big(This was done mainly to illustrate the computer methods, 
since  by then exact analytic 2-monopole solutions had appeared; see Forgacs, 
Horvath and Palla (1981) and Ward (1981).\big)  The numerical methods that  
Piran and I developed were 
described in our {\it Reviews of Modern Physics} article  
Adler and Piran (1984), {\sf R47} 
that marked the completion of the research program on confining models, and     
as a by-product, on monopoles.                                    

\bigskip

\section*{Confinement Models}

Having seen that monopole backgrounds would not confine, I turned my 
attention to another type of semi-classical model, proposed in various 
forms by Savvidy (1977) \big(see also Matinyan and Savvidy (1978)\big) 
and Pagels and Tomboulis (1978).     The basic idea is to 
do electrostatics with Abelianized quark charges, and with the fundamental 
QCD action replaced by a renormalization group improved effective action, 
in which the gauge coupling is replaced by a running coupling, that is  
taken to be a function solely of the field strength squared. Although  
use of the running coupling is only justified by the renormalization group   
in the ultraviolet regime of large field strengths, the model assumes   
that the same functional form can be extrapolated to small field strengths 
as well.  This leads 
to electrodynamics with a nonlinear, field-dependent dielectric constant  
that develops a zero for small squared field strengths.  Because the only 
dynamical input from QCD is the running coupling, 
the model, as Frank Wilczek later remarked to me,  can be considered as 
a very simple embodiment of the idea that ``asymptotic 
freedom'' should be associated with ``infrared slavery''.  Since the   
running coupling involves a scale mass, the model directly incorporates 
the phenomenon of dimensional 
transmutation. Pagels and Tomboulis conjectured, on the basis of 
various evidence, that the nonlinear dielectric 
model would confine, but did not have a proof.  

In the paper Adler (1981), {\sf R48}, I analyzed the effective action 
model in detail and proved that it confines quarks.  The argument starts 
from a Euclidean form of the Feynman path integral, and shows that the 
static potential is the minimum of the effective action in the presence 
of sources.  I then specialized to the leading-logarithm effective 
action, and showed that the action 
minimum is associated with a field configuration 
in which a color magnetic field fills in whenever the color electric 
field is less than the minimum magnitude $\kappa$ at which the effective 
action is minimized.  This reduces the action minimization to 
an  electrostatics 
problem, to which one can apply flux conversation estimates due to 
't Hooft (1974b).  
In the nonlinear dielectric model context, these estimates show  
that the static potential 
is bounded from below by $\kappa Q (R-r)$, with $Q$ the Abelianized 
quark charge, with $r$ a constant, and with $R$ the interquark separation. 
Hence the potential increases linearly for large $R$, and the model 
confines.  In an Appendix to {\sf R48}, I discussed how a one-loop  
renormalization group exact, leading-logarithm 
running coupling can be obtained, by a coupling constant transformation, from  
the more usual two-loop renormalization group exact running coupling (to 
which the confinement argument also applies).  

When I presented this proof of confinement by the nonlinear dielectric model 
at a Department of Energy sponsored workshop in Yerevan, Armenia in 1983, 
an interesting dialogue with the Soviet physicist A. B. Migdal 
ensued.  When I started to talk, and said what I was going to prove, 
Migdal stood up and stated that it was well-known that the Savvidy 
(--Pagels--Tomboulis) model did not confine, and gave some reasons.  
I then presented  
my proof, after which Migdal stood up, and said words to the effect that 
the problem is that there are too many confining models!  As we shall see, 
there is really only one other model, the ``dual superconductor'' model,  
which like the nonlinear dielectric model is motivated by the idea of a 
color magnetic condensate, but describes this with a different dynamics.  

Following publication of {\sf R48}, I wrote a paper 
(Adler, 1982a) formalizing 
the approximations (further discussed below) 
needed to get an Abelianized effective action model 
from the functional integral for QCD.     
I then turned to the problem of 
understanding in detail {\it how} the leading-log model gives a confining 
potential.  Since it was clear that this would, at least in part,  
involve numerical solution of the nonlinear differential equations involved, 
I brought in Tsvi Piran as a postdoc.  Tsvi had worked extensively in 
the numerical solution of the Einstein equations of general relativity, 
and came to the IAS with the understanding that he would continue 
this and other interests he had in astrophysics, but would also collaborate 
with me in the numerical solution of the leading-log model equations.  
Because of my work on the induced gravity program, this collaboration didn't 
start immediately after Tsvi's arrival, but once we began 
work, Tsvi taught me a great deal about setting up an interactive program 
to numerically solve partial differential equations. As is typical in 
doing numerical work, most of our time was spent developing and testing 
our computer codes, which took many months.  To guard against programming 
errors, Tsvi and I each independently wrote our own programs, which once 
debugged gave identical results. The final production runs took a total 
of less than two days running time on the then new IAS VAX 11/780 
computer, using mesh sizes of up to $100 \times 100$ to resolve details of the confinement domain.  

The equations to be solved, in the leading-log model with  
three light fermion flavors and scale mass $\kappa$, are  
\begin{eqnarray}
\vec \nabla \cdot (\epsilon(E) \vec E)&=&j^0~~~,\nonumber\\
j^0&=&Q \delta(x)\delta(y) [\delta(z-a)-\delta(z+a)]~~~, \nonumber\\
\epsilon(E)&=&\frac{1}{4} b_0 \log(E^2/\kappa^2)~~,~~~E=|\vec E|~~~,\nonumber\\
b_0&=&\frac{9}{8\pi^2}~~~. \nonumber
\end{eqnarray}
We also studied the leading-log-log model, in which the two-loop exact form 
of the running coupling is used.   
We originally tried to solve the equations 
directly in terms of the scalar potential 
$A^0$, but found that the numerical programs were unstable.  I then 
introduced a flux function reformulation of the problem (suggested by 
similar methods used in plasma physics), and this gave a stable, rapidly 
convergent iteration showing formation of a flux-confining free boundary.  
To understand the structure of the free boundary, Tsvi suggested that a paper 
of Fichera on elliptic equations that degenerate to parabolic would be 
relevant, and this indeed was the case, as described in Appendix A of 
our review {\sf R47}. Prior to writing the review, we wrote two shorter 
papers. The  first (Adler and Piran, 1982a, {\sf R49})
demonstrated flux confinement and 
gave a numerical determination of the large $R$ asymptotic form of the 
interquark potential, which contains a leading term linear in $R$, and   
a subdominant term proportional to $\log \kappa^{1/2} R$. 
The second (Adler and Piran, 1982b, {\sf R50})
gave compact, accurate functional forms that fit the computed static 
potentials for both the leading-log and the leading-log-log models.  

One nice feature of the leading-log model (as well as the leading-log-log 
extension) is that its small distance 
and large distance limiting cases can be approximated analytically.   
In the small distance limit, I devised an analytic perturbation method 
(Adler, 1982b) which shows that the potential has the standard  
form of a Coulomb potential with a logarithmic 
correction that is expected from  
perturbative QCD, permitting the parameter $\kappa$ of the model to be 
related to the QCD scale mass.  With this identification, the model has no 
adjustable parameters.  In the large distance limit, an ingenious 
analysis by Lehmann and Wu (1984) showed that the confinement domain is 
an ellipsoid of revolution, with maximum diameter growing as $R^{1/2}$ with 
the interquark separation, and gave an analytic expression for the free 
boundary shape for large $R$ as well as the 
subdominant term in the potential.  Thus, the model yields a ``fat'' bag, 
rather than a cylindrical confinement domain of uniform radius; however, 
Lehmann told me at the 
time that he believed the true QCD behavior would show a constant-radius 
cylindrical domain, 
and he appears now (see below) to be right.  As discussed in the articles 
I wrote with Piran, the analytic forms for both small 
and large $R$ agreed   
very well with our numerical results, giving confidence that the numerical 
analysis had been carried out correctly.  

How well do the nonlinear dielectric models agree with QCD?  There are 
two aspects to this question, whether they give satisfactory static 
potentials, and whether they describe the flux confinement domain that 
is realized in QCD.  To assess the static potentials tabulated in {\sf R50},  
one has to do a detailed fit to heavy quark spectroscopic data.  This  
was done in papers of Margolis, Mendel, and Trottier (1986) and 
of Crater and Van Alstine (1988), both of which concluded that the log-log 
model potential is in good agreement with experimental data on heavy 
quark systems, with reasonable values of the quark masses.  The fit of 
Margolis, Mendel, and Trottier used a value of 
$\Lambda_{\bar M\bar S}=0.270\, {\rm GeV}$, while that of Crater and 
Van Alstine used a value of $\Lambda_{\bar M\bar S}=0.215 {\rm GeV}$  (note 
that their $\Lambda$ is the  $\kappa^{1/2}$ of {\sf R50}, which is related 
to  $\Lambda_{\bar M\bar S}$ by $\Lambda_{\bar M\bar S}= 0.959 \kappa^{1/2}$).
These values of $\Lambda_{\bar M\bar S}$ are in reasonable accord with  
the value 
$\Lambda_{\bar M\bar S}=0.218\, {\rm GeV}$ 
that Piran and I had quoted in {\sf R50}, obtained by requiring the 
best fit of our potential to Martin's phenomenological potential 
for heavy quark systems. These values of $\Lambda_{\bar M\bar S}$  
should be compared with the three light quark experimental value 
$\Lambda_{\bar M \bar S}^{(3)}\simeq  0.369\, {\rm GeV}$  (Hinchliffe, 2005). 
For a simple extrapolation from the asymptotically free regime to the 
confining regime of QCD, the nonlinear dielectric model does  
reasonably well in accounting for heavy quark spectroscopy.    

As already noted, the confinement domain in the nonlinear dielectric models 
is an ellipsoid of revolution, with width increasing with the quark 
separation $R$. Let $\rho$ be the cylindrical radial coordinate, and  
$z$ the coordinate along the axis of the cylinder.  
On the medial plane $z=0$, various quantities of interest 
can be computed in the large-$R$ limit directly from the Lehmann--Wu 
asymptotic solution.  The radius of the confinement domain on the medial 
plane is 
\begin{equation}
\rho_m=R^\frac{1}{2} \left(\frac{2Q}{\pi b_0 \kappa}\right)^\frac{1}{4}~~~,
\nonumber\end{equation}
and the value of $|\vec D|$ on the medial plane is 
\begin{equation}
|\vec D|= \frac{1}{R} \left(\frac{2 Q b_0 \kappa}{\pi}\right)^\frac{1}{2}
\big(1-\rho^2/\rho_m^2\big)
~~~,
\nonumber\end{equation}
from which one can check that the flux integral gives 
$2\pi \int_0^{\rho_m}\rho d\rho |\vec D| =Q$.  
The profile of $|\vec D|$ is evidently parabolic, and scales with $\rho_m
\propto R^\frac{1}{2}$. 

To compare this ``fat bag'' confinement domain  
with QCD, one must rely on lattice simulations,   
since in real-world QCD, the confining flux tube breaks up 
through quark-antiquark pair formation before 
the asymptotic regime is reached.  Assuming 
that the lattices used are large enough to accurately approximate the 
continuum theory, the data from simulations that have been carried out  
show a confinement domain of constant diameter in the limit of 
large $R$, as discussed and referenced in the book of Ripka (2004). 
The details of the simulated confinement domain 
favor the ``dual superconductor'' model, in which QCD is regarded  
as a dual of a Ginzburg--Landau superconductor, with magnetic monopole pairs 
replacing the Cooper pairs of superconductivity.  In this picture, in 
addition 
to the color fields, there is a dynamical variable corresponding to the 
monopole condensate.   A numerical analysis of flux confinement in a dual 
superconductor, using the methods described in my review with Piran 
{\sf R47}, has been given by  
Ball and Caticha (1988), who give plots of the confinement domain; for  
further details and references, see both Ripka (2004) and the review 
of Baker, Ball, and Zachariasen (1991).  
For appropriate 
values of the three dual superconductor model 
parameters (a magnetic charge $g$, which can be related to an effective 
QCD coupling $e_{\rm eff}$ by the Dirac quantization condition 
$g=2\pi/e_{\rm eff}$,  a scalar magnetic condensate mass $m_H$, 
and a gauge gluon mass $m_V$), good fits to the lattice 
simulations are obtained, and the  
dual superconductor model also gives a phenomenologically satisfactory 
static potential.  (In a recent preprint, Haymaker and Matsuki (2005) argue that in lattice comparisons, the continuum $m_V$ gives rise to two 
parameters that must be fitted, making four parameters in all including 
$g$.)     However,  
since the dual superconductor gives a Coulomb potential at short distances, 
without logarithmic modifications, the dual superconductor parameters cannot 
be directly related to the QCD scale $\Lambda_{\bar M\bar S}$ as was possible 
for the scale parameter $\kappa$ of the nonlinear dielectric model.  As a 
limiting case, the dual superconductor model gives the standard 
bag model with a 
field discontinuity at the boundary; a    
numerical solution of this model is also discussed in Ball and Caticha (1988). 

Although the nonlinear dielectric model and the dual superconductor model 
successfully describe important aspects of confinement in QCD, major 
steps would be needed to incorporate such classical action models into 
a {\it proof} of confinement.  To do so one would have to prove 
that the true energy of a widely separated quark-antiquark pair 
in QCD is bounded from below by the energy calculated in one or the other 
of the two models. This would require achieving precise control over   
the qualitative approximations involved in the models, which include a 
mean-field approximation to the functional integral as discussed in 
Adler (1982a), the replacement of the 
exact QCD effective action by the model effective action, and replacement 
of the octet of color quark charges by Abelianized effective charges 
lying in the maximal commutative subgroup.  Although, as I argued in the 
case of the nonlinear dielectric model in Adler (1982a), these 
simplifications of the full problem  
are plausible, replacing qualitative approximations by precise mathematical 
statements with error estimates will be no small task.  

In any flux tube picture of confinement based on 
Abelianized charges, such as either  
the nonlinear dielectric model or the bag limit of the Ginzburg--Landau 
dual superconductor model, the string tension scales as the Abelianized 
quark charge, or as the square root of the corresponding Casimir.  In a 
paper with Neuberger (Adler and Neuberger, 1983, {\sf R51}), we 
pointed out that in 
the large-$N_c$ limit of $SU(N_c)$ gauge theory, the string 
tension scales with the Casimir when changing from fundamental to adjoint 
representation quarks; hence 
to the extent that 
flux tube models give a good description of confinement in $N_c=3$ 
QCD, different confinement mechanisms appear to be at work 
in QCD and in its large $N_c$ limit.  

\bigskip
\section*{Chiral Symmetry Breaking}

Not long after I had finished the review paper {\sf R47}  with Piran  
summarizing our work on confinement models, 
Anne Davis suggested looking at another outstanding problem in QCD, that 
of chiral symmetry breaking.  After studying the relevant literature 
\big(reviewed in the Introduction to our paper 
Adler and Davis (1984), {\sf R52}\big), 
we decided to focus on setting up and solving a superconductor-like  
gap equation for fermion pairing in Coulomb gauge, systematically 
imposing the axial-vector current Ward identities to get the correct 
renormalization procedure.  This method 
permitted us to study pairing using a Lorentz vector 
instantaneous confining potential  
with $V \propto r$, getting infrared-finite results for physical quantities   
without imposing {\it ad hoc} infrared cutoffs.  The model gives spontaneous 
breaking of chiral symmetry, but with values of the quark condensate 
$\langle \bar u u \rangle$ and 
the pion decay constant 
$f_{\pi}$ that are considerably too low when the phenomenological  
confining potential (or string tension) is used as input. 
Similar results were also found 
by a group at Orsay, and we learned later that the utility of the 
axial-vector  Ward identities in deriving the gap equation had 
also been noted 
by Delbourgo and Scadron (see the reprinted papers {\sf R52} and 
{\sf R53} for references).  Extensions of the 
model of {\sf R52}  to 
the finite temperature case were later discussed by   
Davis and Matheson (1984), Alkhofer 
and Ammundsen (1987), and Klevansky and Lemmer (1987).

In a subsequent paper (Adler, 1986, {\sf R53})    
that I wrote for the Nambu Festschrift, I reviewed the work of various   
groups on gap equation models, and also noted a problem. In order for there  
to be no explicit breaking of chiral symmetry in the gap equation model, the 
instantaneous potential must be the time component of a Lorentz vector, so 
that it contains factors $\gamma_0$ that anticommute with $\gamma_5$.  
However, 
experimental data on heavy quark spectroscopy show that the confining part 
of the potential is predominantly Lorentz scalar, and using a Lorentz scalar 
potential in the gap equation model would lead to explicit violation of 
chiral symmetry, and therefore invalidate the model.  This suggests that   
the approximations leading to the gap equation model are not valid for 
the confining part of the potential.  In {\sf R53}, I also gave equations 
that I had worked out for a retarded extension of the instantaneous 
potential model. The original intention had been for a graduate student 
in either Princeton or Cambridge to work on solving the extended model, 
but in view of the Lorentz structure problem this was not done 
\big(a covariant   
treatment of the gap equation model was later given by von Smekal, Amundsen, 
and Alkofer (1991)\big).  For various 
proposals for addressing the Lorentz structure issue, 
see  
Laga\"e (1992),  Szczepaniak and Swanson (1997), and Bicudo and Marques 
(2004).

Shortly after the paper {\sf R52} was out, Cumrun Vafa, then a Princeton 
graduate student, had a few conversations with me about his attempts to turn 
the Banks--Casher (1980) eigenvalue density criterion for chiral symmetry 
breaking into a proof that chiral symmetry breaking occurs in QCD. \big(For recent progress in applying the Banks--Casher criterion in the 
large-$N_c$ limit, see Narayanan and Neuberger (2004).\big)
I didn't have much in the way of concrete suggestions to offer, and 
Cumrun started also talking 
to Edward Witten, who very sagely suggested looking at a different problem, 
that of studying whether parity conservation can be spontaneously broken 
in QCD.  This problem proved tractable, and their papers (Vafa and 
Witten, 1984a,b),  proving that parity 
is not spontaneously broken in vector-like gauge theories (and similarly 
for the isospin and baryon number symmetries), became  
part of Vafa's thesis.  The difference between the parity problem and 
the chiral symmetry problem can be understood by considering their 
respective order parameters.  If parity is spontaneously broken, the 
pseudoscalar order parameter $\overline u i\gamma_5 u$ will receive a 
vacuum expectation.  When the fermions are integrated out, one obtains 
a Lorentz invariant, parity-nonconserving operator functional $X$ of the  
gluon fields that 
is real in Minkowski space, but  picks up a factor of $i$ when rotated 
to Euclidean space. This,  together with positivity of the Euclidean 
space Dirac fermionic determinant in a vector-like theory, is the basis 
of the Vafa-Witten proof that adding a small multiple of $X$ to the action 
cannot make the ground state energy lower.  In the chiral symmetry problem, 
the relevant order parameter is the parity conserving but chiral symmetry 
breaking scalar operator $\overline u u$, which when the fermions 
are integrated out 
leads to a functional $X^{\prime}$ of the gluon fields that 
remains real when rotated to Euclidean space.  Hence the Vafa-Witten argument 
suggests that the energy minimum may lie at a nonzero value of $X^{\prime}$, 
but such a local analysis cannot find the global minimum, and hence does     
not give a proof of chiral symmetry breaking. Rigorous lattice inequalities  
given by Weingarten (1983) give a proof of chiral symmetry breaking only 
when additional strong assumptions are made, including the existence of 
the continuum limit and the confinement of color, together with use of  
anomaly matching conditions. 

Over twenty years later, the problem of proving the breakdown of 
chiral symmetry in QCD is still open, as is that of proving confinement.   
In fact, there is considerable evidence 
that chiral symmetry breaking and confinement in QCD are related 
phenomena.  For example, lattice simulations such as D'Elia et al. (2004) 
show that the deconfining and chiral transitions coincide; gap equation  
models of the type studied in {\sf R52} find chiral symmetry breaking for a 
confining potential but not for a Coulomb potential, and lattice inequalities 
of the type studied by Weingarten also need confinement as an ingredient to 
show chiral symmetry breaking.  Thus it appears that both of these 
outstanding problems in QCD are aspects of the larger problem of proving 
that QCD exists and 
has a mass gap, which is one of the seven Clay Foundation Millennium 
Problems in mathematics and mathematical physics.    
Perhaps in this century, with the added incentive of a \$1 million  
reward, rigorous proofs of confinement and chiral symmetry breaking in 
QCD will be found!  

\section*{References for Chapter 7}

\begin{itemize}

\item Adler, S. L. (1978a).  Classical Algebraic Chromodynamics.  
{\it Phys. Rev. D} {\bf 17}, 3212-3224.

\item Adler, S. L. (1978b) {\sf R45}.  Theory of Static Quark Forces. 
{\it Phys. Rev. D} {\bf 18},  411-434.  
Pages 424-429, containing Appendix A, are reprinted here. 

\item Adler, S. L. (1979a) {\sf R46}. Classical Quark Statics.  
{\it Phys. Rev. D} {\bf 19}, 
1168-1187.   Pages 1182-1183, containing Appendix A, are reprinted here. 

\item Adler, S. L.  (1979b). 
Small Deformations of the Prasad-Sommerfield Solution. 
{\it Phys. Rev. D} {\bf 19}, 2997-3007. 

\item Adler, S. L.  (1979c). Global Structure of 
Static Euclidean $SU(2)$ Solutions. 
{\it Phys. Rev. D} {\bf 20}, 1386-1411. 

\item Adler, S. L. (1981) {\sf R48}.  Effective-Action Approach to Mean-Field 
Non-Abelian Statics, and a Model for Bag Formation.  {\it Phys. Rev. D} 
{\bf 23}, 2905-2915. 

\item Adler, S. L. (1982a).  Generalized Bag Models as Mean-Field Approximations 
to QCD.  {\it Phys. Lett. B} {\bf 110}, 302-306. 

\item Adler, S. L. (1982b). Short-Distance Perturbation Theory for the 
Leading Logarithm Models.   {\it Nucl Phys. B} {\bf 217}, 381-394. 

\item Adler, S. L. (1986) {\sf R53}.  Gap Equation Models for Chiral Symmetry 
Breaking.  {\it Progr. Theor. Phys. Suppl.} {\bf 86}, 12-17. 
  
\item Adler, S. L. and A. C. Davis (1984) {\sf R52}.  Chiral Symmetry Breaking in 
Coulomb Gauge QCD.  {\it Nucl. Phys. B} {\bf 244}, 469-491. 

\item Adler, S. L. and H. Neuberger (1983) {\sf R51}.  Quasi-Abelian versus 
Large-$N_c$ Linear Confinement. {\it Phys. Rev. D} {\bf 27}, 1960-1961. 

\item Adler, S. L. and T. Piran (1982a) {\sf R49}.  Flux Confinement in the Leading 
Logarithm Model.  {\it Phys. Lett. B} {\bf 113}, 405-410. 

\item Adler, S. L. and T. Piran (1982b) {\sf R50}. The Heavy Quark Static Potential 
in the Leading Log and the Leading Log Log Models.  {\it Phys. Lett. B} 
{\bf 117}, 91-95. 

\item Adler, S. L. and T. Piran (1984) {\sf R47}. Relaxation Methods 
for Gauge Field 
Equilibrium Equations.  {\it Rev. Mod. Phys.} {\bf 56}, 1-40. Pages 1-12, 
18-21, and 30-38 are reprinted here.  

\item Akhoury, R., J.-H. Jun, and A. S. Goldhaber (1980).  
Linear Deformations of the 
Prasad-Sommerfield Monopole.  {\it Phys. Rev. D} {\bf 21}, 454-465. 

\item Alkhofer, R. and P. A. Amundsen (1987).  A Model for the Chiral Phase 
Transition in QCD. {\it Phys. Lett. B} {\bf 187}, 395-400. 

\item Baker, M., J. S. Ball, and F. Zachariasen (1991).  Dual QCD: A Review.  
{\it Physics Reports} {\bf 209}, 73-127. 

\item Ball, J. S. and A. Caticha (1988).  Superconductivity: A Testing Ground 
for Models of Confinement. {\it Phys. Rev. D} {\bf 37}, 524-535. 

\item Banks, T. and A. Casher (1980).  Chiral Symmetry Breaking in Confining 
Theories.  {\it Nucl. Phys. B} {\bf 169}, 103-125. 

\item Bicudo, P. and G. M. Marques (2004).  Chiral Symmetry Breaking and Scalar 
String Confinement.  {\it Phys. Rev. D} {\bf 70}, 094047. 

\item Bogomol'nyi, E. B. (1976).  The Stability of Classical 
Solutions. {\it Yad. Fiz.} {\bf 24}, 861-870 (English translation: 
{\it Sov. J. Nucl. Phys.} {\bf 24}, 449-454). 

\item Brown, L. S., R. D. Carlitz, D. B. Creamer, and C. Lee (1978). Propagation 
Functions in Pseudoparticle Fields.  {\it Phys. Rev. D} {\bf 17},  
1583-1597. 

\item Callias, C. (1978).  Axial Anomalies and Index Theorems on Open Spaces.  
{\it Commun. Math. Phys.} {\bf 62}, 213-234. 

\item Crater, H. W. and P. Van Alstine (1988).  Two-Body Dirac Equations for 
Meson Spectroscopy.  {\it Phys. Rev. D} {\bf 37}, 1982-2000. 

\item Davis, A. C. and A. M. Matheson (1984).  Chiral Symmetry Breaking at 
Finite Temperature in Coulomb Gauge QCD.  {\it Nucl. Phys. B} {\bf 246}, 
203-220. 

\item D'Elia, M., A. Di Giacomo, B. Lucini, G. Paffuti, and C. Pica (2004). 
Chiral Transition and Deconfinement in $N_f=2$ QCD; arXiv:hep-lat/0408009.

\item Forgacs, P., Z. Horvath, and L. Palla (1981).  Exact Multi-Monopole Solutions 
in the Bogomolny-Prasad-Sommerfield Limit.  {\it Phys. Lett. B} {\bf 99}, 
232-236; erratum, {\it Phys. Lett. B} {\bf 101}, 457 (1981). 

\item  Haymaker, R. W. and T. Matsuki (2005). Consistent 
Definitions of Flux and the Dual Superconductivity Parameters in $SU(2)$ 
Lattice Gauge Theory; arXiv:hep-lat/0505019. 

\item Hinchliffe, I. (2005).  Private communication. For the underlying formulas 
relating the $\Lambda_{\bar M\bar S}^{(n)}$ for different values of $n$, see 
Hinchliffe's QCD review in the 1998 Review of Particle Physics, 
{\it Eur. Phys. J. C} {\bf 3}, 1-794, pp. 81-89. 

\item Houston, P.  and L. O'Raifeartaigh (1980). On the Zeros of the 
Higgs Field for 
Axially Symmetric Multi-Monopole Configurations.  {\it Phys. Lett. B} 
{\bf 93}, 151-154. 

\item Jaffe, A. and 
C. Taubes (1980). {\it Vortices and Monopoles} (Birkh\"auser, Boston). 

\item Klevansky, S. P. and R. H. Lemmer (1987).  Chiral-Symmetry Breaking at 
Finite Temperatures.  {\it Phys. Rev. D} {\bf 38}, 3559-3565. 

\item Laga\"e, J.-F. (1992). Spectroscopy of Light-Quark Mesons and the Nature 
of the Long-Range $q-\bar q$ Interaction.  {\it Phys. Rev. D} {\bf 45}, 
317-327. 

\item Lehmann, H. and T. T. Wu (1984).  Classical Models of Confinement. 
{\it Nucl. Phys. B} {\bf 237}, 205-225. 

\item Margolis, B., R. R. Mendel, and H. D. Trottier (1986).  $Q-\bar q$ Mesons 
in the Leading Log and Leading Log-Log Models. {\it Phys. Rev. D} {\bf 33}, 
2666-2673. 

\item Matinyan, S. G. and G. K. Savvidy (1978).  Vacuum Polarization Induced by   
the Intense Gauge Field. {\it Nucl. Phys. B} 
{\bf 134}, 539-545. 

\item Mottola, E. (1978). Zero Modes of the 't Hooft--Polyakov Monopole.  
{\it Phys. Lett. B} {\bf 79}, 242-244. 

\item Mottola, E. (1979). Normalizable Solutions 
to the Dirac Equation in the Presence 
of a Magnetic Monopole.  {\it Phys. Rev. D} {\bf 19}, 3170-3172. 

\item Nahm, W. (1980).  A Simple Formalism for the BPS Monopole. 
{\it Phys. Lett. B} {\bf 90}, 413-414. 

\item Narayanan, R. and H. Neuberger (2004). Chiral Symmetry 
Breaking at Large $N_c$. {\it Nucl. Phys. B} {\bf 696}, 107-140. 

\item Pagels, H. and E. Tomboulis (1978). Vacuum of the Quantum Yang-Mills   
Theory and Magnetostatics. {\it Nucl. Phys. B} 
{\bf 143}, 485-502. 

\item Polyakov, A. M. (1974).
Particle Spectrum in Quantum Field Theory.  {\it ZhETF Pis. Red.}
 {\bf 20},    430-433 (English translation:  {\it JETP Lett.} {\bf 20},     194-195). 

\item Prasad, M. K. and C. M. Sommerfield (1975). Exact Classical Solution for the 
't Hooft Monopole and the Julia-Zee Dyon.  {\it Phys. Rev. Lett.}  
{\bf 35}, 760-762. 

\item Ripka, G. (2004). {\it Dual Superconductor Models of Color Confinement.}  
Lecture Notes in Physics 639 (Springer, Berlin). 

\item Savvidy, G. K. (1977).  Infrared Instability of the Vacuum State of  
Gauge Theories and Asymptotic Freedom. {\it Phys. Lett. B} {\bf 71}, 133-134. 

\item Szczepaniak, A. P. and E. S. Swanson (1997).  On the Dirac Structure 
of Confinement.  {\it Phys. Rev. D} {\bf 55}, 3987-3993. 

\item Taubes, C. H. (1980). Arbitrary N-Vortex Solutions to the First Order 
Ginzburg--Landau Equations. {\it Commun. Math. Phys.} {\bf 72}, 
277-292. 

\item 't Hooft, G. (1974a). Magnetic Monopoles in Unified Gauge Theories.  
{\it Nucl. Phys. B} {\bf 79}, 276-284 (1974). 

\item 't Hooft, G. (1974b).  Quarks and Gauge Fields, in {\it Recent Progress in Lagrangian 
Field Theory and Applications}, Proceedings of the Marseille 
Colloquium, 1974, C. P. Korthals-Altes, E. de Rafael, and R. Stora, 
eds. (Centre de Physique Th\'eorique -CNRS, Universit\'es d'Aix-Marseille 
I et II, 1975), pp. 58-67. 

\item Vafa, C. and E. Witten (1984a).  Parity Conservation in 
Quantum 
Chromodynamics.  {\it Phys. Rev. Lett.} {\bf 53}, 535-536. 

\item Vafa, C. and E. Witten (1984b).  Restrictions on Symmetry Breaking in 
Vector-Like Gauge Theories.  {\it Nucl. Phys. B} {\bf 234}, 173-188. 

\item von Smekal, L., P. A. Amundsen, and R. Alkofer (1991).  A Covariant Model 
for Dynamical Chiral Symmetry Breaking in QCD. {\it Nucl. Phys. A}   
{\bf 529}, 633-652. 

\item Ward, R. S. (1981).  A Yang-Mills-Higgs Monopole of Charge 2.  {\it Commun. 
Math. Phys.} {\bf 79}, 317-325. 

\item Weinberg, E. J. (1979a). Multivortex Solutions of the Ginzburg--Landau 
Equations.  {\it Phys. Rev. D} {\bf 19}, 3008-3012. 

\item Weinberg, E. J. (1979b).  Parameter Counting for Multimonopole Solutions.  
{\it Phys. Rev. D} {\bf 20}, 936-944. 

\item Weingarten, D. (1983).  Mass Inequalities for Quantum Chromodynamics. 
{\it Phys. Rev. Lett.} {\bf 51}, 1830-1833. 

\end{itemize}

\chapter*{8. Overrelaxation for Monte Carlo and Other Algorithms}

\markboth{Adventures in Theoretical Physics}{8. Overrelaxation for Monte Carlo and Other Algorithms}
\bigskip
\section*{Introduction} 
 
As I have already noted, the investigations described in the previous 
chapter all involved extensive computer work. This got me interested in  
the issue of algorithms more generally, and led to two distinct research 
directions in the years that followed.  One involved  generalizing the  
acceleration methods for solving partial differential equations to the 
related problem of Monte Carlo simulations, as discussed in the first 
section that follows.  The second involved neural networks and pattern 
recognition, and led among other things to  work on image normalization 
methods, described briefly in the second section of this chapter.  
\bigskip
\section*{Overrelaxation to Accelerate Monte Carlo}

In preparation for numerically solving the partial differential equations 
for the leading-log models, I did general reading on numerical 
methods for handling partial differential equations. This taught me about 
the critical slowing down problem -- the fact that as one refines meshes 
to get more accurate numerical solutions, the rate of convergence of the 
iterations slows down.  I also learned about various strategies devised 
for defeating critical slowing down, and in particular about the successive 
over-relaxation (SOR) modification of the standard Gauss--Seidel iteration.  
In a Gauss--Seidel iteration of a positive functional, one replaces 
each successive variable by the value that locally minimizes the functional. 
In SOR, one builds in a systematic overshoot beyond the 
local minimum, with the amount of overshoot tuned to the degree of mesh  
refinement, yielding more rapid convergence as a result. In the work  
of {\sf R47}, Piran and I used SOR in all of our iterative solutions, and 
achieved substantial gains in convergence speed on our finest meshes.    

I became interested in Monte Carlo algorithms because it was clear that 
lattice gauge theory  simulations probably would be 
the only way that one could   
study details of the structure of the flux confinement domain in QCD.  
I knew from talks that I had heard in Princeton that there were two main 
Monte Carlo methods in use, the Metropolis method and the heat bath method, 
and also that the folk wisdom  at the time was 
that heat bath was the best one could do, since it corresponded to 
``nature's way'' of achieving thermal equilibrium.  However, since the 
zero temperature limit of heat bath just corresponds to a Gauss--Seidel 
iteration, which I knew could be accelerated by SOR, I suspected that the 
conventional wisdom was wrong, and that there should be extensions, to 
Monte Carlo thermalizations, of the standard acceleration methods for 
the solution of differential equations.  Since the monopole numerical   
work had brought out the fact that the Yang--Mills action is multiquadratic, 
I decided to study this question in the simple context of multiquadratic 
actions, where the question becomes whether for quadratic actions, the
SOR method for differential equations has an extension to Monte Carlo 
thermalization.  This is the question addressed in Adler (1981), {\sf R54}, 
where I showed that SOR does indeed extend to the thermalization of 
multi-quadratic actions, by explicitly constructing in Eqs. (14a,b) 
the transition probability 
that obeys detailed balance when an overrelaxation parameter is included 
in the iteration. (Note that the normalization factors in these equations   
have the $\pi$ in the correct place, but the other factors inverted; this 
is corrected in Eqs. (9) and (11) of my later paper {\sf R55}.  The 
argument of {\sf R54} does not involve the normalization factors, and is 
unaffected.)   For this overrelaxed thermalization, I showed that the 
means of the thermalized variables iterate according to standard SOR;  
since standard SOR accelerates Gauss--Seidel, this implies that there should 
be a corresponding acceleration of the thermalization process as well.  

The 1981 paper {\sf R54} gave the earliest indication that Monte Carlo methods 
could be accelerated to improve critical slowing down, and for this 
reason was conceptually important, as well as having later applications 
and extensions. To the best of my knowledge, I am supported by the literature on the subject, in stating that  {\sf R54}  first  introduced  
acceleration methods into Monte Carlo. Two compilations of Monte Carlo 
articles edited by Binder contain literature surveys, 
Binder et al. (1987), and Swendsen, Wang, and Ferrenberg  
(1992), relating to the critical slowing down problem.  
In both surveys, the earliest listed reference is from 1983; neither  
survey cites my 1981 paper (or Whitmer's 1984 paper -- see  
below), although  
some of the cited articles  do reference these papers.    

I didn't immediately continue work on Monte Carlo acceleration myself, but 
suggested it as a thesis research area to my Princeton University graduate 
student Charles Whitmer.  He applied the method to $\phi^4$ and Higgs 
actions that are point split on a lattice with unit displacement 
$\hat \mu$ according to $\phi^4(x) \to \phi^2(x) 
\phi^2(x+a\hat \mu)$, which makes them multiquadratic, and in the paper 
Whitmer (1984) reported improvement over conventional heat bath Monte Carlo.  
I got interested in the subject again a few years later, after Goodman 
and Sokal (1986) (who knew about the SOR method of {\sf R54} and Whitmer's 
paper) 
proposed a stochastic extension 
of multigrid methods, and Creutz (1987) and Brown and Woch (1987) gave 
a simple implementation of the SOR idea for lattice gauge theory plaquette 
actions.  This latter development eliminated the need for the problematic 
gauge fixing that I had used in {\sf R54} to keep the latticized gauge    
action multiquadratic, and opened the way to practical applications of SOR 
to gauge theory Monte Carlo studies.  In the spring of 1987, I went 
to Torino, Italy with 
my daughter Victoria, who had been eager to visit Europe after finishing 
her high school requirements.  During this sabbatical term I was a visitor  
at the Institute for Scientific 
Interchange (ISI), at the invitation of Mario Rasetti 
and my former IAS colleague Tullio Regge; I also had an office at the 
University of Torino that I used a couple of days a week.  Although 
I had been spending 
considerable time over the previous few years working on quaternionic 
quantum mechanics (see the next chapter), I decided on this trip to return 
to my old interest in Monte Carlo SOR, stimulated by the fact that experts 
in the Monte Carlo field had started to get interested.  In a paper that 
I wrote while at ISI (Adler, 1988a, {\sf R55}) I gave a much more detailed 
analysis of overrelaxed thermalization for a quadratic action, and also 
gave extensions of the method to non-quadratic actions, including $SU(n)$ 
gauge theory.  

After my return to the IAS from this sabbatical, I continued work on Monte-
Carlo algorithms for several more years. In a paper written in the  
fall of 1987 after I returned from Italy (Adler, 1988b, {\sf R56}),  
I gave an elegant formal analysis showing that the general linear 
iteration $u^{\prime} = Mu +Nf$  corresponding to a splitting 
$1=M+NL$ of the quadratic form $L$ for a Gaussian action, has a corresponding 
stochastic generalization 
\begin{equation}
P(u\to u^{\prime})=(\beta/\pi)^{1/2}  (\det \Gamma)^{1/2} 
\exp[-(u^{\prime}-Mu-Nf)^T \beta \, \Gamma (u^{\prime}-Mu-Nf)]~~~,
\nonumber
\end{equation}
with $\Gamma=\frac{1}{2}(L^{-1}-M L^{-1} M^T)^{-1}$ a modified temperature 
matrix.  This extends the SOR thermalization of {\sf R54} to a general 
linear iterative process.  Later in the same academic year,   
I gave in Adler (1988c) a 
Metropolis variant of the $SU(n)$ method given in {\sf R55}, that 
extended the method for the Wilson action  
used by Creutz and by Brown and Woch to general overrelaxation parameter 
$\omega$.  In collaboration with Gyan Bhanot, a former 
IAS member and a Monte Carlo expert, we made a numerical study of  
the $SU(2)$ version of this algorithm, with results reported 
in Adler and Bhanot (1989), {\sf R57}.   (Growing out of this collaboration,  
Bhanot spent several years as a half-time member of the IAS in the early 
1990's, in the course of which we wrote a number of 
further papers on a variety of Monte Carlo acceleration methods.)   
I also gave talks at lattice conferences; at the biennial Lattice Gauge  
conference Lattice 88, held at Fermilab that year, I gave a plenary 
talk reviewing work on algorithms for pure gauge theory, focusing primarily 
on the theory and application of overrelaxation methods (Adler, 1989, 
{\sf R58}).  Monte Carlo overrelaxation has become a standard part of the 
lattice gauge theorist's tool kit; for a sampling of recent applications,  
see Kiskis, Narayanan, and Neuberger (2003), Holland, Pepe, and Wiese (2004), 
Meyer (2004), Pepe (2004), Shcheredin (2005), and 
de Forcrand and Jahn (2005).

\bigskip
\section*{Image Normalization}

During the 1990s, I interspersed my work on quaternionic 
quantum mechanics and particle physics with  work on aspects of 
neural networks and pattern recognition. My neural network interests 
involved an analog device that I patented (Adler, 1993) 
and an article (Adler, Bhanot, 
and Weckel, 1996) analyzing its algorithmic aspects.  In pattern  
recognition, from lunchtime conversations  
with Joseph Atick and Norman Redlich, I got interested
in the problem of extracting those features of an image that are invariant 
under a symmetry transformation.  This problem is closely analogous to that 
of extracting those features of a gauge potential that are gauge-invariant, 
and in Adler (1998), {\sf R59} I gave a general formal solution, based on 
imposing image normalizing constraints analogous to gauge-fixing 
constraints. I 
have reprinted here only the first two sections of this unpublished article (without 
references), in which the general theory is set up; further sections of 
the article give applications to a variety of viewing transformations of a 
planar object.  Shortly afterwards,    
when one of the IAS string theory postdocs
was interested in switching to a computer-related career, I suggested 
 applying my methods to the problem of the similarity  and affine 
normalization of partially occluded planar curves (such as the boundary 
of a planar object).  We worked this out together and it was published 
as Adler and Krishnan (1998), {\sf R60}.  The excerpt {\sf R59} of the 
general paper that is reprinted here gives the background needed to  
follow the extension of the planar algorithm to curve segments given 
in {\sf R60}.

\section*{References for Chapter 8}

\begin{itemize}

\item Adler, S. L. (1981) {\sf R54}.  Over-Relaxation Method for the Monte Carlo 
Evaluation of the Partition Function for Multiquadratic Actions. 
{\it Phys. Rev. D} {\bf 23}, 2901-2904. 

\item Adler, S. L. (1988a) {\sf R55}.  Overrelaxation Algorithms 
for Lattice 
Field Theories.  {\it Phys Rev. D} {\bf 37}, 458-471. 

\item Adler, S. L. (1988b) {\sf R56}.  Stochastic Algorithm Corresponding to a 
General Linear Iterative Process.  {\it Phys. Rev. Lett.} {\bf 60}, 
1243-1245. 

\item Adler, S. L. (1988c).  Metropolis Overrelaxation for Lattice Gauge Theory 
for a General Relaxation Parameter $\omega$. {\it Phys. Rev. D} {\bf 38},     
1349-1351. 

\item Adler, S. L. (1989) {\sf R58}.  Algorithms for Pure Gauge 
Theory.  
{\it Nucl. Phys. B Proc. Suppl.} {\bf 9}, 437-446. 

\item Adler, S. L. (1993). Neural Network Architecture Based on Summation of  
Phase-Coherent Alternating Current Signals.  U.S. Patent  No. 5,261,035. 

\item Adler, S. L. (1998) {\sf R59}.  General Theory of Image Normalization,  
unpublished; arXiv: cs.CV/9810017.  A slightly abridged and reformatted 
version of the first two sections is reprinted here. 

\item Adler, S. L. and G. V. Bhanot (1989) {\sf R57}.  Study of an Overrelaxation Method 
for Gauge Theories.  {\it Phys. Rev. Lett.} {\bf 62}, 121-124. 

\item Adler, S. L., G. Bhanot, and J. D. Weckel (1996).  Algorithmic Aspects of a 
Neuron for Coherent Wave Synapse Realizations.  {\it IEEE Trans. Neural 
Networks} {\bf 7}, 1262-1271. 
  
\item Adler, S. L. and R. Krishnan (1998) {\sf R60}.  Similarity and Affine Normalization  
of Partially Occluded Planar Curves Using First and Second Derivatives.  
{\it Pattern Recognition} {\bf 31},  
1551-1556. 

\item Binder, K. et al. (1987). Recent Developments, in  
{\it Applications of the Monte Carlo Method 
in Statistical Physics}, K. Binder, ed.  (Springer-Verlag, Berlin),    
Chapter 10, p. 302, refs. [10.15-22]. 

\item Brown, F. R. and T. J. Woch (1987).  Overrelaxed Heat-Bath and Metropolis 
Algorithms for Accelerating Pure Gauge Monte Carlo Calculations.  
{\it Phys. Rev. Lett.} {\bf 58}, 2394-2396. 

\item Creutz, M. (1987).  Overrelaxation and Monte Carlo Simulation.  {\it Phys. 
Rev. D} {\bf 36}, 515-519. 

\item de Forcrand, P. and O. Jahn (2005).  Monte Carlo Overrelaxation for $SU(N)$ 
Gauge Theories; arXiv: hep-lat/0503041. 

\item Goodman, J. and A. D. Sokal (1986).  Multigrid Monte Carlo Method for 
Lattice Field Theories.  {\it Phys. Rev. Lett.} {\bf 56}, 1015-1018. 

\item Holland, K., M. Pepe, and U. J. Wiese (2004).  The Deconfinement Phase 
Transition of $SP(2)$ and $SP(3)$ Yang-Mills Theories in (2+1)-Dimensions 
and (3+1)-Dimensions. {\it Nucl. Phys. B} {\bf 694}, 35-58. 

\item Kiskis, J., R. Narayanan, and H. Neuberger (2003).  Does the Crossover 
from Perturbative to Nonperturbative Physics in QCD Become a Phase 
Transition at Infinite $N$? {\it Phys. Lett. B} {\bf 574}, 65-74. 

\item Meyer, H. B. (2004).  The Yang-Mills Spectrum from a Two-Level Algorithm. 
{\it JHEP} {\bf 0401}, 030. 

\item Pepe, M. (2004).  Deconfinement in Yang-Mills: A Conjecture for a General 
Gauge Lie Group G.  {\it Nucl. Phys. B. Proc. Suppl.} {\bf 141}, 238-243 
(2005). 

\item Shcheredin, S. (2005).  Simulations of Lattice Fermions with Chiral 
Symmetry in Quantum Chromodynamics; arXiv: hep-lat/0502001. 

\item Swendsen, R. H., J.-S. Wang, and A. M. Ferrenberg (1992).  
New Monte Carlo 
Methods for Improved Efficiency of Computer Simulations in Statistical 
Physics, in 
{\it The Monte Carlo Method in Condensed Matter 
Physics}, K. Binder, ed. (Springer-Verlag, Berlin), Sec. 4.1, pp. 75-76. 

\item Whitmer, C. (1984).  Over-Relaxation Methods for Monte Carlo Simulations 
of Quadratic and Multiquadratic Actions.  {\it Phys. Rev. D} {\bf 29}, 
306-311. 

\end{itemize}

\chapter*{9. Quaternionic Quantum Mechanics, Trace Dynamics, 
and Emergent Quantum Theory}

\markboth{Adventures in Theoretical Physics}{Quantum Mechanics, Trace Dynamics, 
and Emergent Quantum Theory}
\bigskip
\section*{Introduction} 
During the twenty years from 1984 to 2004, a large part of my time 
was spent on investigations into foundational areas of quantum mechanics.  
Most of my research from this period was later presented in 
two books that I wrote, {\it Quaternionic Quantum Mechanics and Quantum 
Fields} (Oxford University Press, New York, 1995) and {\it Quantum Theory 
as an Emergent Phenomenon} (Cambridge University Press, Cambridge, 2004). 
I have not included in this reprint volume any research papers  
incorporated (some considerably improved) into the
two books, since this would  be infeasible because of length   
limitations.  So what I discuss in this chapter 
are a few papers dealing with quaternionic topics written during the 
period between the two books, together with a brief description of how  
I got interested in quaternionic quantum theory, and later on, in the 
possibility of a pre-quantum theory.  
\bigskip
\section*{Quaternionic Quantum Mechanics}

My interest in quaternionic quantum mechanics grew out of my interest 
in the Harari (1979)-Shupe (1979) model for composite quarks and leptons. 
They postulated an order-dependence for the preon wave functions (e.g., 
$TTV$, $TVT$, $VTT$ were considered to be three distinct color states), 
which suggested that quantum theory over a noncommutative field might be 
involved.  I was never able to use quaternions or related ideas to make 
a successful preon model, either during the period before my book  
(see Adler, 1979, 1980) or after (Adler, 1994a), but the issues raised,  
and interactions with key people acknowledged in the Preface of the 1995 
volume, led me to undertake a systematic study of quaternionic quantum 
mechanics. Perhaps the most important new result contained in my 
papers (Adler, 1988) and  in my book is the fact that the $S$-matrix 
in quaternionic scattering theory is {\it complex}, not quaternionic, which 
was a surprise to the experts in the field and invalidated proposed searches 
(such as Peres, 1979) for quaternionic effects manifested through noncommuting 
scattering phases.  I also clarified the relationship between time reversal 
symmetry in quaternionic quantum theory (where it is unitary) and in 
complex quantum theory  (where it is antiunitary), proved that positive 
energy quaternionic Poincar\'e group representations are complex and not 
intrinsically quaternionic, and gave a quaternionic generalization     
of projective group representations (to which I shall return shortly). 
These were but a few of the many topics dealt with in my 1995 book.  
My quaternionic investigations also motivated  work I did in new directions 
in standard quantum mechanics, such as a  paper that I wrote 
showing that $SU(3) \times SU(12)$ is the minimal grand unified theory in 
which, species by species for charged fermions, no Dirac sea is required (Adler, 1989).  

After my book on quaternionic quantum mechanics was completed, a number 
of papers that I wrote with collaborators clarified issues that were left 
unresolved, or were inadequately treated, in the book.  One of these 
issues dealt with the non-adiabatic geometric phase in 
quaternionic Hilbert space.  This was discussed in my book, but on a 
visit to the IAS, Jeeva Anandan pointed out that my treatment was incomplete, 
and sketched what was needed to improve it.  I filled in the details and 
drafted a manuscript, which became a joint paper (Adler and Anandan, 1996, 
{\sf R61}) that was published in the Larry Horwitz Festschrift issue of   
{\it Foundations of Physics}.  A second issue that was left hanging was the 
analog of coherent states in quaternionic quantum theory.  My thesis student 
Andrew Millard and I studied this, and wrote a paper (Adler and Millard, 
1996a, {\sf R62}) giving the extension of the Perelomov coherent state 
formalism to quaternionic Hilbert spaces. We also showed that the closure 
requirement forces an attempted quaternionic generalization of standard 
coherent states based on the Weyl group to reduce back to the complex case,  
settling a question raised in discussions with me by John Klauder.  
The other issues that were dealt with after 
publication of the quaternionic book were the structure of quaternionic 
projective representations, and the relationship between standard complex 
quantum mechanics and the dynamics based on a trace variational principle    
that I had introduced in the field theory chapter of the 1995 book.    
These form the subject of the next two sections.  
\bigskip
\section*{Quaternionic Projective Group Representations }

Given two group elements $b,a$ with product $ba$, a unitary operator 
representation 
$U_b$ in a Hilbert space is defined by $U_bU_a=U_{ba}$.  A more general 
type of representation, called a ray or projective representation, is 
relevant to describing the symmetries of quantum mechanical systems.  In 
his famous paper on unitary ray representations of Lie groups, 
Bargmann (1954)  defines a projective representation as one obeying 
$U_bU_a=U_{ba}\omega(b,a),$  with $\omega(b,a)$ a complex phase.  

This 
definition is familiar, and seems obvious, until one asks the following 
question: Bargmann's definition is assumed to hold as an operator identity when acting  
on {\it all} states in Hilbert space.  However, we know that it suffices  
to specify the action of an operator on {\it one} complete set of states 
in Hilbert space to specify the operator completely.  Hence why does one 
not start instead from the definition 
$U_bU_a|f\rangle=U_{ba}|f\rangle \omega(f;b,a) ,$  
with $\{|f\rangle\}$  one complete set of states, as defining a projective  
representation in Hilbert space?  Let us call Bargmann's definition  
a ``strong'' projective representation, and the definition 
with a state-dependent phase  a ``weak'' projective representation.  
Then  the question 
becomes that of finding the relation between weak and strong projective 
representations.  

Although I have formulated this question here in complex Hilbert space, it 
arose and was solved in the context of quaternionic Hilbert space, where 
the phases $\omega(f;b,a)$ are quaternions, which obey a non-Abelian 
 group multiplication law isomorphic to $SO(3) \simeq SU(2)$.   
The strong definition  was adopted for the quaternionic 
case by Emch (1963, 1965), but in Sec. 4.3 of 
my book on quaternionic quantum mechanics  
I introduced the weak definition in order for 
quaternionic projective representations to include  embeddings of nontrivial 
complex projective representations into quaternionic Hilbert space; the 
state dependence of the phase is necessary because even a complex phase 
$\omega$ does not commute with general quaternionic rephasings of the state 
vector $|f\rangle$.  I noted in my 1995 book that the weak definition can be 
extended to an operator relation by defining 
\begin{equation}
\Omega(b,a) =\sum_f|f\rangle \omega(f;b,a)\langle f| ~~~, \nonumber\\
\end{equation}
so that the weak definition then takes the form 
\begin{equation}
U_bU_a=U_{ba}\Omega(b,a)~~~,\nonumber\\
\end{equation}
which gives the general operator form taken by projective representations 
in quaternionic quantum mechanics.  I also introduced in Sec. 4.3 of my book 
two specializations 
of this definition, motivated by the commutativity properties of the phase 
factor in complex projective representations.
I defined  a {\it multicentral} 
projective representation as one for which
\begin{equation}
[\Omega(b,a),U_a]= [\Omega(b,a),U_b]= 0 ~~~\nonumber\\
\end{equation} 
for all pairs $b,a$ 
(note that in Eq.~(4.51a) of my book, $U_{ab}$ should read $U_{ba}$,  
so that the two conditions just given suffice),  
and I defined a {\it central} projective representation 
as one for which 
\begin{equation}
[\Omega(b,a),U_c]=0~~~\nonumber\\
\end{equation}
 for all triples $a,b,c$.

Subsequent to the completion of my book, I read Weinberg's first volume 
on quantum field theory (Weinberg, 1995)  and realized, from his 
discussion in Sec. 2.7 of the associativity condition for complex 
projective representations, that there must be an analogous associativity 
condition for 
weak quaternionic projective representations.  I worked this out 
(Adler, 1996, {\sf R63}), and 
showed that it takes the operator form 
\begin{equation}
U_a^{-1}\Omega(c,b)U_a=\Omega(cb,a)^{-1}\Omega(c,ba)\Omega(b,a)~~~,\nonumber\\
\end{equation}
which by the definition of $\Omega(b,a)$ shows that $U_a^{-1}\Omega(c,b)U_a$ 
is diagonal in the basis $\{|f\rangle\}$, with the spectral representation  
\begin{equation}
U_a^{-1}\Omega(c,b)U_a=\sum_f |f\rangle \overline{\omega(f; cb,a)}
\omega(f;c,ba)\omega(f;b,a)  \langle f|~~~.\nonumber\\
\end{equation}
On the basis of some further identities, I also raised the question  of 
whether one can construct a multicentral representation that is not central, 
or whether a multicentral representation is always central.  

Subsequently, I discussed the issues of quaternionic projective 
representations with Andrew Millard.  He explained them to his roommate Terry Tao, a mathematics 
graduate student working for Elias Stein, and at my next conference with 
Andrew, Tao came along and presented the outline of a beautiful theorem  
that he had devised.  This was written up as 
a paper of Tao and Millard (1996), 
and consists of two parts.  The first part is an algebraic analysis based 
on the 
spectral representation given above, which leads to the following theorem  

{\bf Structure Theorem:} {\it Let $U$ be an irreducible projective 
representation of a connected Lie group $G$.  There then exists a reraying 
of the basis $|f\rangle$ under which one of the following three possibilities 
must hold.} 
\begin{enumerate}
\item $U$ is a real projective representation.  That is, $\omega(f;b,a)
=\omega(b,a)$ is independent of $|f\rangle$ and is equal to $\pm 1$ for 
each $b$ and $a$. \hfill\break  
\item $U$ is the extension of a complex projective representation.  
That is, the matrix elements $\langle f|U_a|f^{\prime}\rangle$ are complex 
and $\omega(f;b,a)=\omega(b,a)$ is independent of $|f\rangle$ and is 
a complex phase. \hfill\break 
\item $U$ is the tensor product of a real projective representation and 
a quaternionic phase.  That is, there exists a decomposition 
$U_a=U_a^{\cal B}\sum_f|f\rangle \sigma_a \langle f|$, where the unitary 
operator $U_a^{\cal B}$ has real matrix elements, $\sigma_a$ is a 
quaternionic phase, and $U_{ba}^{\cal B}=\pm U_b^{\cal B}U_a^{\cal B}$ for 
all $b$ and $a$. 
\end{enumerate}

{}From the point of view of the Structure Theorem, case (1) corresponds to 
the only possibility allowed by the strong definition of quaternionic 
projective representations, as demonstrated earlier by Emch (1963, 1965), 
while case (2) corresponds to an embedding of a complex projective 
representation in quaternionic Hilbert space, the consideration of which was 
my motivation for proposing the weak definition. 
Specializing the Structure Theorem to a complex Hilbert space, where  
case (3) cannot be realized, we see that in complex Hilbert 
space the weak projective representation defined above  
{\it implies} 
the strong projective representation; hence no generality
is lost by starting from the strong definition, as in Bargmann's paper.  

The second part of the Tao--Millard paper is a proof, by real analysis 
methods, of a Corollary to the structure theorem, stating 

{\bf Corollary 1:}  {\it Any multicentral projective representation of a 
connected Lie group is central}.  

This thus solved the question of the relation of centrality to 
multicentrality that I raised in my paper {\sf R63}.  

Subsequent to this work, I had an exchange with Gerard Emch in the {\it Journal 
of Mathematical Physics}  debating the merits of the strong and 
weak definitions.  After a visit to Gainesville where we reconciled 
differing notations, we wrote a joint paper 
(Adler and Emch, 1997, {\sf R64}) clarifying the situation, 
and reexpressing the strong and weak definitions 
in the language and notation often employed in 
mathematical discussions of projective group representations.  

\bigskip

\section*{Trace Dynamics and Emergent Quantum Theory}

My  work on emergent quantum theory arose from the merging of two lines 
of thought.  The first line of thought 
arose from answering the question of whether 
quaternionic quantum mechanics ameliorates the measurement problem of 
standard quantum mechanics; the answer is ``no'', because quaternionic 
quantum theory still has a unitary time evolution, and so the usual  
problems persist.  However, in the course of working this through I read 
some of the literature on the measurement problem in standard quantum 
theory, and came away convinced that there were real issues to be addressed. 
The second line of thought arose from my attempts to 
construct quaternionic quantum field 
theories.  I found that the canonical quantization method could not be 
extended to the quaternionic case, and so I had to resort to an  
alternative formalism, 
which I variously called ``generalized quantum dynamics'', ``total trace 
dynamics'', or finally, simply ``trace dynamics'', to generate operator 
equations without ``quantizing'' a classical theory.  This was done  
by using a variational principle based on a Lagrangian constructed 
as a trace of noncommuting operator variables, making systematic use 
of cyclic permutation under the trace operation. These ideas were  
developed in the paper Adler (1994b) and were described in 
Chapter 13 of my 1995 book; in Chapter 14, I suggested that the 
nonlinearity of trace dynamics could make it relevant for resolving the 
measurement problem in quantum theory.  However, the problem of relating 
the trace dynamics 
formalism to the standard canonical formalism of complex quantum 
field theory remained unsolved.  

One of the questions I had posed to Andrew Millard was that of better  
understanding trace dynamics, in the hope of finding a connection to   
standard quantum theory. After I arrived in Aspen in the summer of 
1995, Andrew sent me a memo containing his discovery that in trace dynamics  
with a Weyl symmetrized Hamiltonian and noncommuting  
boson degrees of freedom $q_r, \, p_r$, 
the operator $\sum_r[q_r,p_r]$ is conserved. I soon found that the  
generalization to include fermions is the conserved operator that we 
denoted by $\tilde C$, defined by 
\begin{equation}
\tilde C \equiv \sum_{r,\,{\rm boson}}[q_r,p_r] - \sum_{r, \, {\rm fermion}}
\{q_r,p_r\}~~~,\nonumber\\
\end{equation}
and that this operator is conserved as long as the trace Hamiltonian has 
no fixed operator coefficients, which is equivalent to saying the the 
trace Hamiltonian has a global unitary invariance.  It then seemed natural 
to suggest that the equipartitioning of $\tilde C$ in a statistical 
thermodynamical treatment would provide the missing connection between 
trace dynamics and standard quantum mechanics.  

The implementation of this 
idea was published in Adler and Millard (1996b), and I developed it 
further over the following years with many collaborators, as described 
in Sec. 5 of the ``Introduction and Overview'' that opens my 2004 book on       
emergent quantum theory.  This book, which is set within the framework  
of complex Hilbert space, gives a complete, self-contained 
development of trace dynamics as a  
(noncommutative)  dynamics underlying  quantum theory.  From the  
statistical mechanics of this underlying theory there emerge, in   
a mutually complementary way, both the unitary and the nonunitary parts of 
orthodox quantum theory. The unitary part of quantum theory (the 
canonical algebra and the Heisenberg representation time evolution of 
operators) comes from an application of generalized equipartition theorems  
in the statistical thermodynamics of trace dynamics.  The nonunitary part of 
quantum theory, in the form of stochastic state vector reduction models  
from which the Born rule for probabilities can be derived, comes from the 
Brownian motion corrections to this thermodynamics.   
Thus, trace dynamics provides a unified framework from which both the 
unitary dynamics of quantum systems, and the nonunitary evolution describing  
state vector reduction associated with measurements, emerge in a natural 
way.  

Although quantum mechanics and quantum field 
theory have been the undisputed basis for all progress in fundamental 
physics during the last 80 years, the extension of the current 
theoretical frontier to Planck scale physics, and recent enlargements 
of our experimental capabilities, 
may make the 21st century the period in which possible limits of quantum 
theory will be probed.  My 2004 book suggests a concrete framework  
for  exploration  
of the proposition that quantum mechanics may not be the final layer 
of fundamental theory.  It also addresses the phenomenology 
of modifications to quantum theory, specifically as implemented 
through stochastic 
additions to the Schr\"odinger equation.  I have continued with these 
phenomenological studies since completion of the book; my most 
recent papers  (Bassi, 
Ippoliti, and Adler, 2005; Adler, Bassi, and Ippoliti, 2005; Adler, 2005) 
have dealt with analyzing possible tests of stochastic  
localization theories in nanomechanical oscillator 
and gravitational wave detector experiments.

\section*{References for Chapter 9}

\begin{itemize}

\item Adler, S. L. (1979).  Algebraic Chromodynamics.  {\it Phys. Lett. B} 
{\bf 86}, 203-205. 

\item Adler, S. L. (1980).  Quaternionic Chromodynamics as a Theory of Composite 
Quarks and Leptons. {\it Phys. Rev. D} {\bf 21}, 2903-2915. 

\item Adler, S. L. (1988).  Scattering and Decay Theory for Quaternionic Quantum 
Mechanics, and the Structure of Induced $T$ Nonconservation.  {\it Phys. 
Rev. D} {\bf 37}, 3654-3662. 

\item Adler, S. L. (1989).  A New Electroweak and Strong Interaction Unification 
Scheme.  {\it Phys. Lett. B} {\bf 225}, 143-147. 

\item Adler, S. L. (1994a).  Composite Leptons and Quarks Constructed as Triply 
Occupied Quasiparticles in Quaternionic Quantum Mechanics.  {\it Phys. Lett. 
B} {\bf 332}, 358-365. 

\item Adler, S. L. (1994b).  Generalized Quantum Dynamics.  {\it Nucl. Phys. B} 
{\bf 415}, 195-242. 

\item Adler, S. L. (1996) {\sf R63}. 
Projective Group Representations in Quaternionic 
Hilbert Space  {\it J. Math. Phys.} {\bf 37}, 2352-2360. 

\item Adler, S. L. (2005).  Stochastic Collapse and 
Decoherence of a Non-Dissipative 
Forced Harmonic Oscillator.  {\it J. Phys. A: Math. Gen.} {\bf 38}, 
2729-2745. 

\item Adler, S. L. and J. Anandan (1996) {\sf R61}.  Nonadiabatic Geometric Phase in 
Quaternionic Hilbert Space.  {\it Found. Phys.} {\bf 26}, 1579-1589. 
  
\item Adler, S. L., A. Bassi, and E. Ippoliti (2005).  Towards Quantum 
Superpositions of a Mirror: An Exact Open Systems Analysis -- Calculational 
Details.  {\it J. Phys. A: Math. Gen.} {\bf 38}, 2715-2727. 

\item Adler, S. L. and G. G. Emch (1997) {\sf R64}. 
A Rejoinder on Quaternionic Projective 
Representations.  {\it J. Math. Phys.} {\bf 38}, 4758-4762. 

\item Adler, S. L. and A. C. Millard (1996a) {\sf R62}.  Coherent States in Quaternionic 
Quantum Mechanics.  {\it J. Math. Phys.} {\bf 38}, 2117-2126. 

\item Adler, S. L. and A. C. Millard (1996b).  Generalized Quantum Dynamics as 
Pre-Quantum Mechanics.  {\it Nucl. Phys B} {\bf 473}, 199-244. 

\item Bargmann, V. (1954).  On Unitary Ray Representations of Continuous Groups.  
{\it Ann. Math.} {\bf 59}, 1-46. 

\item Bassi, A., E. Ippoliti, and S. L. Adler (2005).  Towards Quantum 
Superpositions of a Mirror:  An Exact Open Systems Analysis.  {\it Phys. Rev. 
Lett.} {\bf 94}, 030401. 

\item Emch, G. (1963).  M\'ecanique Quantique Quaternionienne et Relativit\'e 
Restreinte.  I, {\it Helv. Phys. Acta} {\bf 36}, 739-769; 
II, {\it Helv. Phys. Acta} {\bf 36}, 770-788. 

\item Emch, G. (1965). Representations  
of the Lorentz Group in Quaternionic Quantum Mechanics (presented at the 
summer 1964 Lorentz Group Symposium), in {\it Lectures in Theoretical 
Physics} Vol. VIIA, W.E. Brittin and A. O. Barut, eds. 
(University of Colorado Press, 
Boulder), pp. 1-36. 

\item Harari, H. (1979).  A Schematic Model of Quarks and Leptons.  {\it 
Phys. Lett. B} {\bf 86}, 83-86. 

\item Peres, A. (1979).  Proposed Test for Complex versus Quaternion Quantum 
Theory.  {\it Phys. Rev. Lett.} {\bf 42}, 683-686.

\item Shupe, M. A. (1979)  A Composite Model of Leptons and Quarks.  
{\it Phys. Lett. B} {\bf 86}, 87-92. 

\item Tao, T. and A. C. Millard (1996).  On the Structure of Projective Group 
Representations in Quaternionic Hilbert Space. {\it J. Math. Phys.}
 {\bf 37},  5848-5857. 

\item Weinberg, S. (1995) {\it The Quantum Theory of Fields}, Volume I: Foundations
(Cambridge University Press, Cambridge), Sec. 2.7. 

\end{itemize}

\chapter*{10. Where Next?}

\markboth{Adventures in Theoretical Physics}{Where Next?}

In looking back at my work, I see one pattern that is repeated over 
and over.  Many of the most interesting research results  that I have 
obtained were unanticipated consequences of other, quite different 
research programs.  In the course of detailed calculations, or 
speculative explorations, I noticed something that seemed worth pursuing, 
even though tangential to my original motivations, and this new direction 
ended up being of much greater interest.   This happened with my calculations 
of weak pion production, which led as spin-offs to the forward lepton theorem, 
the neutrino sum rule, and soft pion theorems.  It happened again with my 
exploration of gauging of the axial-vector current as an explanation 
for the muon mass, which led to anomalies.  My interest in an eigenvalue  
condition in QED led to the calculation of photon splitting, and later 
on to an improved method for analyzing collider data. My attempts at a 
composite  graviton led to an investigation of Einstein gravity 
as a symmetry breaking effect.  
My interest in the (spurious) Argonne threshold events induced  me 
to extend my weak pion work to neutral currents, which contributed to the 
first unique determination of the electroweak couplings by Abbott and Barnett.   
My  attempt to relate  monopole background fields to  
confinement played a role in the multimonopole existence proof of Taubes. 
My computational experience  
in solving effective action confinement models led to 
overrelaxation as an acceleration method for Monte Carlo.  
And most recently, my interest in composite models for quarks and 
leptons led 
to a long exploration of the fundamentals of quantum theory, first through 
my study of quantum theory in quaternionic Hilbert space, and growing out 
of that, through my development of trace dynamics as a possible pre-quantum 
theory.  

I think this pattern is no accident, but rather a reflection of 
my guiding philosophy in doing research, which 
has been that it is more important to start  
somewhere, even with a speculative idea or an apparently routine calculation, 
than to sit around waiting for an ``important'' idea.  Once immersed in the 
nitty-gritty of an investigation, things have a way of appearing, that  
often lead off in very fruitful directions.  
So given this, when I look ahead, I can only say the following: I have some 
rough ideas as to where I would like to start in new 
explorations in fundamental theory and particle phenomenology,  
but I cannot say where these may 
ultimately lead, in the course of my continuing adventures in theoretical 
physics.  


\end{document}